\documentclass[fleqn,usenatbib,useAMS]{mnras}

\usepackage{graphicx}	
\usepackage{amsmath}	
\usepackage{amssymb}	
\usepackage{multicol}        
\usepackage{bm}		
\usepackage{pdflscape}	
\usepackage{epsfig}
\usepackage{subcaption}
\usepackage[utf8]{inputenc}
\usepackage[export]{adjustbox}
\usepackage{wrapfig}

\def\aj{AJ}
\def\araa{ARA\&A}
\def\apj{ApJ}
\def\apjl{ApJL}
\def\apjs{ApJS}
\def\ao{Appl.~Opt.}
\def\aap{A\&A}
\def\aaps{A\&AS}
\def\mnras{MNRAS}
\def\memras{MmRAS}
\def\nar{New A Rev.}
\def\pasp{PASP}
\def\nat{Nature}
\def\memsai{Mem.~Soc.~Astron.~Italiana}

\catcode`\@=11
\def\gsim{\ifmmode{\mathrel{\mathpalette\@versim>}}
    \else{$\mathrel{\mathpalette\@versim>}$}\fi}
\def\lsim{\ifmmode{\mathrel{\mathpalette\@versim<}}
    \else{$\mathrel{\mathpalette\@versim<}$}\fi}
\def\@versim#1#2{\lower 2.9truept \vbox{\baselineskip 0pt \lineskip
    0.5truept \ialign{$\m@th#1\hfil##\hfil$\crcr#2\crcr\sim\crcr}}}
\catcode`\@=12

\newcommand{\beq}{\begin{equation}}
\newcommand{\eeq}{\end{equation}}

\usepackage[T1]{fontenc}
\usepackage{ae,aecompl}

\usepackage{newtxtext,newtxmath}

\begin{document}

\title[ULIRG]
{Tracing the Evolution of Ultraluminous Infrared Galaxies into Radio Galaxies with Low Frequency Radio Observations}
\author[Nandi et al.]
{S. Nandi$^{1,2}$ \thanks{e-mail: snandi@ncra.tifr.res.in}
M. Das$^{2}$, K.S.Dwarakanath$^{3}$
\\
 $^1$National Centre for Radio Astrophysics, TIFR, Ganeshkhind, Pune 411 007\\
 $^2$Indian Institute of Astrophysics, Koramangala II Block, Bangalore 560034, India\\
 $^3$Raman Research Institute, C. V. Raman Avenue, Sadashivanagar, Bangalore 560080, Karnataka, India\\
 }


\pagerange{\pageref{firstpage}--\pageref{lastpage}} \pubyear{}

\maketitle

\label{firstpage}

\begin{abstract} 
We present radio observations of ultraluminous infrared galaxies (ULIRGs) using the Giant Metrewave Radio Telescope (GMRT) and combine them with archival multi-frequency observations to understand whether ULIRGs are the progenitors of the powerful radio loud galaxies in the local Universe. ULIRGs are characterized by large infrared luminosities ($L_{IR}>$10$^{12}$L$\odot$), large dust masses ($\sim10^{8}M_{\odot}$) and vigorous star formation (star formation rates $\sim$10-100 $M_{\odot}~$yr$^{-1}$). Studies show that they represent the end stages of mergers of gas-rich spiral galaxies. Their luminosity can be due to both starburst activity and active galactic nuclei (AGN). We study a sample of 13 ULIRGs that have optically identified AGN characteristics with 1.28~GHz GMRT observations. Our aim is to resolve any core-jet structures or nuclear extensions and hence examine whether the ULIRGs are evolving into radio loud ellipticals. Our deep, low frequency observations show marginal extension for only one source. However, the integrated radio spectra of 9 ULIRGs show characteristics that are similar to that of GPS/CSS/CSO/young radio sources. The estimated spectral ages are 0.4 to 20 Myr and indicate that they are young radio sources and possible progenitors of radio galaxies. Hence, we conclude that although most ULIRGs do not show kpc scale extended radio emission associated with nuclear activity, their radio spectral energy distributions do show signatures of young radio galaxies.
\end{abstract}

\section{Introduction}
\label{section1}

Ultraluminous infrared galaxies (ULIRGs) were first detected as extremely bright sources of infrared emission ($L_{IR}>10^{12}L_{\odot}$) in the images from the Infrared Astronomical Satellite (IRAS) \citep{sanders.mirabel.1996}. Later studies revealed that they are gas rich merger remnants that have large amounts of dust \citep{rigopoulou.etal.1999}. Photometric studies show that the light profiles of ULIRGs are similar to that of ellipticals \citep{genzel.etal.2001,dasrya.etal.2006}. These observations combined with simulations \citep{delucia.2006} indicate that ULIRGs represent the final stages of the merging process of two gas rich galaxies of comparable masses into an elliptical galaxy. 
Although the major energy source in these systems appears to be the intense star formation associated with the galaxy merger process \citep{veilleux.etal.2009}, active galactic nuclei (AGN) have been detected in nearly half of all ULIRGs that have been observed \citep{alonso.etal.2006,nardini.etal.2010,vivian.etal.2012}. In some cases ULIRGs are predominately powered by central AGNs \citep{mao.etal.2014} as well. The radiation from starburst and AGN activity is reprocessed by the surrounding dust, which makes them extremely bright at mid and far infrared wavelengths. Hence, studying AGNs in ULIRGs is an opportunity to understand the transformation of gas rich mergers into powerful quasars at low redshifts \citep{2017ApJ...843...18V}. It is also a way to understand how winds and outflows in galaxy mergers enrich the circum-galactic medium (CGM) surrounding galaxies \citep{rupke.etal.2005,anderson.etal.2016}. However, the AGNs in ULIRGs are optically obscured by large amounts of dust; hence radio and X-ray wavelengths are the most effective wavelengths at which to look into their evolving 
centers \citep{2003A&A...409..115N, 2005ApJ...633..664T}. 

As an ULIRG evolves, the nuclei of the individual galaxies come closer. Hence it is not surprising that double nuclei are often observed in ULIRGs. One of the earliest observations of such sources was the X-ray observation of the nearest ULIRG NGC~6240 \citep{2003ApJ...582L..15K}. Since then, several optical and infrared observations of ULIRGs have revealed double nuclei \citep{2006NewAR..50..720D}. Double nuclei are also detected in unresolved ULIRGs using optical spectroscopy where they can appear as double peaked emission lines in the optical spectra. However, sometimes the double peaked AGN emission lines (DPAGN) can be due to rotating nuclear disks, AGN outflows or collimated jets \citep{2006MmSAI..77..733K, 2019ApJ...871..249K, 2019MNRAS.484.4933R, 2020arXiv201206290N}. The final stage of the evolution is the merging of the supermassive black holes (SMBHs) \citep{khan.etal.2013} and the triggering of radio jets \citep{sajina.etal.2007}. This evolutionary path is similar to that of the powerful high redshift quasars. Hence, ULIRGs are often thought to represent the low redshift counterparts of high redshift radio galaxies  \citep{2012MNRAS.422.1453N}. Submillimeter and far-IR surveys indicate that submillimeter bright sources (SMGs) are the high redshift ($z\gtrsim1$) equivalents of ULIRGs. Though these SMGs or distant ULIRGs are  more numerous at high redshift, study of their central region is not easy as they are faint and distant. However, the study of local universe ULIRGs can help us to understand the physical processes of these distant objects \citep{2018MNRAS.475.2097C, 2006ApJ...641L..17F}.

Galaxy mergers induce  intense starburst activity along with quasar mode accretion  of mass onto a pre-existing quiescent black hole \citep{2016AN....337...36C}. It has also been shown that the triggering of radio loud AGN and relativistic jet emission can be associated with merger events irrespective of redshift or AGN power \citep{2015ApJ...806..147C}. The most extreme ULIRG, IRAS 00182$-$7112 provides strong observational evidence of this process \citep{2012MNRAS.422.1453N}. The integrated radio spectrum of IRAS 00182$-$7112 shows a turnover at low frequencies. Very long baseline interferometry (VLBI) imaging of IRAS 00182$-$7112 revealed a 1.7 kpc size classical double-lobed radio galaxy at the center of the galaxy. This represents the transitional stage of a Gigahertz Peaked Spectrum (GPS) source (source size $\leq$ 1 kpc  and radio spectrum turnover frequency $\sim$ 1 GHz) into a Compact Steep Spectrum (CSS) source (source size $\sim$ 1 to 10 kpc and radio spectrum turnover frequency $\lsim$ 500 MHz).  
GPS and CSS sources with small-scale miniature structures similar to radio galaxies are referred to as compact symmetric objects (CSOs) \citep{1994ApJ...432L..87W, 2017MNRAS.465.4772R}. These sources are generally associated with low-redshift galaxies \citep{2015ApJ...809..168C, 2014ApJ...780..178M}. Thus, in general it is thought that GPS, CSS and CSO sources represent the  early phase of radio galaxy evolution \citep{1998PASP..110..493O, 2009AN....330..120F}. However, a dense medium may impede their normal growth and make the jet frustrated \citep{2015ApJ...809..168C}.

In this paper we present new observations of a sample of ULIRGs at 1.28 GHz with the Giant Metrewave Radio Telescope (GMRT). Our main aim is to obtain a clearer picture of the evolution of ULIRGs into radio galaxies by investigating how the radio emission from the centers of these galaxies changes with frequency. We have supplemented our radio observations with available archival radio data to study their spectral energy distribution over a wide radio frequency band. To understand the merger activity in our sample, we also studied existing optical archival data from the Sloan Digital Sky Survey (SDSS) Data Release 14 (DR14)\footnote{https://www.sdss.org/dr14/} and the Panoramic Survey Telescope and Rapid Response System (Pan-STARRS)\footnote{https://panstarrs.stsci.edu/}. In Section \ref{sec:sample_setection} we describe the sample selection. The radio data and optical spectra analysis are discussed in Sections \ref{sec:obs_data} and \ref{sec:opt_data} respectively. The details  of the spectral age estimations are given in Section \ref{sec:spect_aging}. The results of individual sources are presented in Section \ref{sec:result}. We have discussed the results in Section \ref{sec:discussion} and presented our conclusions in Section \ref{sec:Conclusions}.

\section{SAMPLE SELECTION} 
\label{sec:sample_setection} 

The parent sample for the ULIRGs in our study is the IRAS 1 Jy sample which consists of 118 ULIRGs drawn from the IRAS Faint Source Catalog \citep{1998ApJS..119...41K, 1990IRASF.C......0M}. 
\citet{2003A&A...409..115N} further studied 83 sources of the parent sample at 15 GHz with the VLA. Their nuclear radio core detection rate in Seyfert and LINER type ULIRGs is $\sim$ 75\%  with a flux limit of $\sim$0.8 mJy. Our sub-sample consists of 13 ULIRGs that have optical spectral signatures of AGN activity and have been detected by \citet{2003A&A...409..115N} at 15 GHz.
The optical spectroscopy of these ULIRGs show the nuclear emission lines characteristic of Seyfert 1, Seyfert 2 or LINER type AGN \citep{1999ApJ...522..113V}. In addition we have studied one more ULIRG, IRAS 13305$-$1739, which is not included in \citet{2003A&A...409..115N}, but has a Seyfert 2 type nucleus \citep{2002ApJS..143..315V}. The 15 GHz VLA observations have an angular resolution of 150 mas. At this resolution \citet{2003A&A...409..115N} could not confirm that AGN are the dominant power source in these systems. However, the compact high power radio cores, association of AGN-type nuclei, 
the low soft X-ray (0.5-2 keV) to nuclear monochromatic radio luminosity ratio (at 15 GHz), the uncorrelated radio power of the nuclei and the $H_{\alpha}$ luminosity indicate that the radio emission origin is AGN activity rather than starburst activity, or radio supernovae \citep{2003A&A...409..115N}.
The sample is listed in Table \ref{table1}. Due to the time constraints during observations, we could not observe 2 of the sources, IRAS 12112$+$0305 and IRAS 17179$+$5444. However, they have enough archival low frequency radio data to complete our low frequency study. Hence, we have included these sources for our analysis as well.

 \section{Radio Data Analysis} 
 \label{sec:obs_data}
 We carried out GMRT radio continuum observations to image the radio emission 
 from 11 ULIRGs. The details of these observations are 
 given in Table \ref{table1}. All the sources were observed at 1.28 GHz
 for 2 hours each. 
 Deeper follow-up radio continuum observations (6 hrs for each source) were carried out for the sources IRASF 13305-1739, IRAS 15001+1433 and IRASF 23389+0300. We used the NRAO Astronomical Image Processing System (AIPS) package for data reduction \citep{2003ASSL..285..109G}. The flux density calibrator was observed at the beginning and at the end of each observing run, while the phase calibrators were observed after each observation of the  target sources. Strong radio frequency interference (RFI),  non-working antennas and other obvious defects were edited out  after visual inspection. The AIPS tasks UVFLG, TVFLG, SPFLG, FLGIT and IBLEAD were also used to remove the bad data. Around 20$\%$ data were flagged out. The amplitude calibrators are 3C48, 3C 286 and 3C 147 while the flux density calibration at 1.28 GHz uses the scale of \citet{2013ApJS..204...19P}. Different point sources are used for phase calibration (see Table \ref{table1}). The flux density of the phase calibrator was computed using the task GETJY. The flux density of the phase calibrator were cross checked with values given in the VLA calibrator list. We use one of the flux density calibrators as a bandpass calibrator. Due to the large field of view we applied the multi facets imaging process. For each source 25 facets were created. Also,  5 to 7 cycles of phase only self calibration with one amplitude calibration in the final round were carried out to produce the final images. To obtain the diffuse extended structure the  robustness parameter was set between 2 to 3. Primary beam correction was applied using the task PBCOR. The flux densities were measured from the fitted Gaussian components using task JMFIT. 
 The estimated flux densities and rms noise of the 11 ULIRGs at 1.28 GHz from these observations are given in Table \ref{table2}.
 
\begin{table*}
 \centering
  \caption{The ULIRGs observed with the GMRT at 1.28 GHz.}
   \begin{tabular}{@{}lcccccccccccccc@{}}
  \hline
     Name  &RA           & Dec		        & $z$      &Obs. date	  &Primary& Phase cal& Beam size& PA& T                            \\
           &hh:mm:ss.ss  & dd:mm:ss.ss		&          &    	  &        &        & ($\arcsec \times \arcsec$) &($^\circ$) & (hr) & \\
      (1)  &(2)	  &(3)			&(4)       &(5)	          & (6)    & (7)      & (8)          & (9) & (10)\\
 \hline
IRAS 00188$-$0856   & 00:21:26.50  &	$-$08:39:26.00	&  0.128  & 26-08-2007 &3C286,3C147     & 0022+002         &	4.84$\times$4.84&00        &1.25&   \\
IRAS 01572$+$0009   & 01:59:50.20  &	$+$00:23:41.00	&  0.163  & 02-09-2007 &3C48, 3C147     & 1634+627         &	5.14$\times$2.14&43        &1.70&   \\
IRAS 13305$-$1739   & 13:33:16.50  &    $-$17:55:11.00  &  0.148  &  19-07-2015&3C286           & 1351-148         &    4.22$\times$3.19 & 38      &4.00&   \\
IRAS 14070$+$0525   & 14:09:31.20  &    $+$05:11:31.00  &  0.264  &  26-08-2007&3C286, 3C147    & 1354-021         &    8.67$\times$7.01&-53       &0.81&   \\ 
IRAS 14394$+$5332   & 14:41:04.40  &    $+$53:20:09.00  &  0.104  &  27-08-2007&3C286           & 1445+099         &    9.24$\times$4.93&8         &1.70&   \\
IRAS 15001$+$1433   & 15:02:31.90  &    $+$14:21:35.00  &  0.162  &  18-07-2015&3C286, 3C48     & 1445+099         &    2.81$\times$2.22&70       &4.00&   \\
IRAS 16156$+$0146   & 16:18:09.40  &    $+$01:39:21.00  &  0.132  &  26-08-2007&3C286, 3C147    & 1557-000         &    5.13$\times$3.88&-23       &0.83&   \\
IRAS 17028$+$5817   & 17:03:41.90  &    $+$58:13:45.00  &  0.106  &  26-08-2007&3C286, 3C147     & 1634+627         &     7.84$\times$2.93&13       &0.83&   \\
IRAS 17044$+$6720   & 17:04:28.40  &    $+$67:16:23.00  &  0.135  &  02-09-2007&3C48, 3C147     & 1634+627         &     6.38$\times$1.95&-80      &1.70&   \\
IRAS 23060$+$0505   & 23:08:33.90  &    $+$05:21:30.00  &  0.173  &  26-08-2007&3C286, 3C147    & 2136+006         &     4.64$\times$4.64&0        &1.06&   \\
IRAS 23389$+$0300   & 23:41:30.30  &    $+$03:17:26.00  &  0.145  & 28-12-2015 &   3C48    & 0022+002         &    3.45$\times$1.96&59       &4.00&   \\
IRAS 12112$+$0305$\dagger$   & 12:13:46.00  &   $+$02:48:38.00  &  0.073  &   Archival data        &                &                  &                    &          & &    \\
IRAS 17179$+$5444$\dagger$   & 17:18:54.40  &    $+$54:41:48.00  &  0.147  &  Archival data        &                &                  &                     &         & &   \\
\hline 
\end{tabular}
\\
Column 1: source name; Columns 2 and 3: right ascension and declination of the optical objects in J2000 co-ordinates; Column 4: redshift; Column 5: the dates of GMRT observations; Columns 6 and 7: the names of the calibrators used for each observation; Columns 8 and 9: the major and minor axes of the restoring beam in arcsec and its position angle (PA) in degrees; Column 10: observing time on source in units of hr. $\dagger$For these sources we did not get any GMRT time.
\label{table1}
\end{table*}
\begin{figure*}
\vbox{
   \hbox{
      \psfig{file=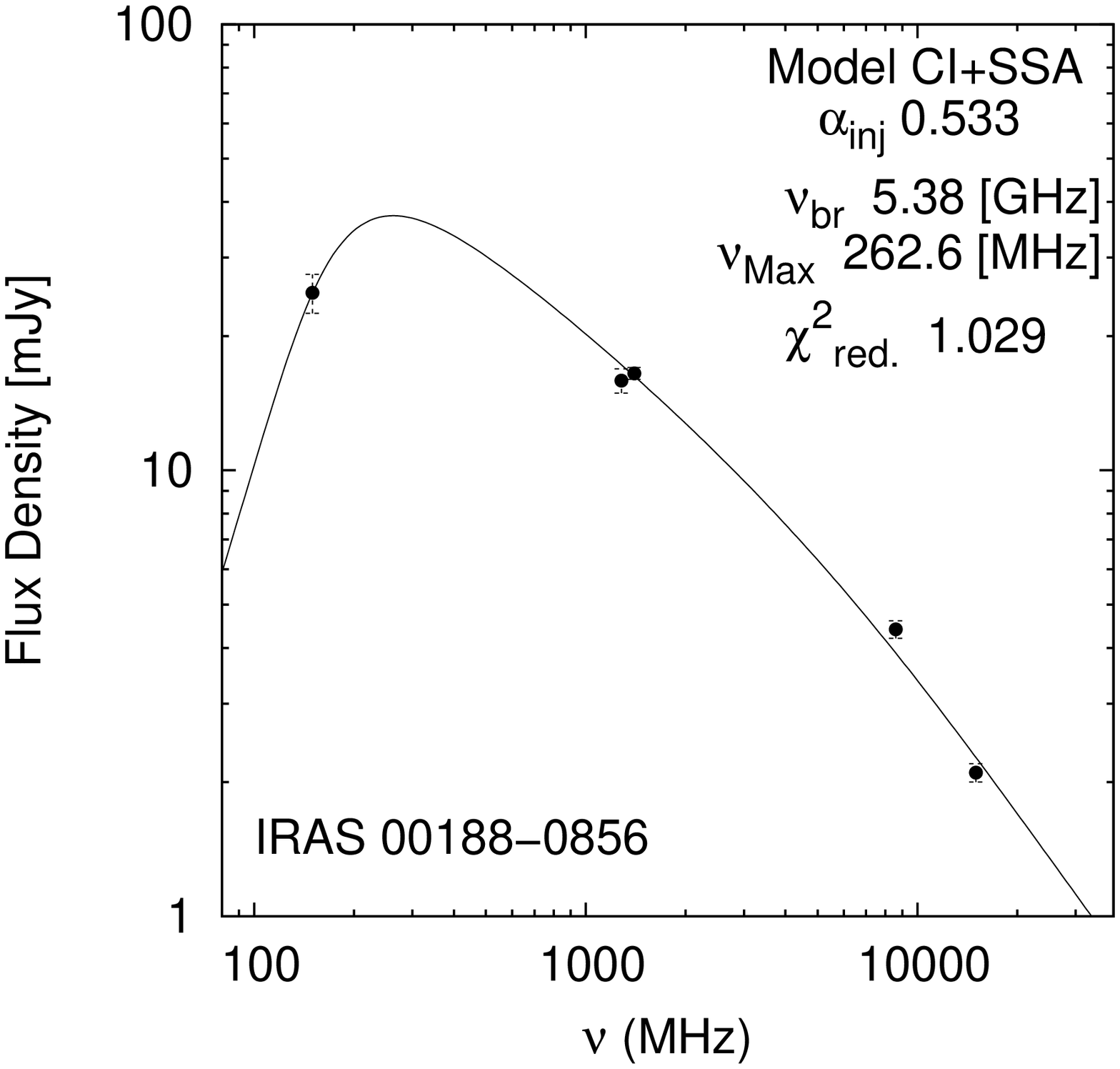,width=2.4in,angle=270}%
      
      \psfig{file=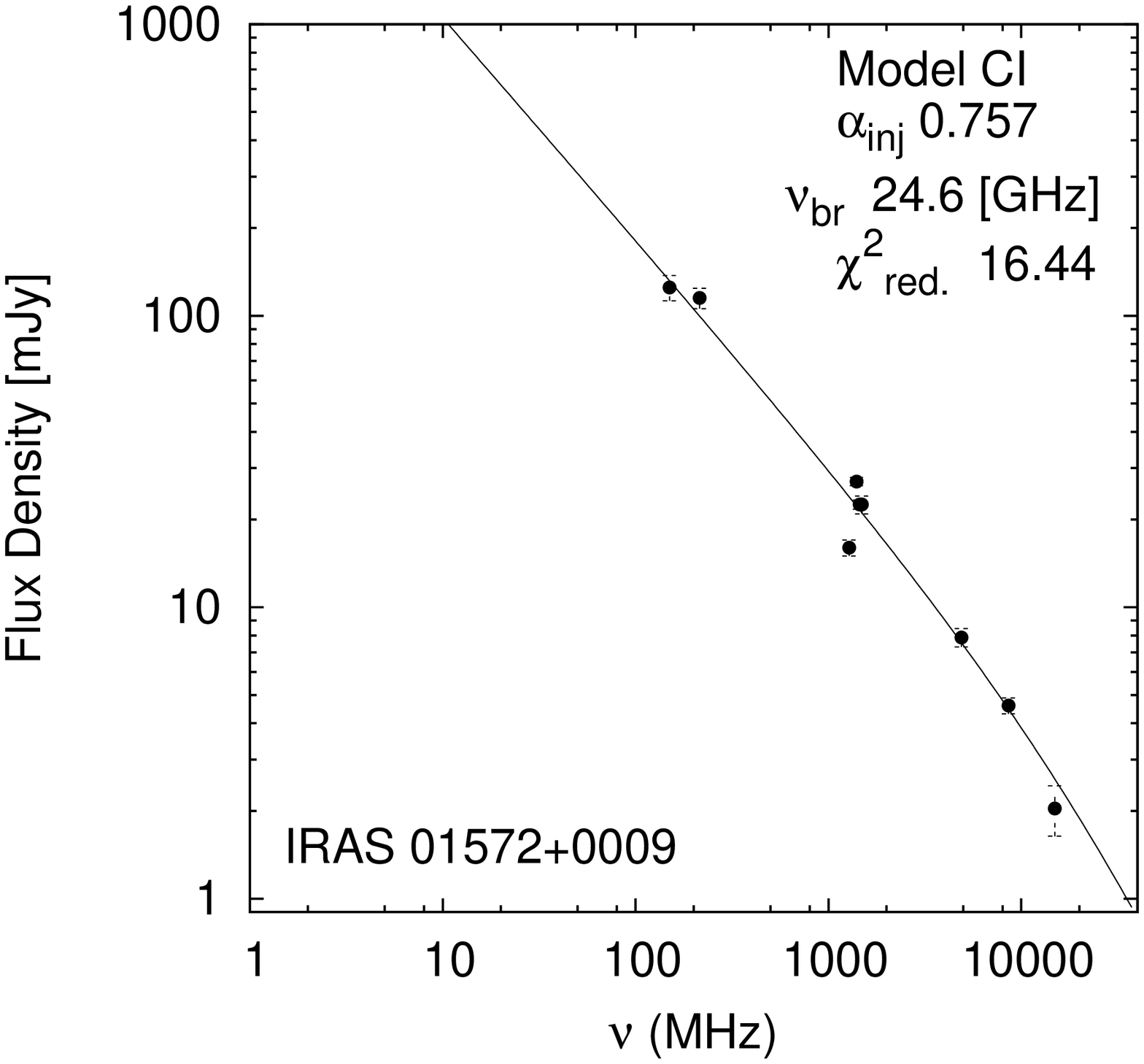,width=2.4in,angle=270}%
    }
    \hbox{
    \vspace{0.1cm}  
      \psfig{file=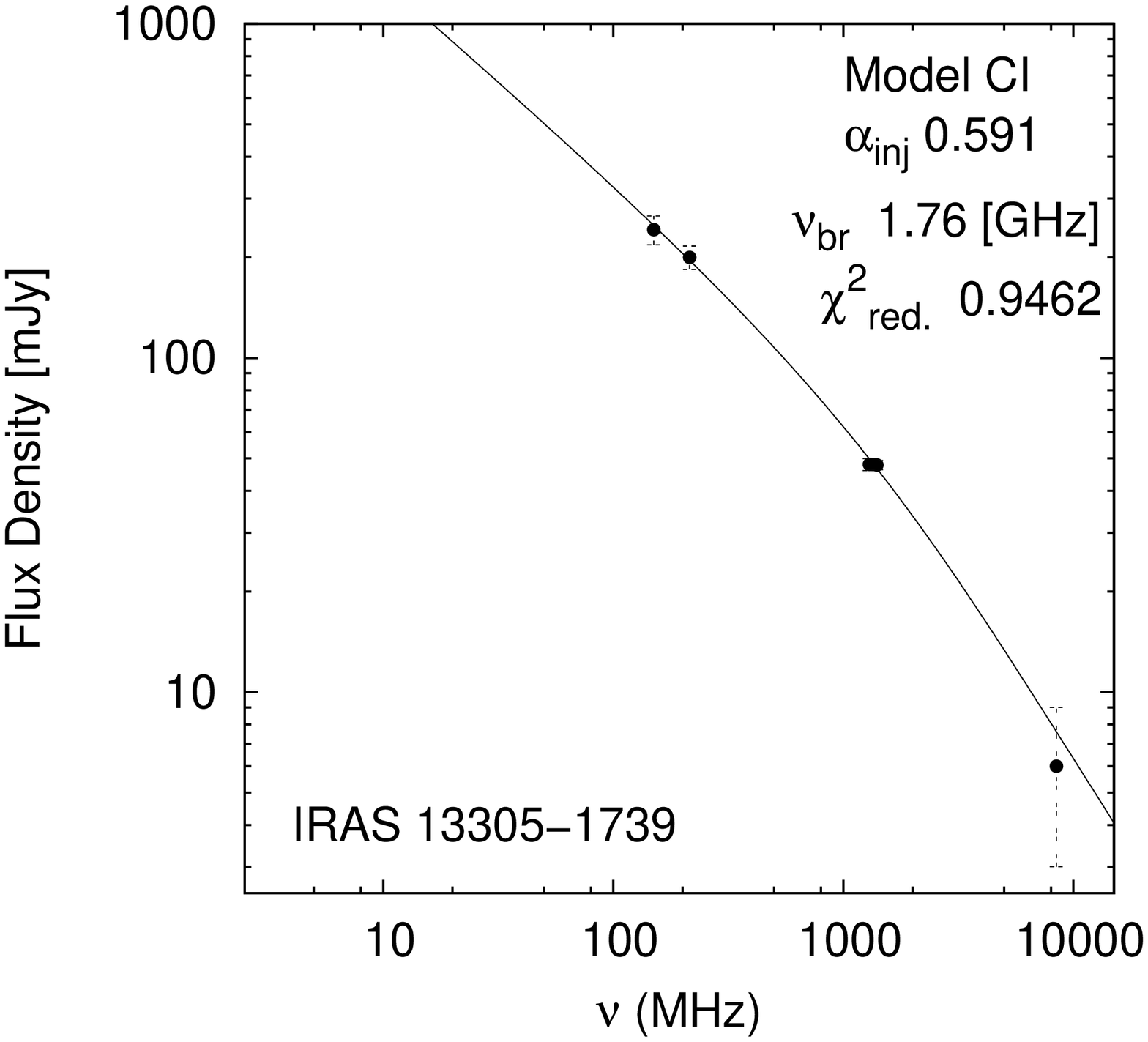,width=2.4in,angle=270}%
        
      \psfig{file=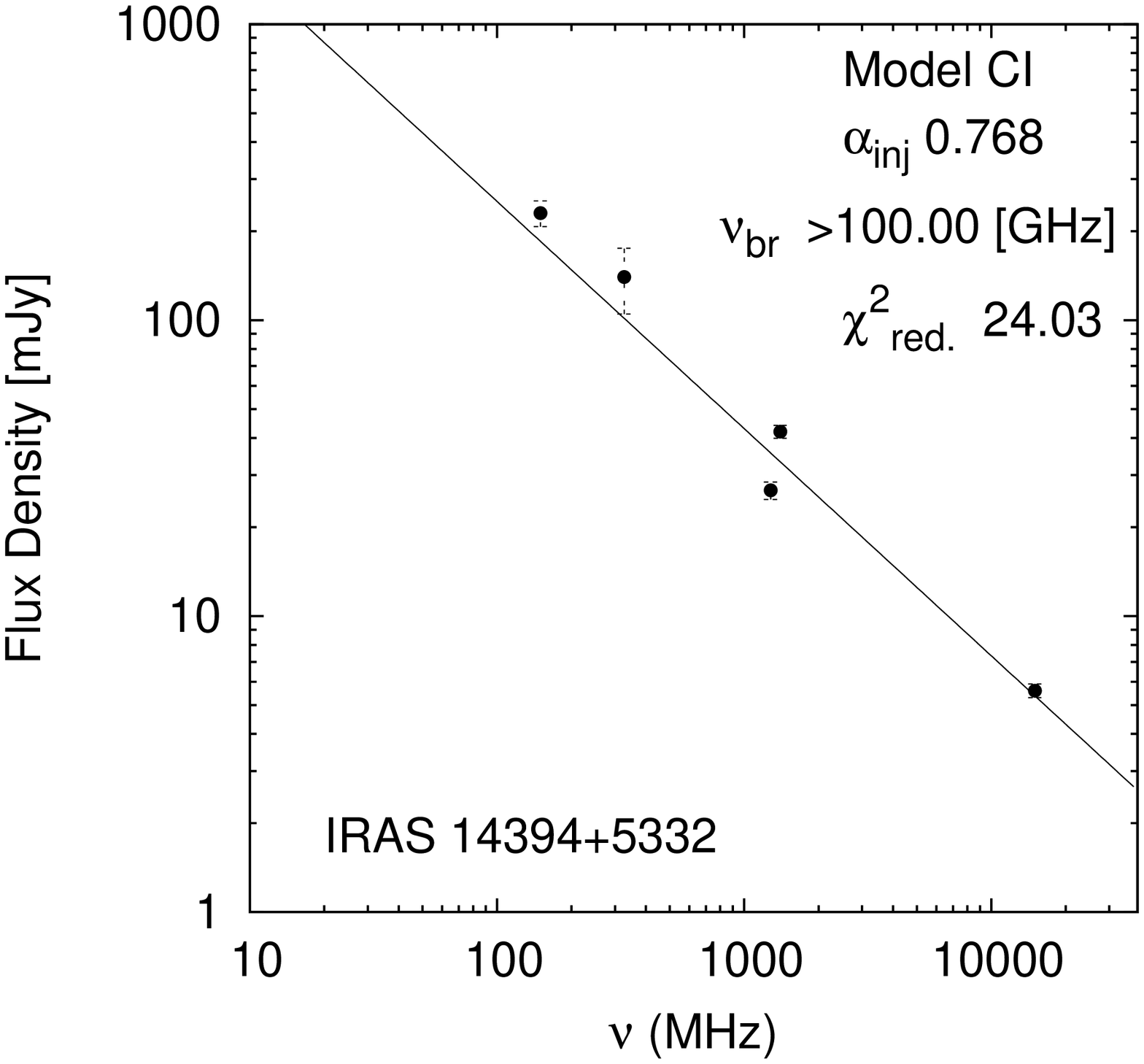,width=2.4in,angle=270}%
      }
      \vspace{0.1cm}
    \hbox{
      \psfig{file=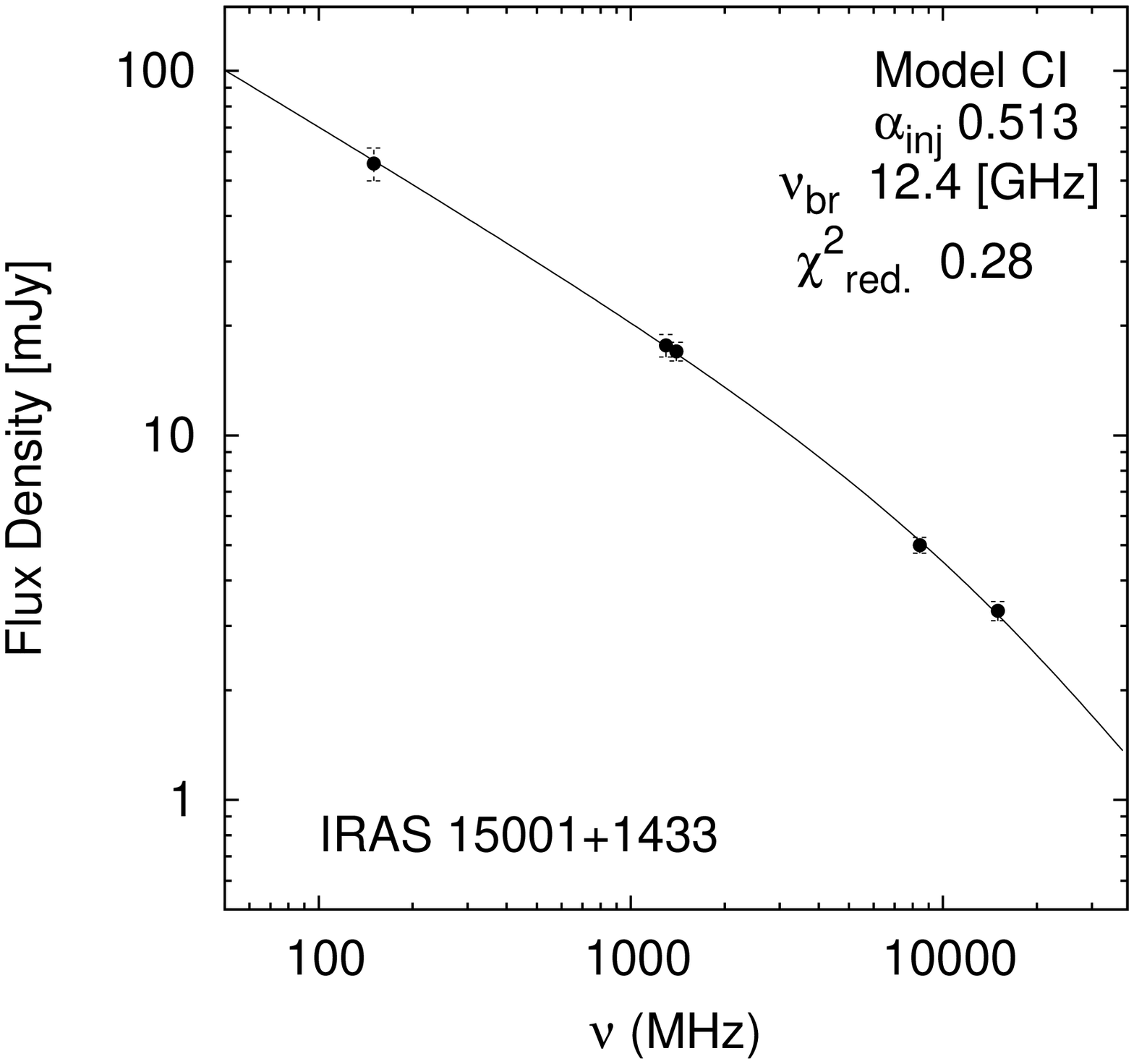,width=2.4in,angle=270}%
      \psfig{file=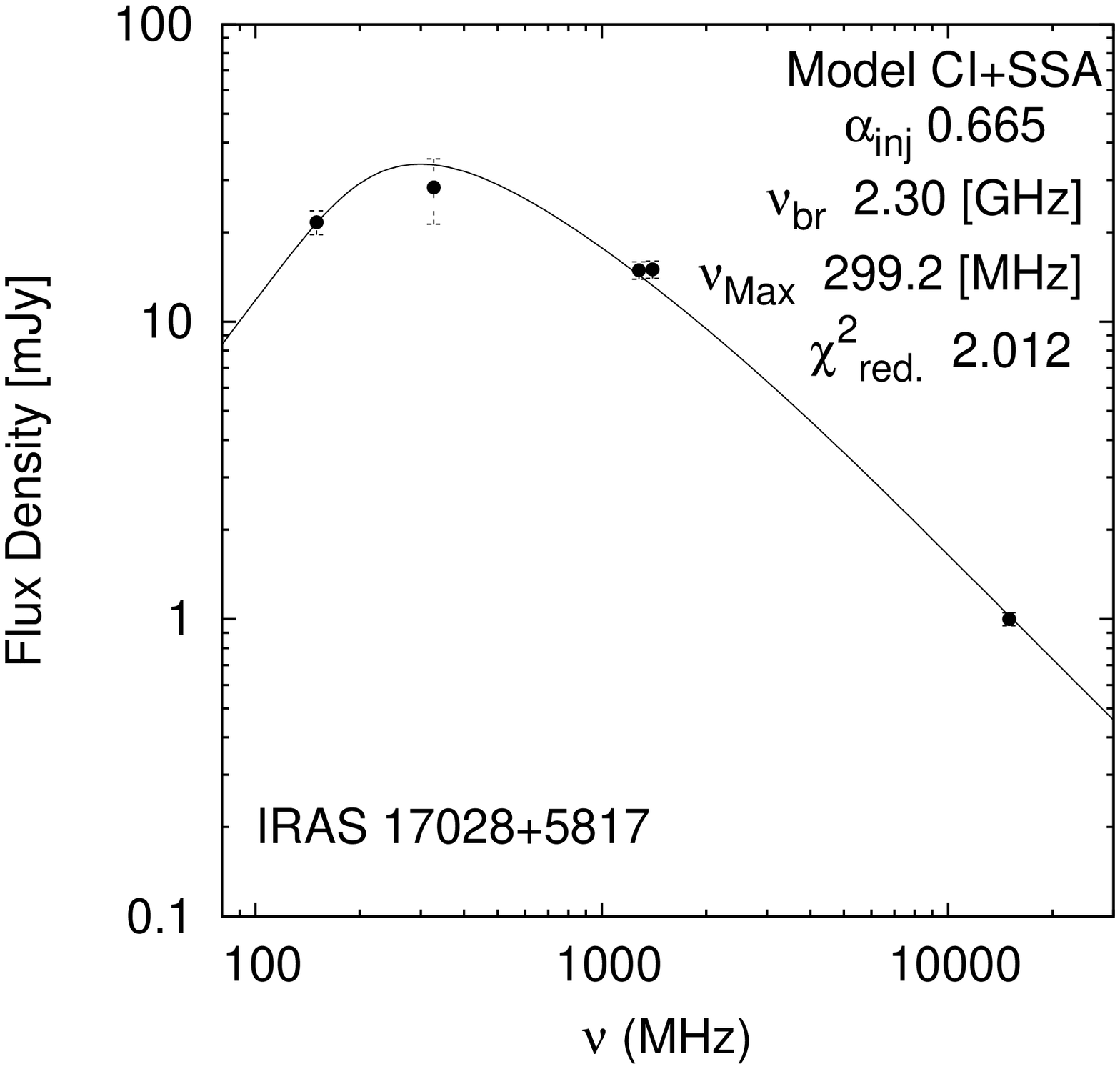,width=2.4in,angle=270}%
     }
}
\caption[]{ The integrated radio spectra of 9 ULIRGs from our sample and their best fit models.}
\label{int_spect}
\end{figure*}
\begin{figure*}
\vbox{
   \hbox{

      \psfig{file=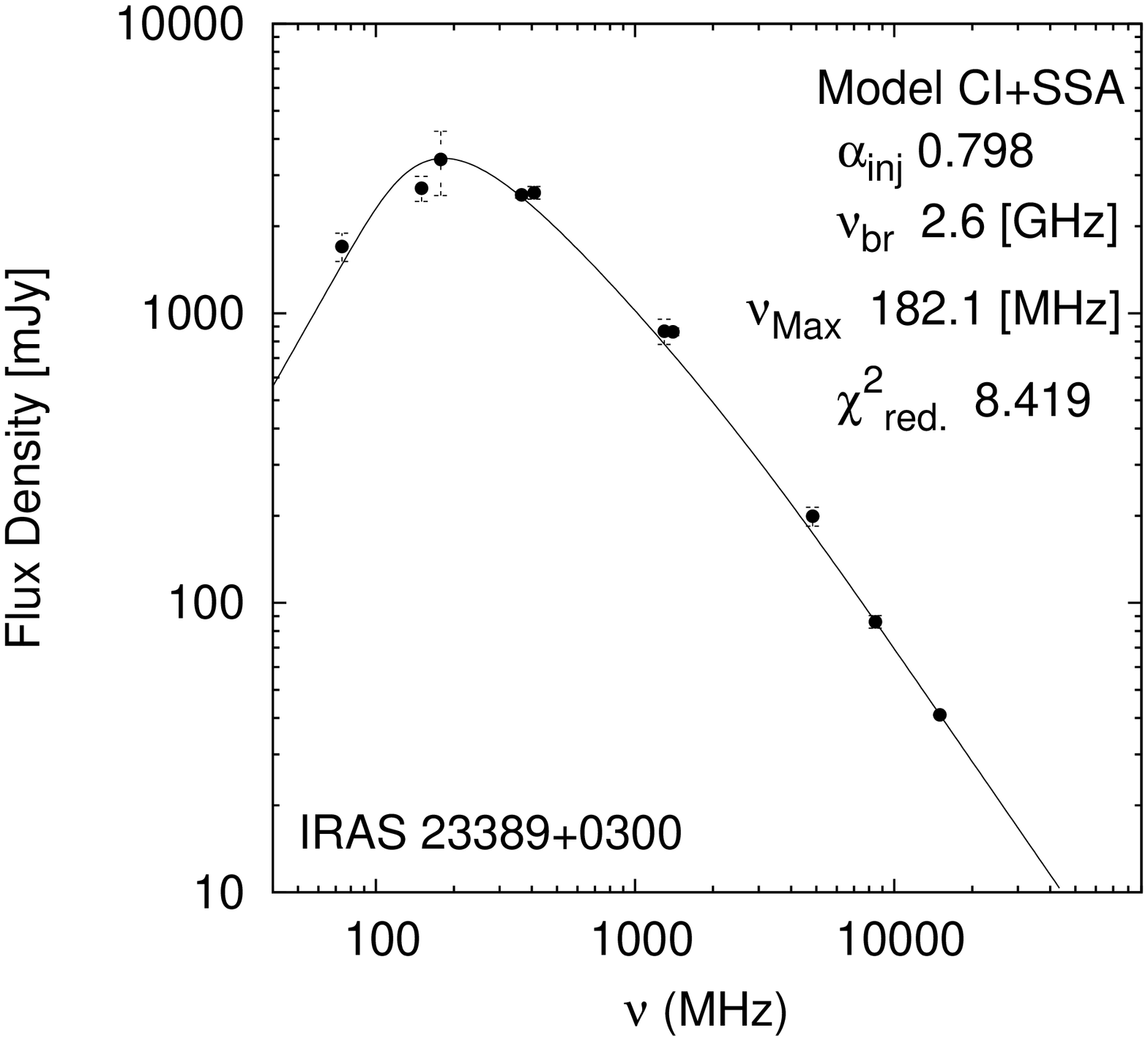,width=2.4in,angle=270}%
      \psfig{file=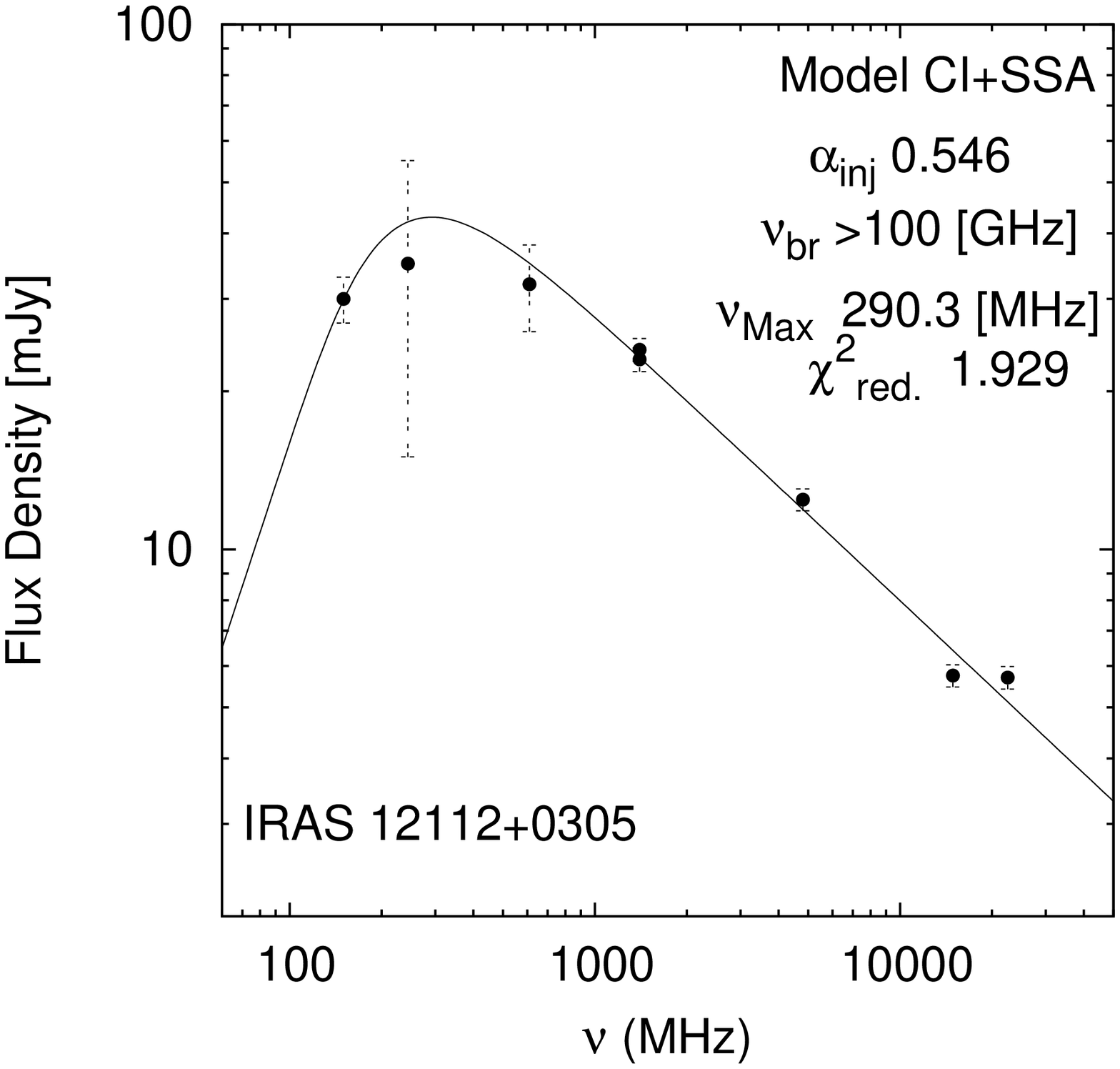,width=2.4in,angle=270}%
    }
    \hbox{
    \vspace{0.1cm}  
        \psfig{file=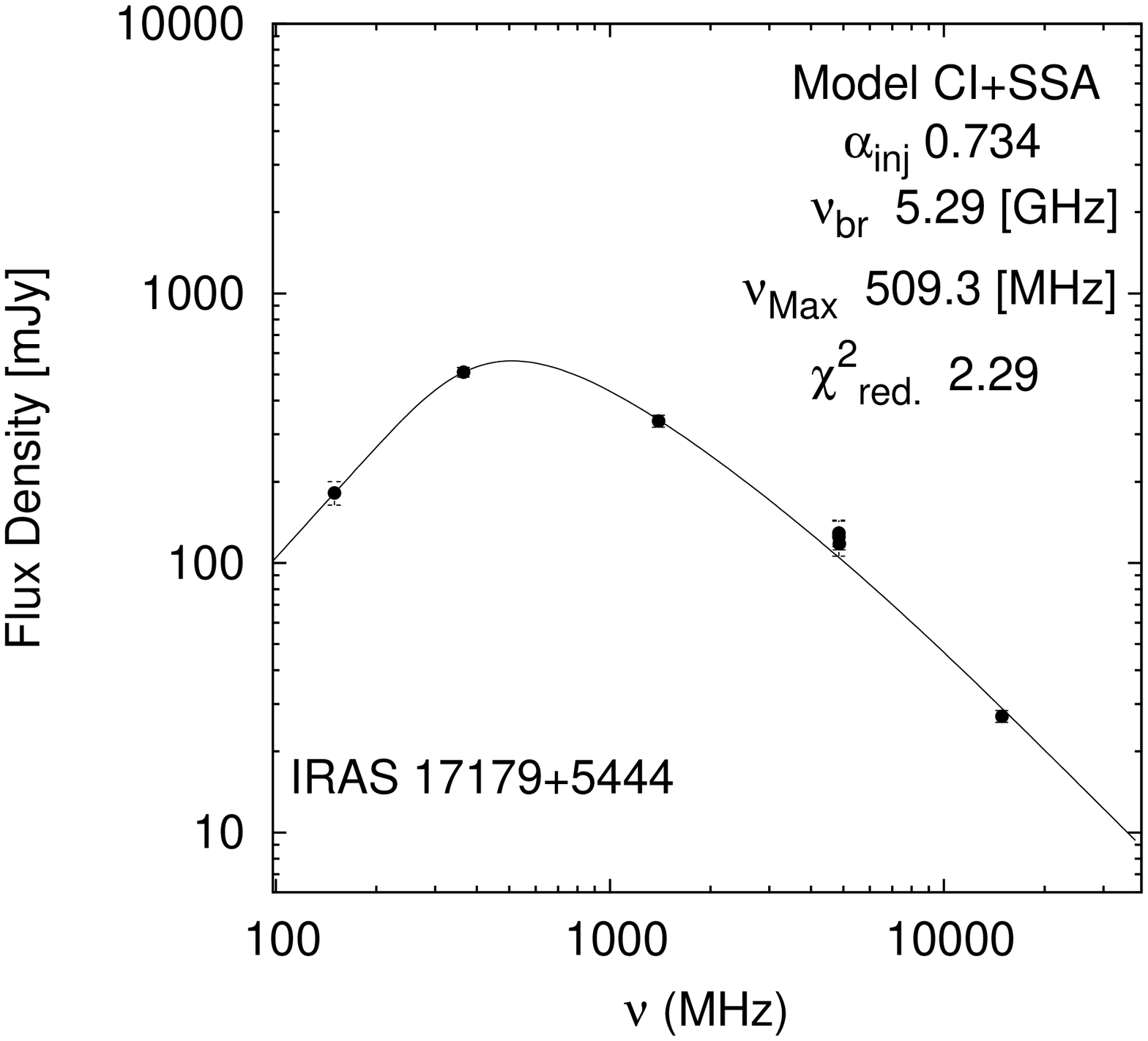,width=2.4in,angle=270}%
      }
      
}
\contcaption{The integrated radio spectra of 9 ULIRGs from our sample and their best fit models.}
\end{figure*}
\begin{table*}
 \caption{Estimated flux densities of ULIRGs at 1.28 GHz.}
 \centering
   \begin{tabular}{@{}lccccccccc@{}}
  \hline
     Name  &host&E(B-V) &g-i&IR     &rms    &        S$_p$ 	        & S$_t$           & P&\\
           &type&       &   &log(L$_{IR}/L_{\odot}$) &(mJy/b)  & (mJy/b)		&  (mJy)    &  (W Hz$^{-1}$) &   \\
      (1)  &(2)&(3)	  &(4)&(5)&(6)  &(7) &(8)	&		(9)            \\
 \hline
IRAS 00188$-$0856  & L  & 0.0315 &  0.8973 &12.33     & 0.41         &	14.0  	      & 15$\pm{1.05}$  &23.81   \\
IRAS 01572$+$0009  & S1 & 0.0248 &  0.1546 &12.53      & 0.09         &	15.0          & 16$\pm{1.12}$  &24.26   \\
IRAS 13305$-$1739  & S2 & 0.0612 &  0.8266 & 12.21     & 0.04         &    47           & 48$\pm{3.36}$  &24.41   \\       
IRAS 14070$+$0525  & S2 & 0.0210 &  0.6896 &12.76      & 0.20         &    4             & 4$\pm{0.28}$   &23.59   \\ 
IRAS 14394$+$5332  & S2 &  0.0096&  1.174 & 12.04      & 0.70         &    26             & 26$\pm{1.82}$  &23.87   \\
IRAS 15001$+$1433  & S2 &  0.0239&  1.1212 & 12.38     & 0.03         &    16            & 18$\pm{1.26}$ &24.04   \\
IRAS 16156$+$0146  & S2 &  0.0728&  1.3894 & 12.04     & 0.15         &    5             & 5$\pm{0.35}$  &23.31   \\
IRAS 17028$+$5817  &  L &  0.0196&  1.5867 & 12.10     & 0.27         &    15            & 14$\pm{1.05}$ &23.44   \\
IRAS 17044$+$6720  & L  &  0.0256&  1.0041 & 12.13     & 0.16         &    2             & 3$\pm{0.14}$  &23.12   \\
IRAS 23060$+$0505  &  S1&  0.0542&  1.5753 & 12.44     & 0.09         &    8             & 9$\pm{0.63}$  &23.38   \\
IRAS 23389$+$0300  &  S2&  0.0456&  1.6896  & 12.09     & 0.04         &    871           & 867$\pm{60.96}$ &25.70 \\
IRAS 12112$+$0305  &  L &  0.0188&  0.9334 & 12.28   & &  &                                   &23.48    \\
IRAS 17179$+$5444  &  S2&  0.0256&  1.192  & 12.20    & & &                                    &25.26    \\

\hline 
\end{tabular}
\\
Column 1: source name; Column 2: host galaxy type - S1, S2, L represent Seyfert 1, Seyfert 2 and LINER type AGN respectively;  Columns 3: color excess; Column 4: g-i color; Column 5: infrared luminosity;
Column 6: rms noise of the image; Column 7: peak flux densities in units of mJy/beam; Column 8: integrated flux densities in units of mJy; Column 9:  log of radio power at 1.4 GHz. 
\label{table2}
\end{table*}

\begin{table}
   \caption{Integrated flux densities}
   \centering
   \begin{tabular}{cccccc}
  \hline
Object &        Obs. Freq.    &S             &error  & Ref & \\
name   &         (MHz)        &(mJy)         &(mJy)&    used & \\
(1)    &          (2)         &(3)            &(4) &      (5)&  \\
 \hline
  IRAS 00188$-$0856&150 &   25     &2.5       &(1)\\             
                 &1280 &    15.9 &1.0         &(2)\\
                 &1400 &   16.50      &0.5         &(3)\\
                 &8600 &   4.40       &0.2         &(4)\\
                 &15000&   2.1        &0.1         &(5)\\ 
IRAS 01572$+$0009& 150 &        129.00&12.9    &(1)\\
                 &1280 &        15.85 &1.11    &(2)\\
                 &1400 &            27.00 &0.90    &(3)\\
                 &1450 &            22.50 &0.80    &(6)\\ 
                 &1490 &            22.50 &1.60    &(7)\\
                 &4890 &            7.88  &0.57    &(7)\\
                 &8600 &            4.60  &0.29    &(8)\\
                 &14900&            2.04  &0.40    &(7)\\
IRAS 13305$-$1739&150  &             260  &26.0    &(1)\\
                 &215  &         200.00   & 16.00  &(9)\\
                 &1298 &          48.37 &3.38    &(2)\\
                 &1400 &           48.00  &1.50    &(3)\\
                 &8440 &            6.30  &0.31    &(4)\\                 
IRAS 14070$+$0525 &1280&         3.98& 0.3         &(2)\\
                  &1400&         5.00&0.25         &(3)\\                
                  &15000&        0.70&0.03         &(5)\\              
IRAS 14394$+$5332 &150&           230&23.0        &(1)\\ 
                  &327&          140 &35.11         &(10)\\
                  &1280&       26.6  &1.8         &(2)\\ 
                  &1400&            42&2.1         &(3)\\
                  &15000&         5.60&0.3         &(5)\\ 
IRAS 15001$+$1433&150  &          57.68&5.76    &(1)\\
                 &1298 &       17.66 &1.24    &(2)\\
                 &1400 &          17.00   &1.00    &(3)\\
                 &8439 &           5.00   &0.25    &(4)\\
                 &15000&           3.30   &0.20    &(5)\\               
IRAS 16156$+$0146 &150&            25&2.5         &(1)\\
                  &1280&          5.28&0.3         &(2)\\
                  &1400&          8.30&0.4         &(3)\\ 
                  &15000&         3.50&0.2         &(5)\\
IRAS 17028$+$5817 &150&          21.6&2.0         &(1)\\
                  &327&          28.29&7          &(10)\\
                  &1280&         14.9&1.0         &(2)\\
                  &1400&          15.0&1.0         &(3)\\ 
                  &15000&        1.00&0.05         &(5)\\
IRAS 17044$+$6720 &1280&         2.97&0.2         &(2)\\
                  &1400&        5.00& 0.25         &(3)\\
                  &15000&       2.00& 0.10         &(5)\\
IRAS  23060$+$0505&1280&        8.97&0.63         &(2)\\             
                  &1400&         6.50&0.32         &(3)\\  
                  &15000&       2.40&0.12          &(5)\\
IRAS 23389$+$0300&74   &       1751  &195.76  &(11)\\
                 &150  &       2783  &278.30  &(1)\\
                 &178  &       3400  &850.00  &(12)\\
                 &365  &       2650  & 65.20  &(13)\\
                 &408  &       2558  &128.00  &(14\\
                 &1298 &        845   & 59.15  &(2)\\
                 &1400 &         863      & 26.00  &(3)\\
                 &4850 &            199   & 15.00  &(15)\\
                 &15000&             41   &  4.10  &(5)\\
\hline 
\end{tabular}
\\
\label{table3}
\end{table}
\begin{table}
   \contcaption{Integrated flux densities}
   \centering
   \begin{tabular}{cccccc}
  \hline
Object &  Obs. Freq.  &S             &error          & Ref \\
name&  (MHz)       &(mJy)         &(mJy)             & used\\
(1)&  (2)         &(3)            &(4)               & (5) \\
 \hline
 
IRAS 12112$+$0305&150  &   30.00 &  3.00&(1) \\
                 &244  &   35.00& 20.00&(16)   \\
                 &610  &   32.00&  6.00&(16)   \\
                 &1400 &   24.00&  1.20&(4)    \\
                 &1400 &   23.00&  1.20&(4)    \\
                 &4800 &   12.44&  0.60&(4)    \\
                 &14900&    5.75&  0.28&(4)    \\
                 &22500&    5.70&  0.28&(17)   \\
IRAS 17179$+$5444&150  &  182.00 & 18.20&(1) \\
                 &365  &  510.00 & 20.50&(13) \\
                 &1400 &  336.00& 17.00&(3)    \\
                 &4860 &  117.88&  6.00&(4)    \\
                 &4850 &  125.00& 19.00&(18)   \\
                 &4850 &  129.00& 14.00&(19)   \\
                 &14900&   27.00&  1.35&(4)    \\
\hline
\end{tabular}
\\
Column 1: source name; Column 2: observing frequency; Columns 3 and 4: total flux densities of the source and the error; Column 5: references - (1) TGSS, (2) current observation, (3)\citet{1998AJ....115.1693C} (4) VLA, (5) \citet{2003A&A...409..115N}, (6)\citet{1989ApJS...70..257B}, (7) \citet{1996AJ....111.1431B}, (8) \citet{2005ApJ...618..108B}, (9) GLEAM, (10) \citep{1997A&AS..124..259R}, (11)\citet{2007AJ....134.1245C}, (12) \citet{1967MmRAS..71...49G}, (13) \citet{1996AJ....111.1945D}, (14) \citet{1981MNRAS.194..693L}, (15) \cite{1995ApJS...97..347G}, (16) \citet{2008A&A...477...95C}, (17)\citet{2008A&A...477...95C}.
(18) \citet{1991ApJS...75....1B}, (19) \citet{1991ApJS...75.1011G}.
\end{table}
\section{SDSS spectra analysis}
 \label{sec:opt_data}
  We searched for the SDSS nuclear spectra of these ULIRGs. However, only 4 sources of this sample had double peaked emission lines. This means that some of their emission lines had two kinematic components signifying doppler shifts between the two line components. Such features could be due to the presence of two AGNs in the nucleus of the ULIRG or due to a rotating nuclear disk, or it could also be due to outflows from a single AGN. 
 In order to obtain accurate measurements of the AGN emission line properties the stellar components should be removed from the redshift corrected spectrum of the corresponding galaxy. Two of these four sources IRAS 1439$+$5332 and IRAS 23060$+$0505 have outflow signatures in their spectrum while IRAS 15001$+$1433 and IRAS 12112$+$0305 do not have any such signature. 
 
 We first removed the stellar absorption features of the host galaxies IRAS 15001$+$1433 and IRAS 12112$+$0305 using the pPXF code (Penalized  Pixel-Fitting  stellar kinematics  extraction) \citep{2004PASP..116..138C} which generated a model of the underlying stellar population.
 While fitting, the AGN emission components have been masked to preserve their flux in these process.  Each of these continuum-subtracted normalized spectra represents AGN emission lines of corresponding galaxy. The spectra of both sources show signatures of double peaks for the lines H$_\alpha$ $\lambda$ 6563, [NII] $\lambda$$\lambda$ 6548, 6584 and [SII]  $\lambda$$\lambda$ 6717, 6731. These double peak emission lines can be de-blended with two narrow Gaussian functions. The width of the narrow components are set to be less than 1000~km s$^{-1}$.  We have also constrained equal widths for the two components arising from the same emission line in velocity space.   
 We have fitted the multi component Gaussian functions for  H$_\alpha$ $\lambda$ 6563, [NII] $\lambda$$\lambda$ 6548, 6584 and [SII]  $\lambda$$\lambda$ 6717, 6731 to measure the velocity separation between two probalbe components. 
 
 For IRAS 1439$+$5332 and IRAS 23060$+$0505 we found double peak signatures in [OIII] $\lambda$$\lambda$ 4960, 5007. 
 These narrow meta stable lines are normally observed as an AGN emission component, and very rarely found in stellar absorption spectrum. Since the outflow signatures of the AGN emission lines are much stronger than their correnponding stellar components. The estimation of the eission line kinametic can also be done without removing the stellar components. So,we have fitted multi component Gaussian of [OIII] $\lambda$$\lambda$ 4960, 5007 lines of IRAS 1439$+$5332 and IRAS 23060$+$0505 without fitting of PPXF and estimated the velocity separation.
\section{Radio spectra and spectral age}
\label{sec:spect_aging}
 To study the integrated spectra of these sources we have compiled the flux densities at different wavelengths from the literature using the NASA/IPAC Extragalactic Database, Very large Array (VLA) archival images, different surveys (e.g. VLA Lowfrequency Sky Survey Redux (VLSSr),  GaLactic and Extragalactic
All-sky MWA (GLEAM) survey, TIFR GMRT Sky Survey (TGSS)), 
and the current observations. For most of the sources the multi frequency radio data was found (Table \ref{table3}). 
However all these flux density measurements are not on a consistent flux density scale. So, the absolute flux density scale of all sources was set to the Baars scale \citep{1977A&A....61...99B} for this study.

To quantify each radio spectrum  we used the SYNAGE software \citep{1999A&A...345..769M} for spectral fitting. It is well known that after several Myr to Gyr of evolutionary time, the synchrotron spectrum in radio sources develops curvatures at low and high frequencies. The low-frequency turnovers are interpreted  as due to synchrotron self-absorption. However, free–free absorption can also absorb the power-law synchrotron spectrum and show a low frequency turnover \citep{2010MNRAS.405..887C}. The high frequency steepening is due to particle energy losses. The initial injected power-law slope with injection spectral index ($\alpha_{inj}$; flux density $S_\nu\propto\nu^{-\alpha}$) deviates above the spectral break frequency $\nu_{br}$. The radiative synchrotron age, $\tau_{\rm syn}$, is related to $\nu_{\rm br}$
and the magnetic field strength, $B$, through the following relation.

\begin{equation}
\tau_{\rm syn}[{\rm Myr}]=50.3\frac{B^{1/2}}{B^{2}+B_{\rm iC}^{2}}[\nu_{\rm br}(1+z)]^{-1/2},
\end{equation}

\noindent
where $B_{\rm iC}$=0.318(1+$z)^{2}$ is the magnetic field strength equivalent to the
inverse-Compton microwave background radiation. The parameters $B$ and $B_{\rm iC}$ are in
units of nT and $\nu_{\rm br}$ is in GHz. All ULIRGs have compact radio 
emission and have continuous flow of fresh relativistic particles. So, we applied the
continuous injection (CI) model for them. 
For fitting we consider $\alpha_{inj}$ as a free parameter.
For sources which do not have any low frequency curvature, we only fitted the CI model, 
while for sources which show low frequency turn over, we fitted the CI model modified by low-frequency synchrotron-self absorption \citep{1970ranp.book.....P, 1999A&A...345..769M}. 
To estimate the magnetic field strength we followed \citet{2008MNRAS.383..525K} and \citet{2010MNRAS.404..433N}. We assumed equipartition conditions between the magnetic field and the particles. A cylindrical geometry has been assumed for each source and their sizes have been estimated  reliably from the available highest resolution maps.
\begin{figure}
\center
\hbox{
\psfig{file=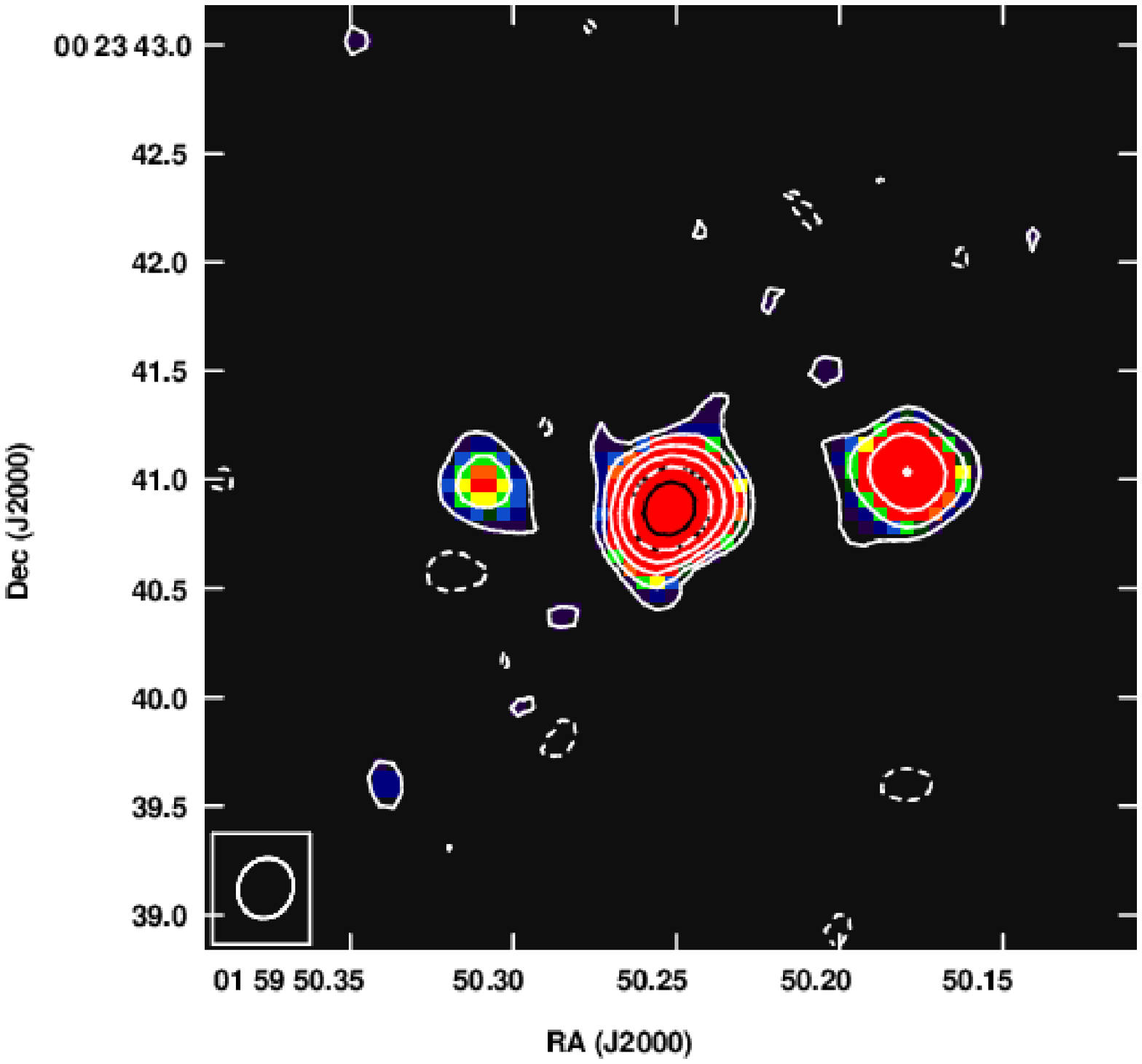,width=7.0cm,angle=90}
     }
\hbox{
 \psfig{file=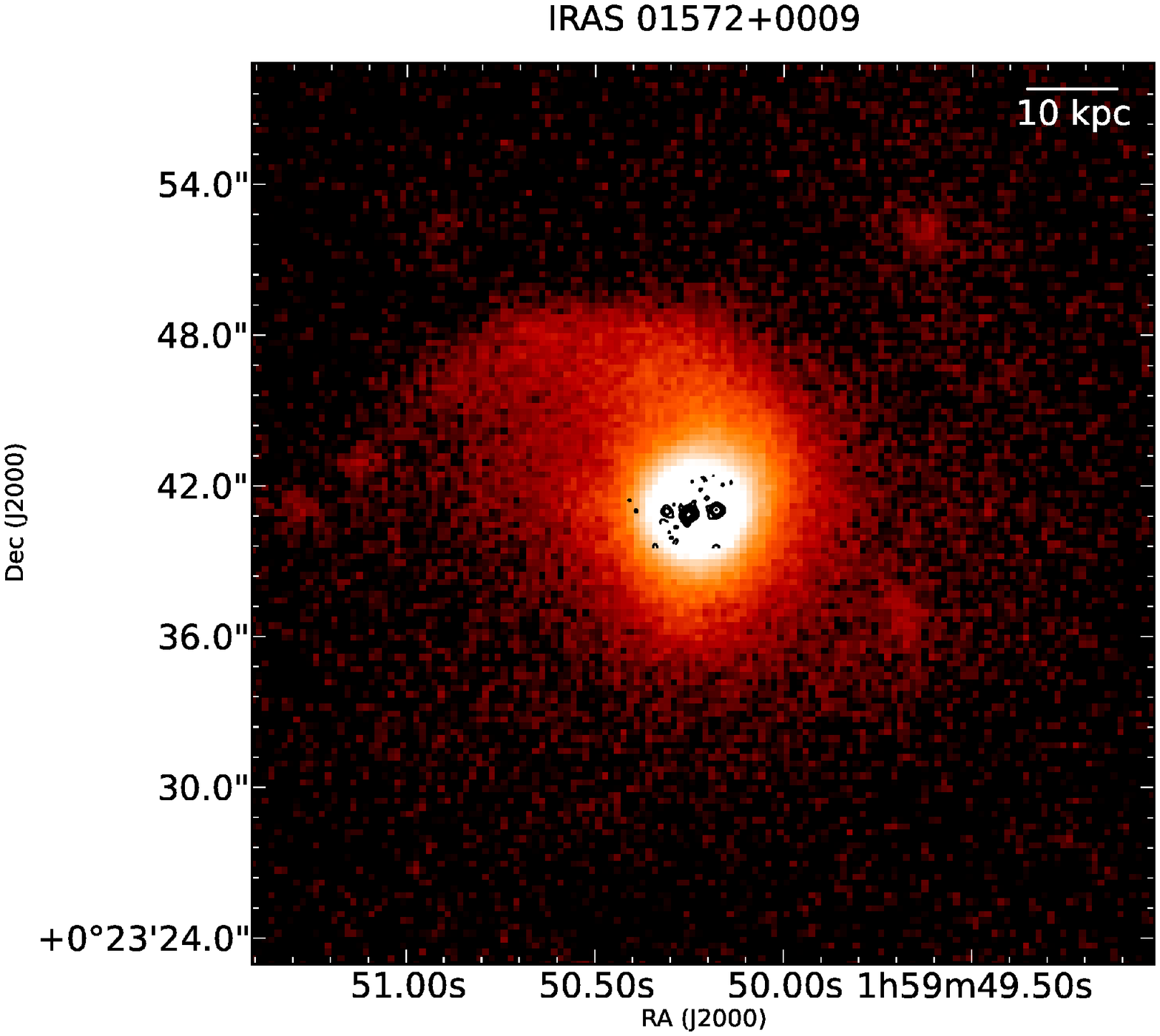,width=7.0cm,angle=90}
     }

\caption[]{Upper panel: VLA 8.4 GHz radio map of IRAS 01572$+$0009. The contours are plotted at 3$\sigma$[-1, 1, 2, 4, 8, 16, 32, 64, 128, 256], where the rms noise $\sigma$=0.01 mJy/beam. Lower panel: The 8.4 GHz contours are overlaid on the Pan-STARRS r-band image.}
\label{IRAS01572_fig}

\end{figure}

\section{Results}
\label{sec:result} 
We carried out GMRT observations in 2 phases. In the first phase (in 2007) we observed all these galaxies (Table \ref{table2}, except IRAS 12112$+$0305 \&  IRAS 17179$+$5444) at 1.28~GHz for 2~hrs each. 
Three sources IRAS 15001$+$1433, IRAS 23389$+$0300 and IRAS 13305$-$1739 showed signatures of extended emission. So we did follow-up observations of these 3 sources in the second phase of our study (in 2015), during which, six hour GMRT observations per source were done.
We could not observe the two ULIRGs, IRAS 12112$+$0305 and IRAS 17179$+$5444 with the GMRT due to time constraints and so we used archival radio data for these two sources. The observational parameters of the GMRT observations are presented in Table \ref{table2}.
The typical uncertainty on GMRT flux density measurements to be 7\% at 1.28 GHz \citep{2012MNRAS.424.1061K}. 
This includes the uncertainties in the measurement of calibrator flux density and uncertainties introduced through the telescope hardware. 
We have also estimated the radio power at 1.4 GHz for all these ULIRGs. Two primary morphological classes of radio sources are Fanaroff–Riley type I (FRI) and type II (FRII). The break between FRI and type II FRII in terms of radio power occurs at a power value of $P_{1.4GHz}\sim 10^{25}$ W Hz$^{-1}$,  where the FRI sources have radio power below $P_{1.4GHz}$, and the FRII sources have radio power higher than this value \citep{2009MNRAS.392..617D}. However, this break depends on the optical host galaxy magnitude \citep{1996AJ....112....9L}. Recent study also shows that large overlap in luminosity for the FRI and FRII sources and we can not reliably predict their morphology just estimating their luminosity \citep{2019MNRAS.488.2701M}. The infrared luminosity are taken from \citep{1999ApJ...522..113V}. The estimated g-i colors of the ULIRGs indicate that these are highly reddened objects.

We also found that a wide coverage of multi-frequency archival radio data exists for the 9 sources. 
In Table \ref{table3} we present all the archival flux density values and their references.  
The integrated spectra of these sources are shown in Fig. \ref{int_spect}. 
The estimated values of break frequencies, turnover frequencies, spectral ages and radio luminosity at 1.4 GHz for those 9 sources are given in Table \ref{table4}.
The optical r band images of these ULIRGs are shown in Fig. \ref{host_galaxy1}, Fig.\ref{host_galaxy2} and Figs \ref{host_galaxy3} . Our GMRT observations show that most of the ULIRGs are unresolved at 1.28 GHz. The only one exception is IRAS 15001$+$1433, which shows weak signatures of extended emission. 
In the following paragraphs we present a detailed description of the results for each of the individual sources.
\begin{figure} 
\center
\hbox{
 \psfig{file=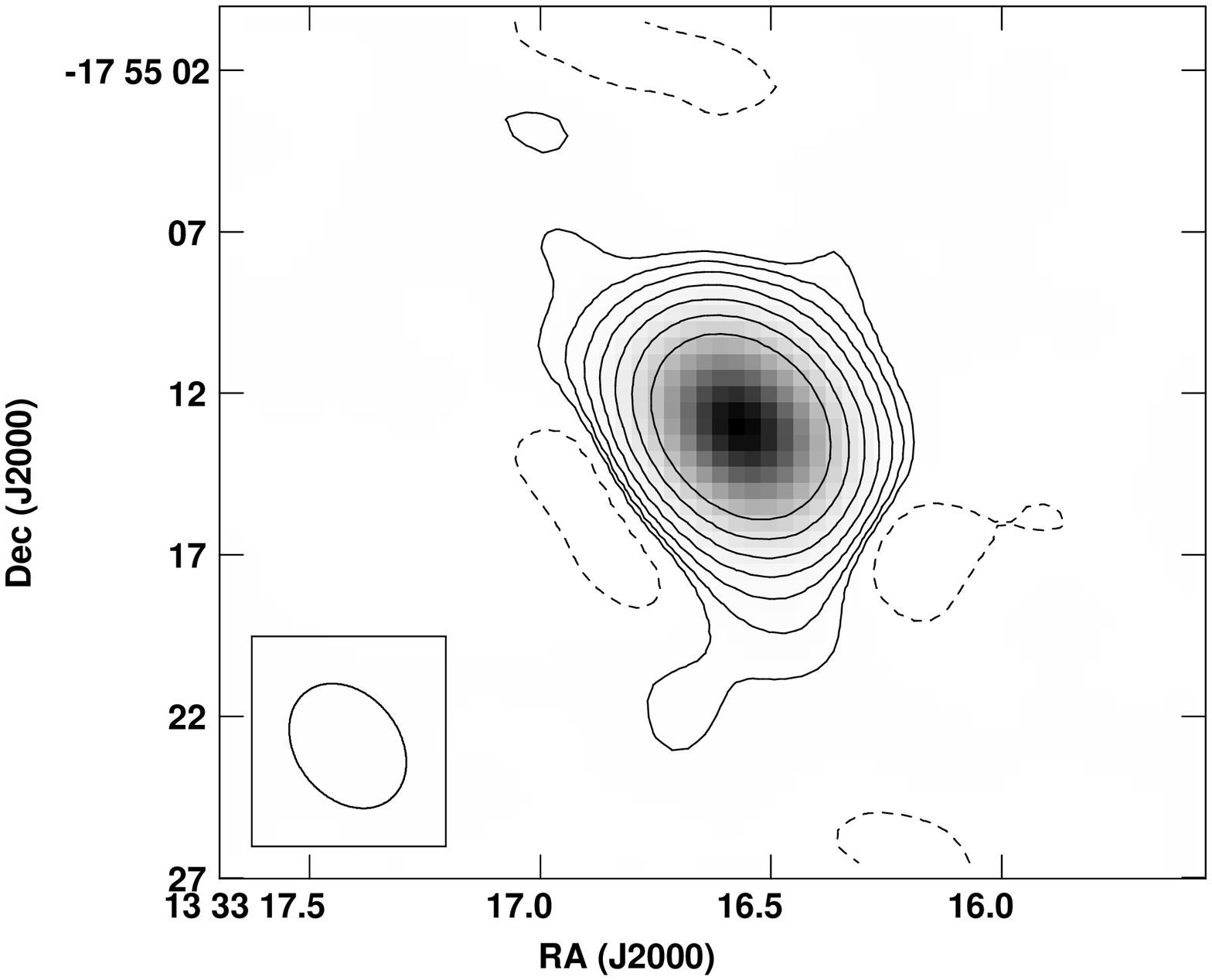,width=5.8cm,angle=-90}
     }
 \hbox{
 \psfig{file=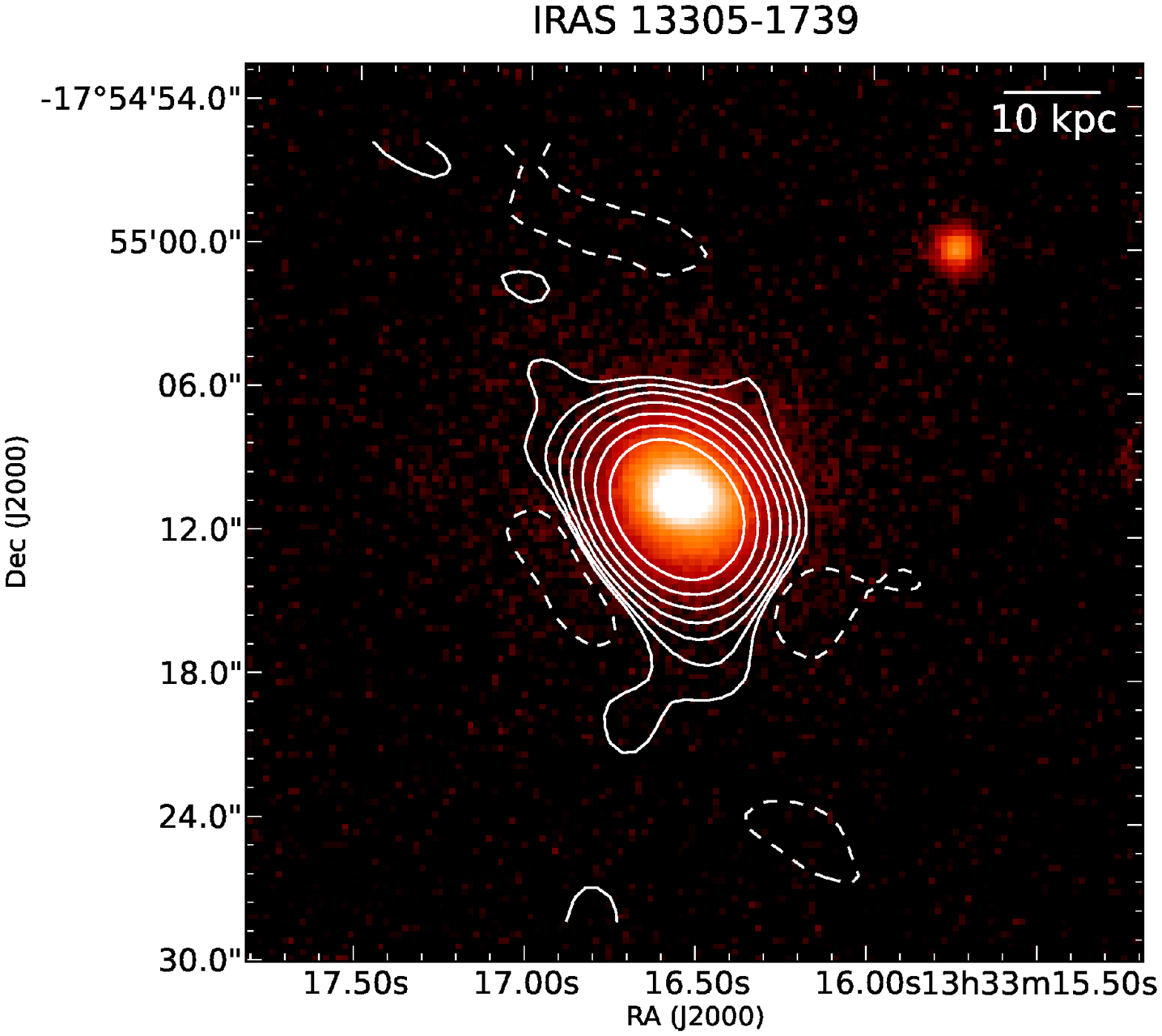,width=7.0cm,angle=90}
     }    
     
\caption[]{Upper panel: GMRT 1.28 GHz image of  IRAS 13305$-$1739. The contours are plotted at 3$\sigma$[-1, 1, 2, 4, 8, 16, 32, 64], where the rms noise is $\sigma$=0.04 mJy/beam. Lower panel: The GMRT radio contours of IRAS 13305$-$1739 are overlaid on the Pan-STARRS r-band optical image.}
\label{IRAS13305_fig}
\end{figure}

\noindent
{\bf IRAS 00188$-$0856:} This is a  compact single-nucleus system. The southern nearby bright source (Fig. \ref{host_galaxy1}) is a star
\citep{2002ApJS..143..315V}. No extended emission is detected in the GMRT observation, however the available multifrequency radio data shows a low frequency turnover at 263 MHz and the a break at $\sim$5 GHz (Fig. \ref{int_spect}). Since IRAS 00188$-$0856 is unresolved at 15~GHz, the source size must be $<$ 0.94$\arcsec$ (which is the resolution of the 15~GHz VLA image). The estimated spectral age 4.3 Myr. 
%


\noindent
{\bf IRAS 01572$+$0009:} The central optical compact component of this ULIRG is a quasar and shows a filamentary structure in its optical image \citep{2002ApJ...574L.105B} (Fig. \ref{host_galaxy1}). The 8.4~GHz radio image of this source \citep{2006A&A...455..161L} shows a triple radio component. This triple radio source has a classical double-lobed structure  with a bright central core which is well within the optical structure (Fig. \ref{IRAS01572_fig}). The distance between the two hotspots is 1.3~kpc.
The integrated spectrum of this source covers a large frequency range (Fig. \ref{int_spect}).
There is no signature of a low frequency turn over for this source and the estimated break frequency is $\sim$10.35 GHz. The spectral age is 1.41 Myr. 

\noindent
{\bf IRAS 13305$-$1739:}
The optical images show that this is a
highly nucleated galaxy with slight isophotal distortions (Fig. \ref{host_galaxy1}) \citep{2002ApJS..143..315V}. Both the optical and near-IR nuclear spectra indicate that this source hosts a Seyfert~2 type AGN. The broad emission lines, especially in the near-IR spectrum suggests that  IRAS 13305$-$1739 is a ULIRG that is well on its way to becoming a quasar \citep{rodriguez.etal.2009}. However, our GMRT 6 hour deep observation of this source does not show any significant extended features (Fig. \ref{IRAS13305_fig}). 
The integrated spectrum (Fig. \ref{int_spect}) does not show any low frequency turn over. The break frequency and spectral age of this source are 1.98 GHz and 7.8~Myr respectively.   

\noindent
{\bf IRAS 14070+0525:}  The optical image shows (Figure \ref{host_galaxy1}) that it is a compact object with disturbed isophotes.  It has no tidal feature as well \citep{2002ApJS..143..315V}. Since this source does not have good radio frequency coverage we did not apply SYNAGE fits to the data. Hence the spectral age estimation was not possible for this source.

\begin{figure}
\center
\psfig{file=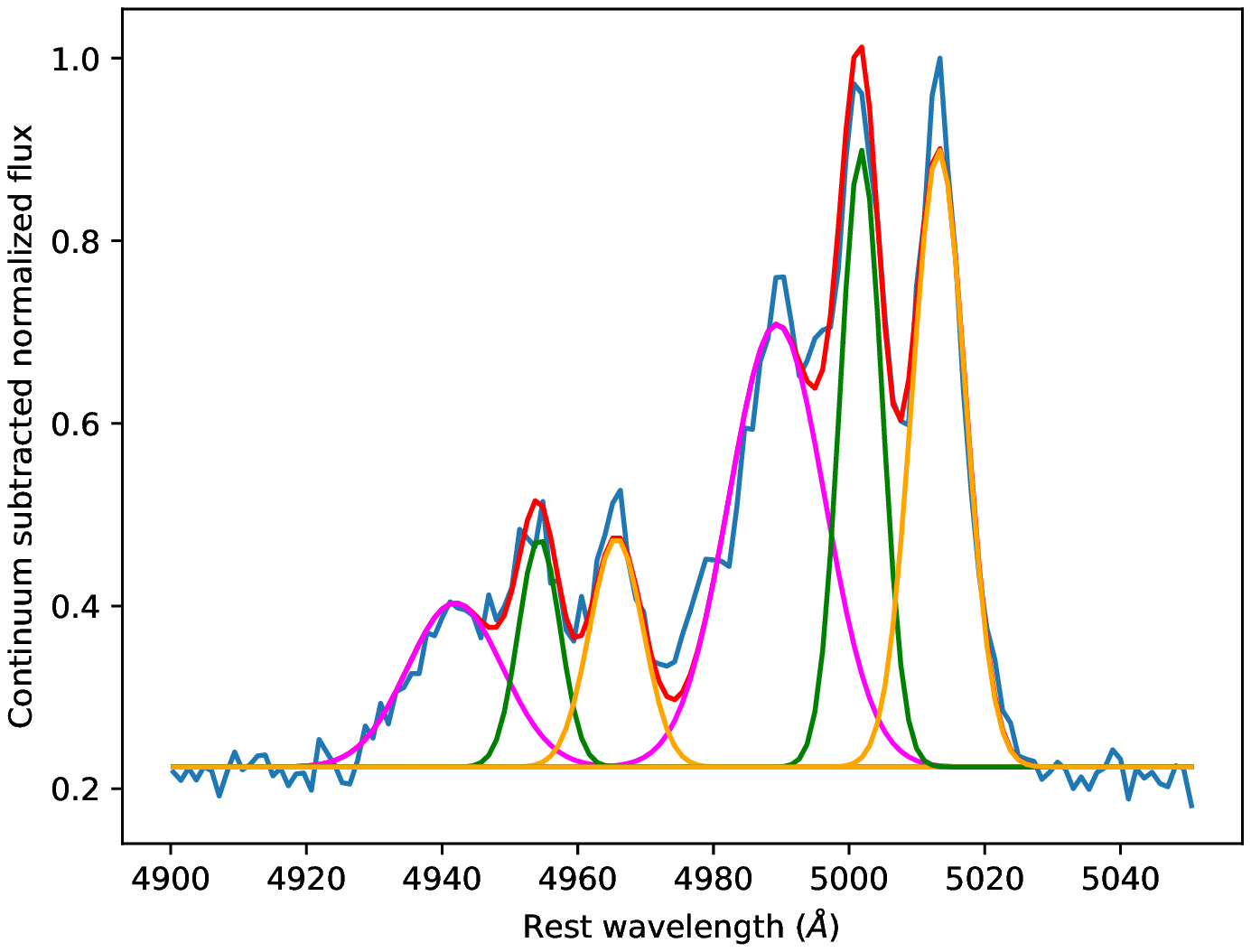,,width=8.0cm,angle=0}
\hbox{
 \psfig{file=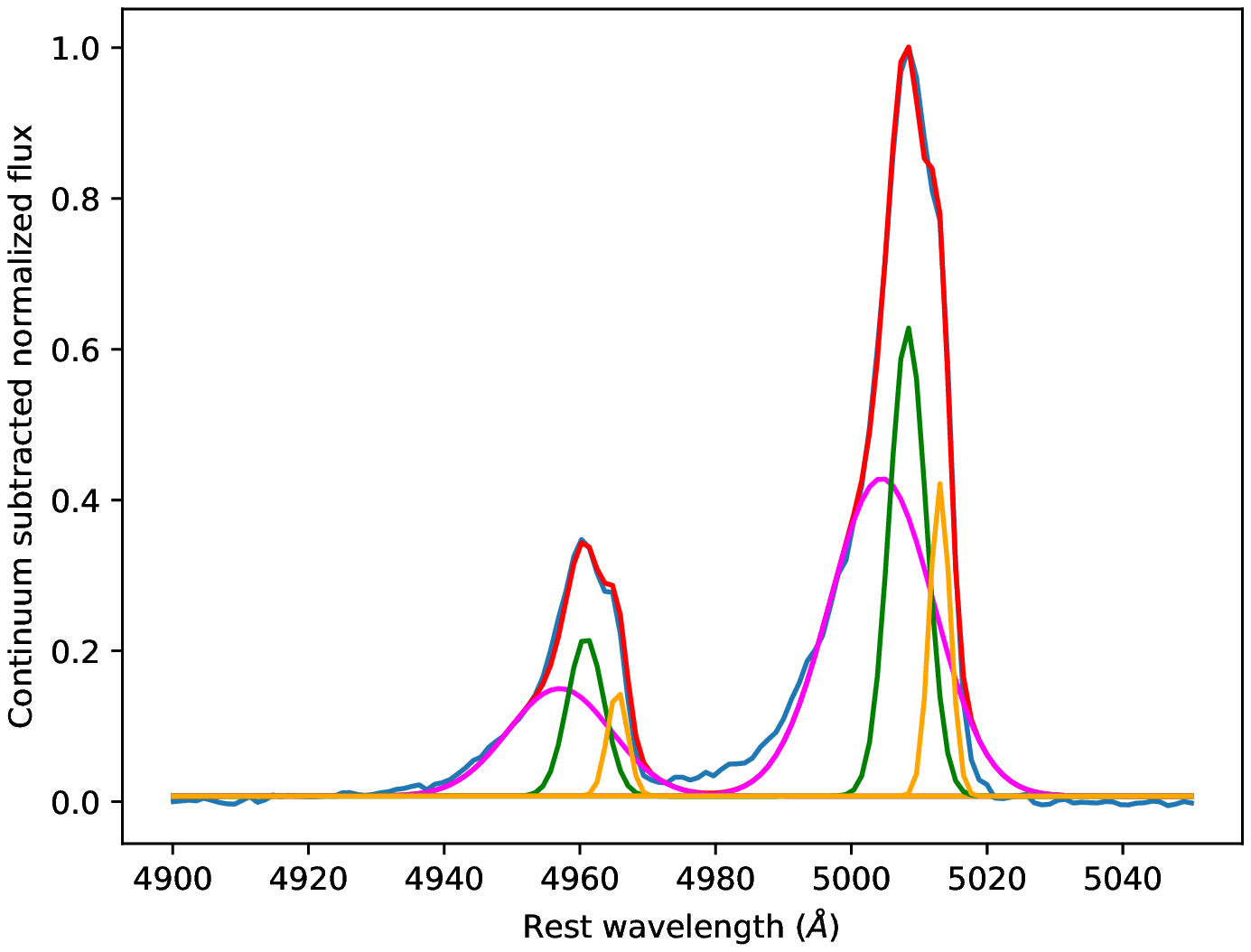,width=8.0cm,angle=0}
     }
\caption[]{ The double peaks in [OIII] emission lines are shown for the sources IRAS 1439$+$5332 (upper panel) and IRAS 23060$+$0505 (lower panel). 
The green and yellow color gaussian fits represent two kinematic components. 
}
\label{OIII_fig1}

\end{figure}

\noindent
{\bf IRAS 14394$+$5332:} This is a merging system of three components. The east component has two close nuclei at a separation 2.6 Kpc (1.29'') \citep{2002ApJS..143..315V}. The third west component is connected with a 35 kpc tidal tail (Fig. \ref{host_galaxy1}) with the east component. The east component is more luminous than the west one and shows strong AGN emission lines. GMRT observations detect radio emission from this component. The radio spectrum does not show any low frequency curvature while the analysis with SYNAGE gives a high break frequency. The highest integrated flux density that we found for this source is at $\sim$ 15~GHz, hence the lower limit of the break frequency is at frequencies $>$15~GHz. Therefore, the upper limit of the age is 1.5~Myr. The SDSS optical spectrum of the east component shows double peak emission lines for the [OIII] doublet (Fig. \ref{OIII_fig1}). This line splitting may be due to the presence of two close nuclei.

\begin{figure}
\center
\hbox{
\psfig{file=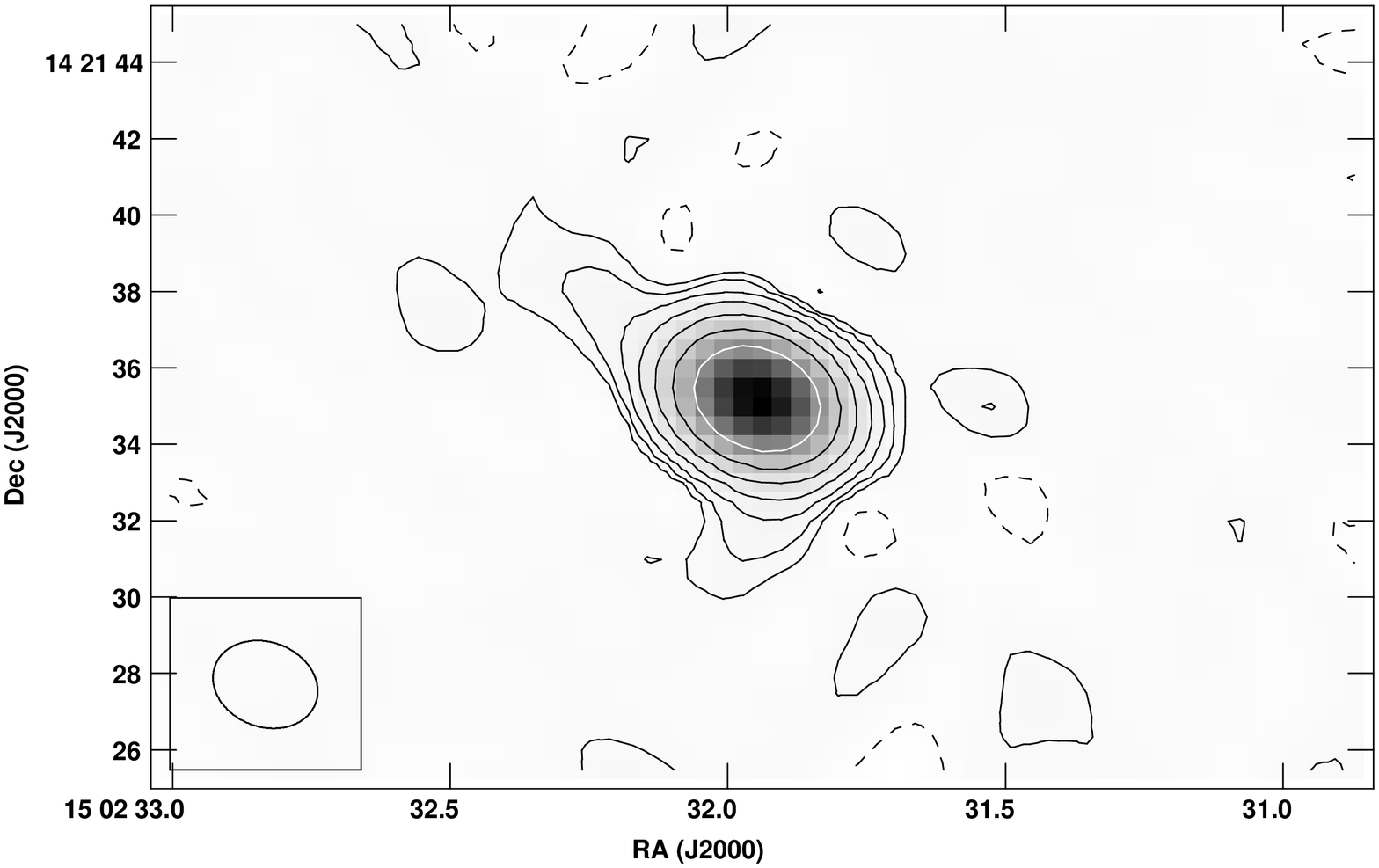,width=5.0cm,angle=270}
     }
\hbox{
 \psfig{file=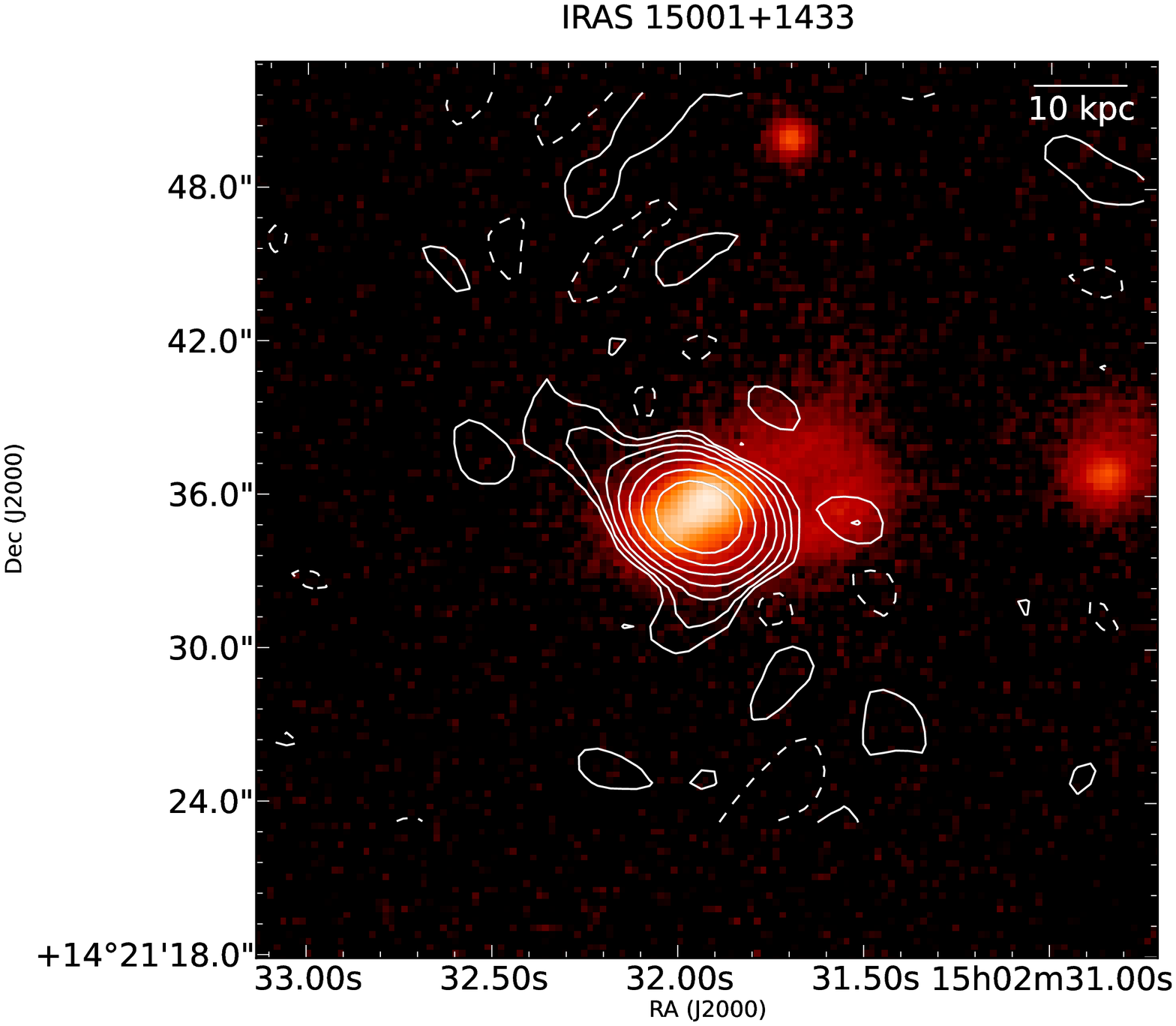,width=8.8cm,angle=0}
 }
\caption[]{Upper panel: GMRT 1.28 GHz image of  IRAS 15001$+$1433. The contours are plotted at 3$\sigma$[-1, 1, 2, 4, 8, 16, 32, 64], where rms noise $\sigma$=0.03 mJy/beam; Lower panel: GMRT radio contours of IRAS 15001$+$1433 are overlaid on the Pan-STARRS r-band optical image.}
\label{IRAS1500_fig1}

\end{figure}
\begin{figure}
\center
\hbox{
 \psfig{file=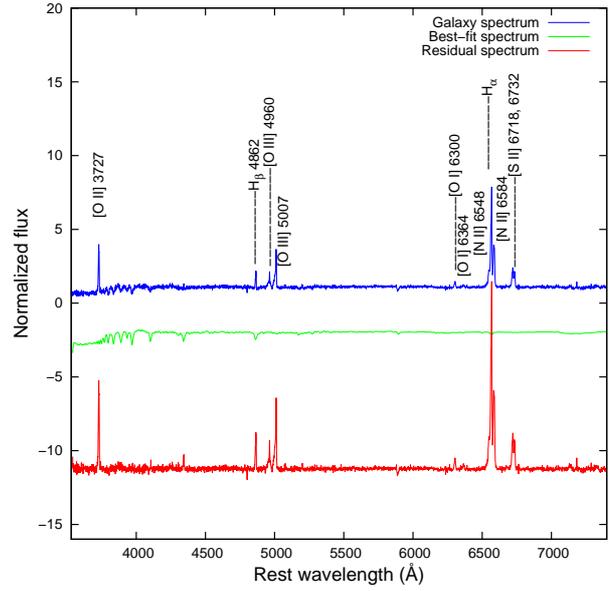,width=7.7cm,angle=-90}
     }
\hbox{
 \psfig{file=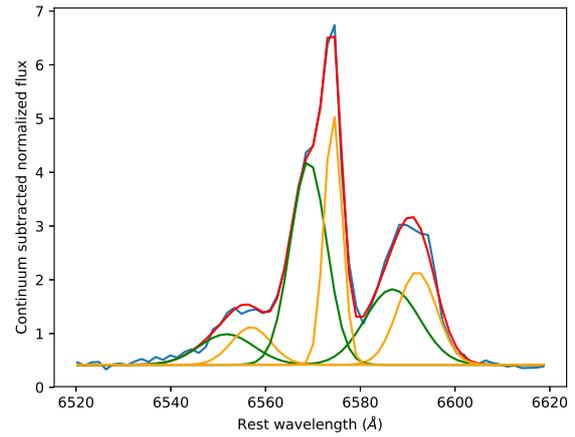,width=8.7cm,angle=0}
      }
  \hbox{
 \psfig{file=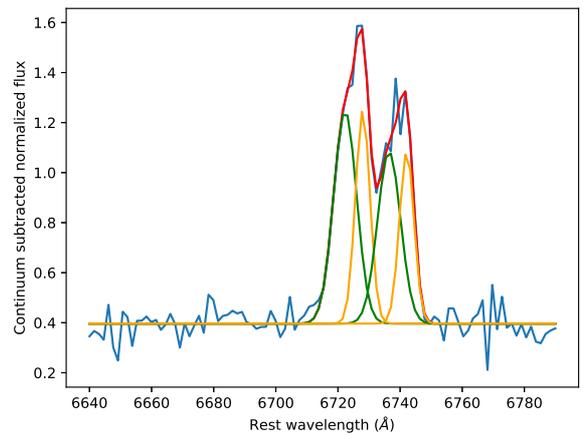,width=8.7cm,angle=0}
      }    
\caption[]{Upper panel: The redshift-corrected SDSS spectrum of IRAS 15001$+$1433 (in blue) with the strongest emission lines labelled. The best-fit model spectrum for the underlying stellar population of the host produced by pPXF (in green). The residual spectrum is in red. Middle panel: The double peaks in  H$_\alpha$ and [NII] emission lines; Lower panel: The double peaks in [SII] emission line.
For middle and lower panel the green and yellow gaussian fits correspond to two kinematic components.}
\label{IRAS1500_fig2}

\end{figure}

\noindent
{\bf IRAS 15001+1433:} The optical r band image shows that this ULIRG has a disturbed morphology and is more extended on one side of the nucleus (Fig. \ref{host_galaxy2}). It maybe interacting with two nearby companions. The nucleus is known to host a Seyfert~2 type AGN. In Fig. \ref{IRAS1500_fig1} (upper and lower panels) we show the GMRT 1.28 GHz radio contour map of IRAS 15001$+$1433 and its overlay on the optical image respectively. We notice a marginal signature of extended radio emission. The extended radio emission can be due to radio jets or AGN outflows/winds. Alternatively it can be due to outflows associated with the intense star formation taking place in such merger remnants, many of which may have dual AGN as well. We present the integrated radio spectral energy distribution using SYNAGE in Fig. \ref{int_spect}. 
The estimated break frequency and corresponding spectral age are 11.30 GHz and 0.42 Myr respectively for IRAS 15001+1433. Fig. \ref{IRAS1500_fig2} (upper panel) shows the SDSS optical spectrum of the nuclear region in IRAS 15001+1433. Here both H$\alpha$, [NII], and the [SII] doublets show signatures of double peak emission (Fig. \ref{IRAS1500_fig2}, middle and lower panels). As mentioned in Section~1 this maybe due to a rotating disk of ionized gas, nuclear outflows or the presence of two nuclei. 

\noindent
{\bf IRAS 16156$+$0146:}  Figure \ref{host_galaxy2} shows that this is a double nucleus system and the nuclear separation is 8 kpc \citep{rodriguez.etal.2009}. \citet{2002ApJS..143..315V} also reported a tidal bridge like structure that connects the two nuclei. The 
available radio data is sparse and having irregular cadence in frequency space. Therefore, we did not fit the SYNAGE for this source.

\noindent
{\bf IRAS 17028$+$5817:} This source is a widely separated( $\sim$ 25 kpc) two component system \citep{rodriguez.etal.2009}. Diffuse tidal features also exist in the optical image. A wide range of multi-frequency radio data is available for this source and they show a turnover at 299 MHz. The high frequency break is at 2.3 GHz and the corresponding spectral age is 8 Myr.

\noindent
{\bf IRAS 17044$+$6720:} The source is a single nucleus and it has  tidal features towards the north west direction. The bright, southern object is a star (Figure \ref{host_galaxy2}) . The flux density at 12.8 GHz is $\sim$3 mJy. The spectral age estimation is not possible because of limited multi-frequency radio data. 

\noindent
{\bf IRAS 23060$+$0505:}  According to SDSS this source is hosted by a compact QSO. The estimated flux density at 1.28 GHz is 8.97 mJy.  Due to unavailability of multi frequency radio data we could not estimate the spectral using the SYNAGE code. 
Normally quasars have a blue-continuum-dominated spectra, but for this object the spectrum is highly obscured. Such
extinction  and  obscuration is similar  to  that  of the red  quasars which represents an early stage of AGN evolution \citep{2014ApJ...789...16N, 2012ApJ...757...51G}. We also note that the nucleus in this galaxy has double peak [OIII] emission lines (Figure \ref{OIII_fig1}, lower panel).




\begin{figure*}
\vbox{
   \hbox{
     \psfig{file=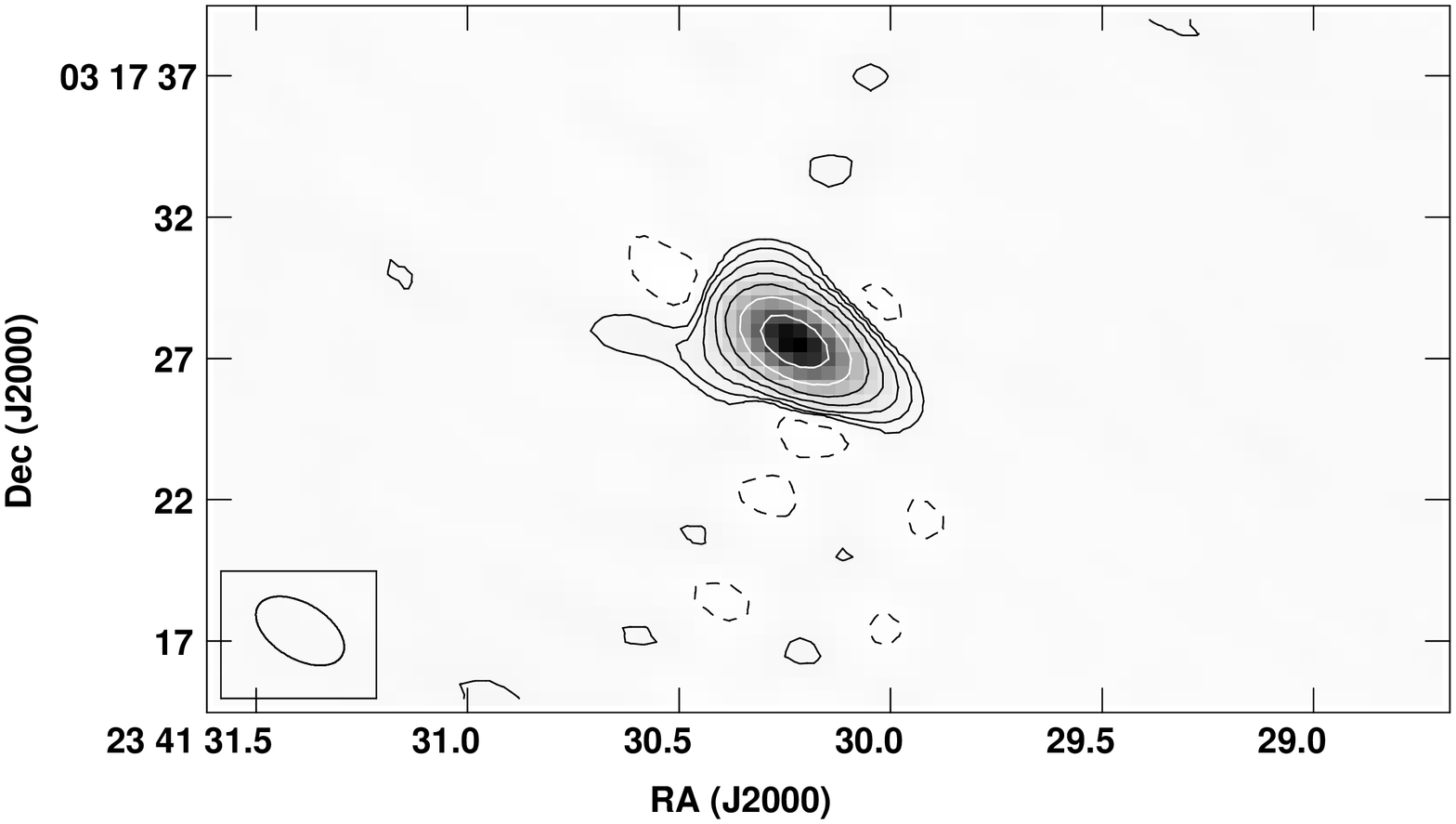,width=2.0in,angle=270}%
 \hskip 0.2cm
 \psfig{file=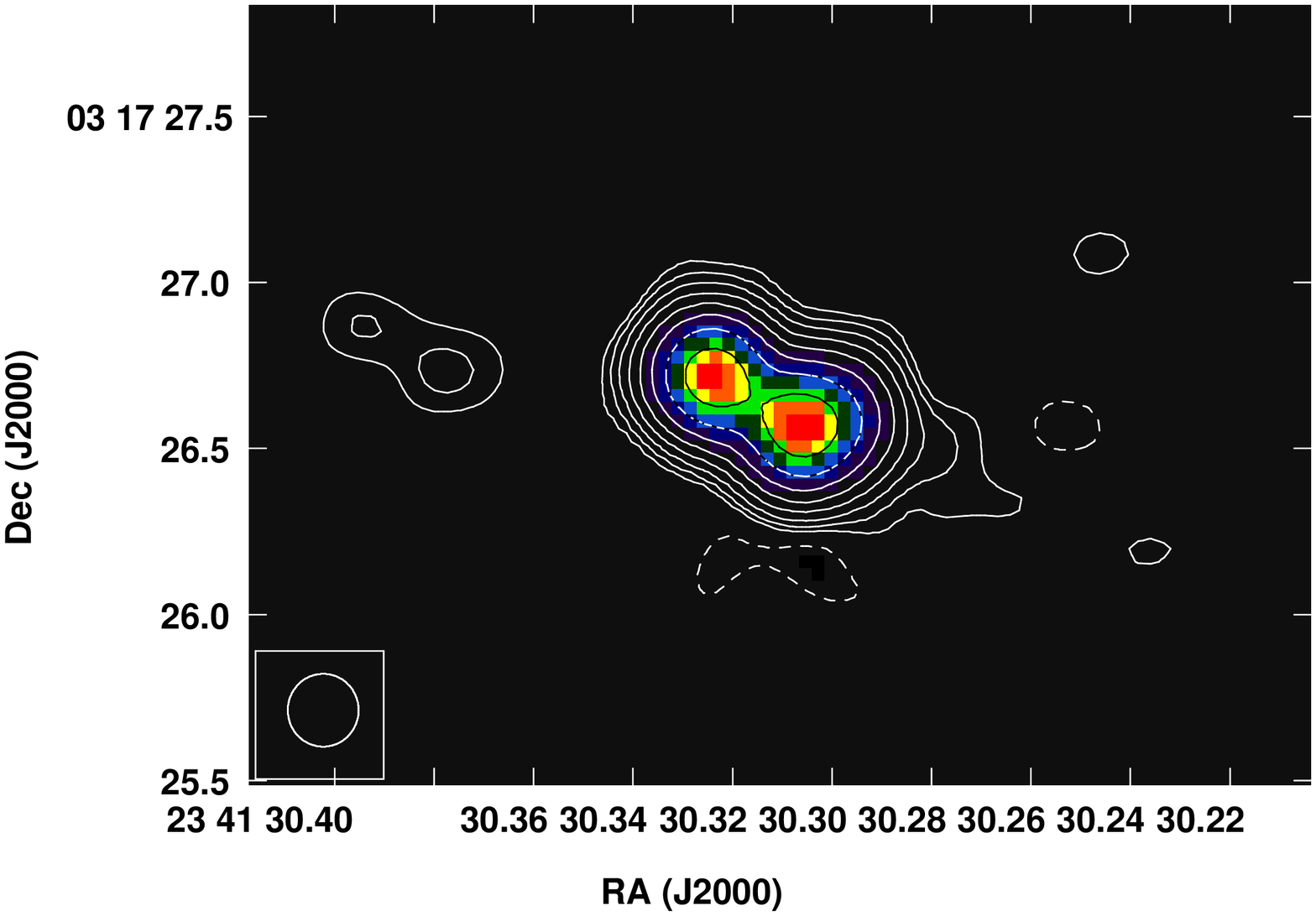,height=2.9in,angle=270}%
        }
}
\caption[]{ Left panel: The GMRT 1.28 GHz image of IRAS 23389$+$0300. The contours are plotted at 3$\sigma$[-1, 1, 2, 4, 8, 16, 32, 64, 128], where the rms noise is $\sigma$=3.0 mJy/beam; Right panel: The VLA 8.4 GHz image of IRAS 23389$+$0300. For this image the contours are plotted at levels of 3$\sigma$[-1, 1, 2, 4, 8, 16, 32, 64, 128], where the rms noise is $\sigma$=0.06 mJy/beam.}
\label{IRAS2338_fig}

\end{figure*}

\begin{figure*}
\vbox{
   \hbox{
     \psfig{file=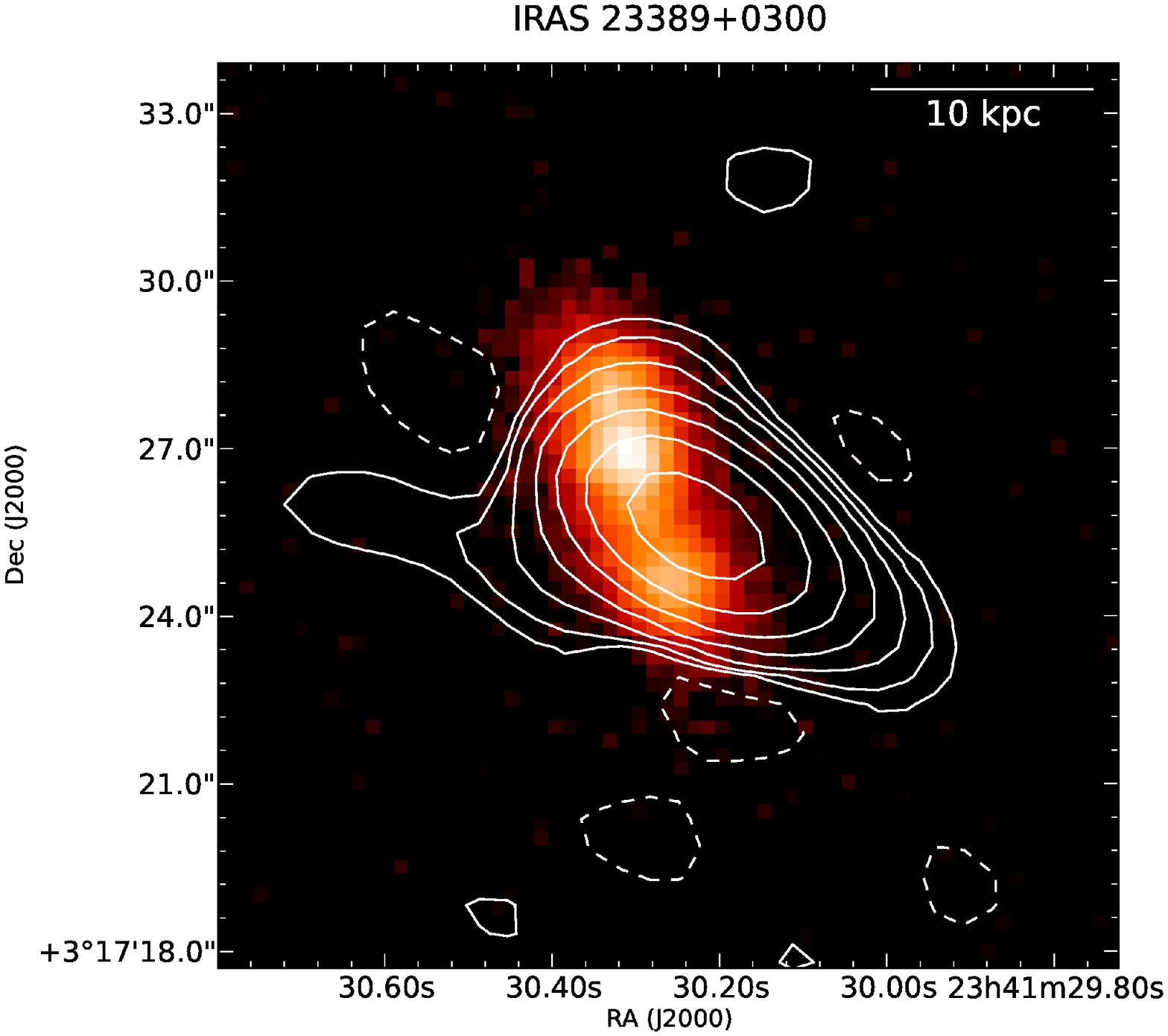,width=2.8in,angle=90}%
 \hskip 0.2cm
 \psfig{file=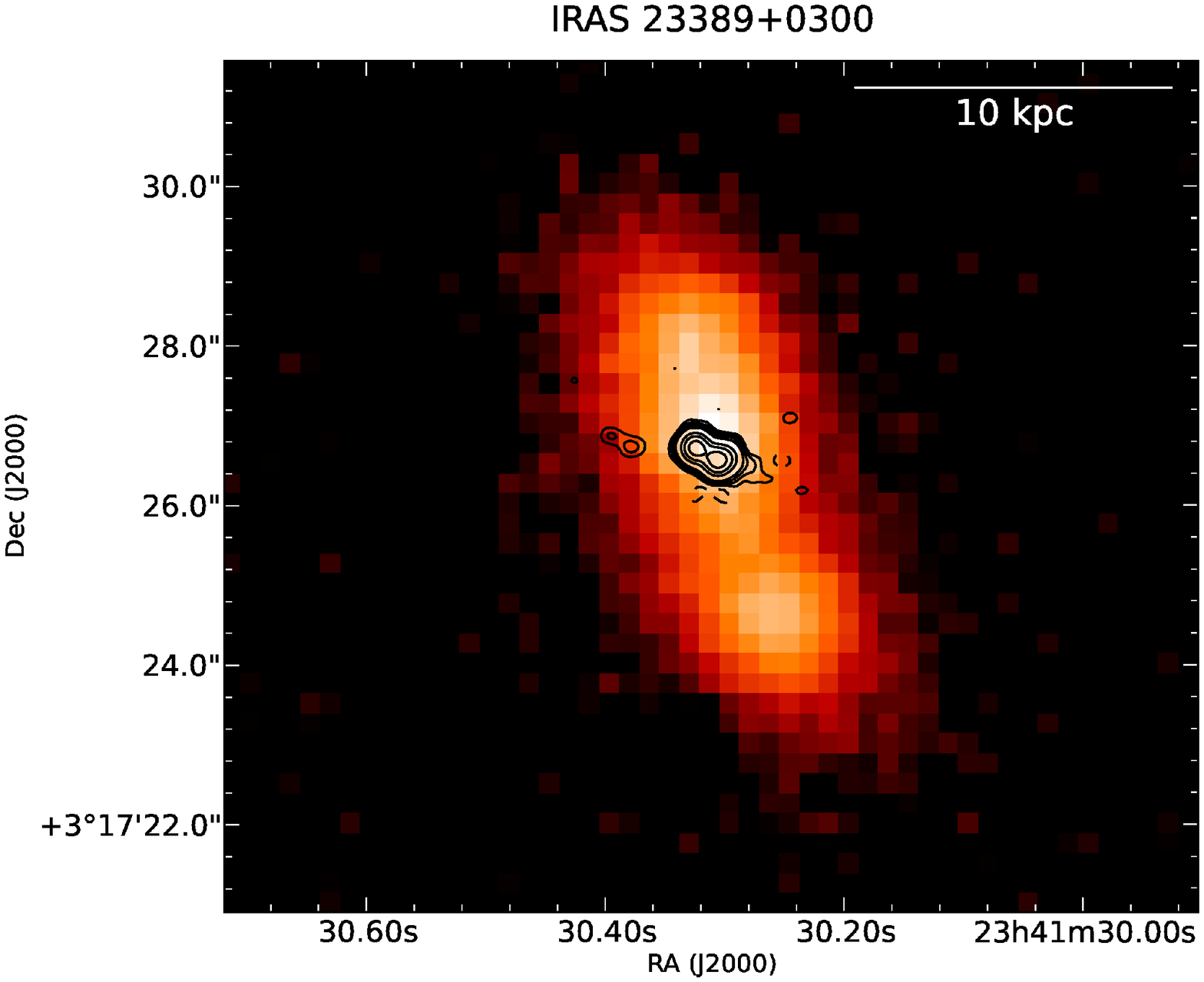,height=3.4in,angle=90}%
         }
}
\caption[]{ Left panel: The GMRT 1.28 GHz contours of IRAS 23389$+$0300 are overlaid on the Pan-STARRS r-band image.
Right panel: The VLA 8.4 GHz contours of IRAS 23389$+$0300 are overlaid on the Pan-STARRS r-band optical image.}
\label{IRAS2338_fig1}

\end{figure*}

 \begin{figure}
\center
\hbox{
 \psfig{file=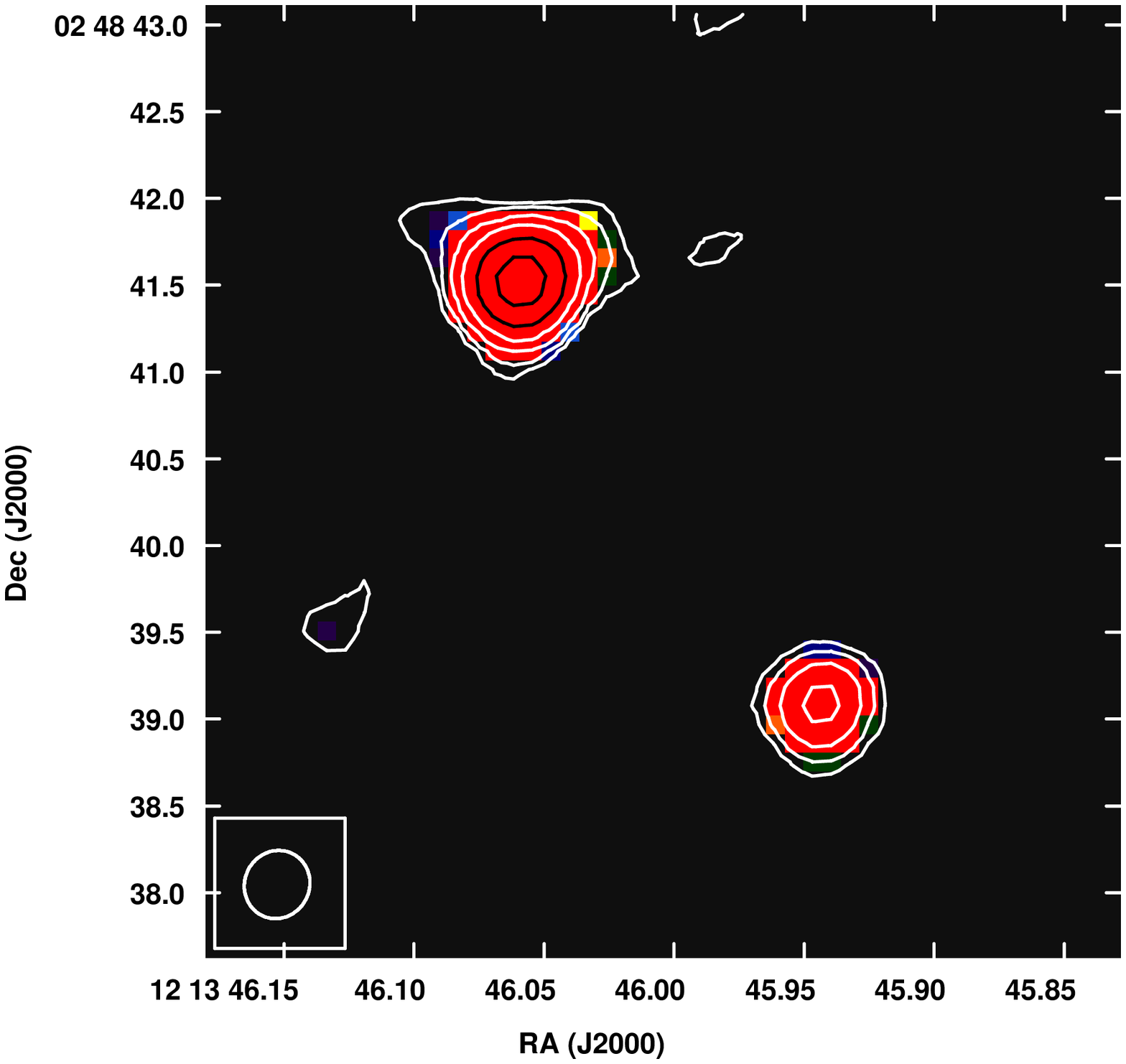,width=7.6cm,angle=0}
     }
\hbox{
 \psfig{file=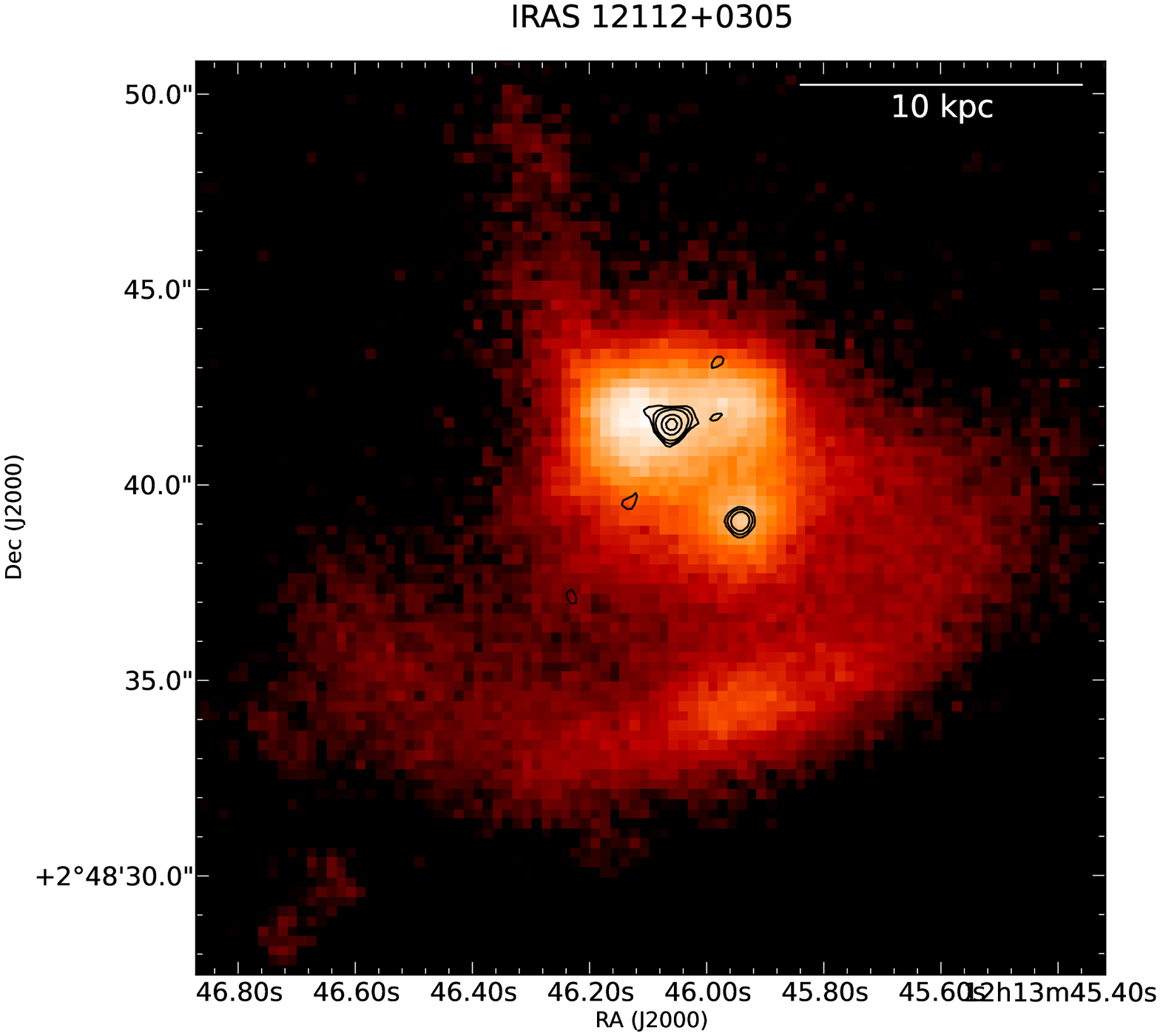,width=9.0cm,angle=0}
     }
\caption[]{Upper panel: The VLA 4.8 GHz image of IRAS 12112$+$0305. The contours are plotted at levels of 3$\sigma$[-1, 1, 2, 4, 8, 16, 32, 64, 128, 256], where the rms noise is $\sigma$=0.06 mJy/beam; Lower panel: The same 4.8 GHz contours are overlaid on the Pan-STARRS r-band optical image.}
\label{IRAS12112_fig1}

\end{figure}
\begin{figure}
\center
\hbox{
\psfig{file=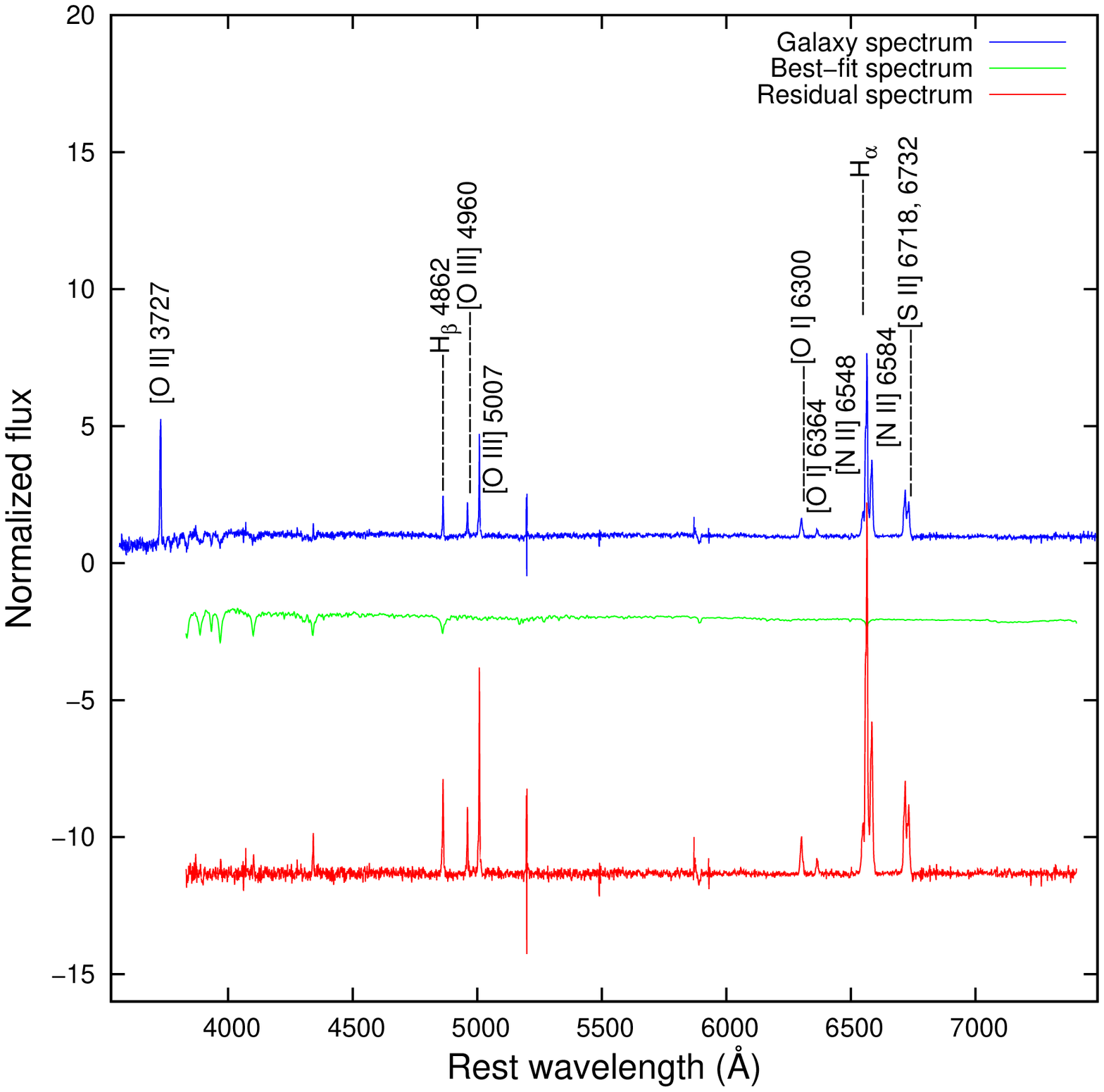,width=7.6cm,angle=-90}
     }
\hbox{
 \psfig{file=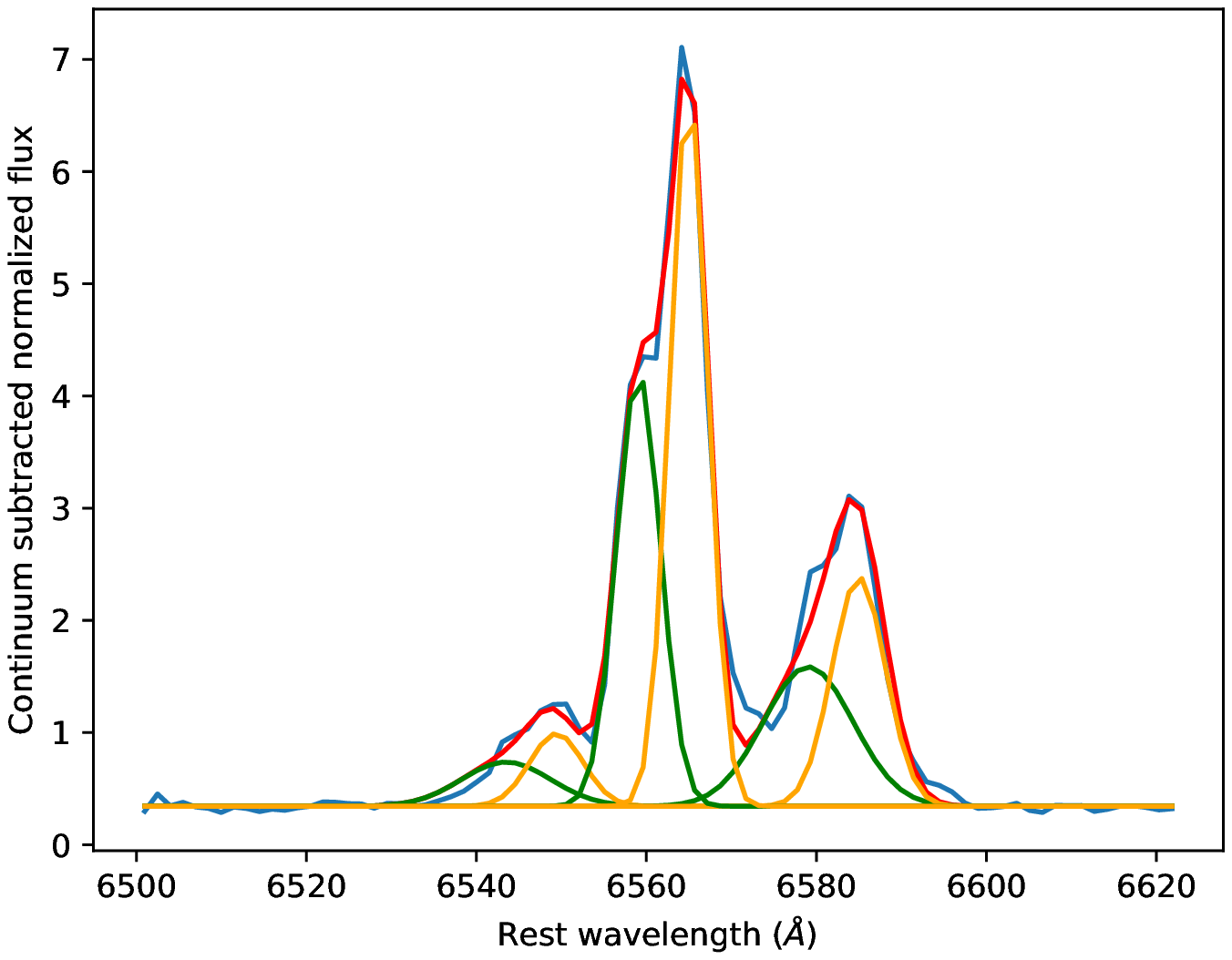,width=9.0cm,angle=0}
      }
  \hbox{
 \psfig{file=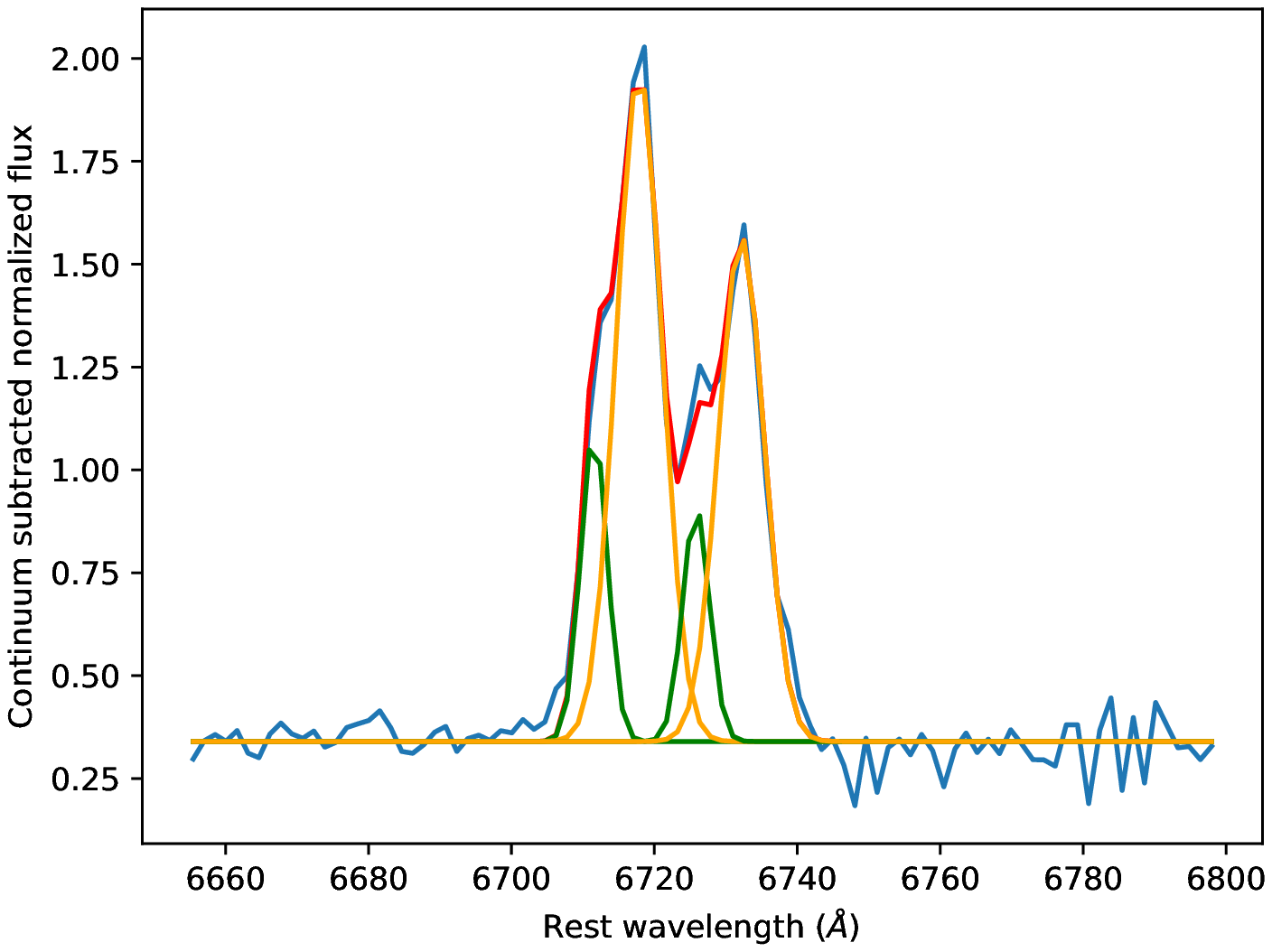,width=9.0cm,angle=0}
      }    
\caption[]{ Upper panel: Redshift-corrected SDSS spectrum of IRAS 12112$+$0305  (in  blue) with the strongest emission lines labelled. The best-fit model spectrum for the underlying stellar population of the host produced by pPXF (in green). The residual spectrum in red. Middle panel: Double peaks in  H$_\alpha$ and [NII] emission lines; Lower panel: Double peaks in [SII] emission line.
For middle and lower panel the green and yellow gaussian fits correspond to two kinematic components.}
\label{IRAS12112_fig2}

\end{figure}
\noindent
{\bf IRAS 23389$+$0300}: The Pan-STARRS r-band optical image of IRAS 23389$+$0300 is shown in Fig. \ref{host_galaxy2}. 
The host galaxy appears to be compact, and more red in color than blue. However, the IR luminosity of IRAS 23389$+$0300 is high. This suggests that the star formation in this ULIRG is heavily enshrouded in dust. The GMRT 1.28 GHz image (Fig. \ref{IRAS2338_fig}, left panel)  does not show any extended  feature for  IRAS 23389$+$0300. The VLA 8.4GHz radio image 
(Fig. \ref{IRAS2338_fig}, right panel) shows the resolved double lobe structure of IRAS 23389$+$0300. The low frequency and high frequency radio map of IRAS 23389$+$0300 are overlaid on the optical image in Fig. \ref{IRAS2338_fig1}.
\citet{2003A&A...409..115N} also reported the existence of twin radio lobes separated by 0.7 kpc for this ULIRG. Fig. \ref{int_spect} shows the integrated radio spectrum using our measurements and 
the measurements from the literature (See Table \ref{table3} ). We find that the low frequency turn off is at 181 MHz and the break frequency is equal to 2.69 GHz. 
The estimated spectral age is equal to 0.66 Myr. 


\noindent
{\bf IRAS 12112$+$0305:} This is an interacting system of two spiral galaxies \citep{2009MNRAS.400.1139R}. The optical observations reveal a pair of well separated nuclei and a pair of tidal tails for this ULIRG. The northern nucleus is much brighter than the southern one (Fig. \ref{host_galaxy2}). The northern tidal tail is 18~kpc in size. Another 30 kpc arc like tail extends along the southern  
part \citep{2000ApJ...529..170S}. 
 The higher frequency radio image reveals a double structure for this source (Fig. \ref{IRAS12112_fig1}, upper panel).
The separation between the two radio components is $\sim$ 4.4~kpc.  We note that they coincide with the optical nuclei (see Fig. \ref{IRAS12112_fig1}, lower panel). The optical position is within $\sim$0.8$\arcsec$ of the radio peak position. We have estimated the spectral indices for the two radio components 
using the radio images at 1.4~GHz, 4.5~GHz and 14~GHz. To estimate the spectral indices for both components we convolved all three maps to a common resolution of 1.51$\arcsec$. The spectral index for the northern component is 0.58$\pm$0.11, while for the southern component it is 0.80$\pm$0.01. We also studied the SDSS optical spectrum of IRAS 12112$+$0305 (see Fig. \ref{IRAS12112_fig2}, upper panel). For this source both the H$\alpha$, [NII] lines as well as the [SII] doublet have double peaked emission lines (see Fig. \ref{IRAS12112_fig2}, middle and bottom panels). For IRAS 12112$+$0305, the integrated radio spectrum 
(Fig. \ref{int_spect}) shows a low frequency turn over at $\sim$ 296~MHz while the break frequency is very high.
Here the integrated flux densities of this source are available up to $\sim$ 14.5~GHz, so we consider the lower limit of the break frequency to be $>$14.5~GHz. Hence we note that the upper limit of the age is 2.86~Myr.     

\noindent
{\bf IRAS 17179$+$5444:} The Pan-STARRS r-band optical image of this ULIRG shows that the nucleus has extended emission which is most likely tidal in origin and associated with the merging process \citep{2002ApJS..143..315V} (Fig. \ref{host_galaxy3}). We did not get any GMRT observation time for this source. However, we used a large range of multi-frequency archival data to study its radio properties. There is no outflow signature in the TGSS map and there is no report of any double nuclei at 15~GHz by \citet{2003A&A...409..115N}, which could be a remnant of the merging process. In the integrated radio spectral energy distribution (Fig. \ref{int_spect}) we note a low frequency turnover similar to that observed in CSS sources. The estimated turnover frequency is 519~MHz while the break frequency is 5.29~GHz. IRAS 17179$+$5444 is also unresolved at 15~GHz, hence the source size must be less than the resolution of the 15~GHz VLA image. The estimated source age is 1.4~Myr.       

\begin{table*}
   \caption{Estimates of break frequency, spectral age turnover frequency and radio power}
   \begin{tabular}{cccccccccc}
  \hline
Object &  $\alpha_{inj}$&$\nu_{\rm br}$&${\chi}^2_{red}$&   B     &$\tau_{\rm syn}$& $\nu_{\rm max}$& $\delta{l}$&$\delta{v}$&P\\
       &                   &(GHz)         &  &(nT)          & (Myr)& (MHz)&  (kpc)& (km s$^{-1}$)& (W Hz$^{-1}$)\\
(1)&  (2)         &(3)            &(4)               & (5)& (6) & (7)& (8)& (9)& (10)\\
 \hline
IRAS 00188$+$0856& 0.533 & 5.38    &1.03 & 2.8        &4.3   &263&   unresolved      & &   23.81\\  
IRAS 01572$+$0009& 0.757 &24.6     &16 & 4.8        &0.87  &&           1.3       &&   24.26\\
IRAS 13305$-$1739& 0.591 & 1.76    &0.95& 1.4       &20    &&     unresolved      &&   24.41\\
IRAS 14070$+$0525$\dagger$&&&&&&&unresolved&&                                            23.59\\
IRAS 14394$+$5332& 0.768 & $>$100  &24 &  4.07      &1.5   &&     unresolved &688$\pm$8.8& 23.87\\ 
IRAS 15001$+$1433& 0.513 & 12.40   &0.28  &10.39    &0.40  &&unresolved&245.5$\pm$17     & 24.04\\
IRAS 16156$+$0146$\dagger$&&&&&&&unresolved&&                                            23.31\\
IRAS 17044$+$6720$\dagger$&&&&&&&unresolved&&                                            23.12\\
IRAS 17028$+$5817& 0.665 & 2.30    &2.01&    2.43    &   8.10   &299&    unresolved       &&       23.44\\            
IRAS 23060$+$0505$\dagger$&&&&&&&unresolved&284$\pm$5.7&                                 23.38\\
IRAS 23389$+$0300& 0.795 &  2.69   &8.47&12.4    &0.66  &182&0.7     &&                 25.70\\
IRAS 12112$+$0305& 0.546 &  $>$100   &3.90&4.34   &$<$1.12  &290&4.4  &111$\pm$24&       23.48\\
IRAS 17179$+$5444& 0.734 &  5.29   &2.29& 5.9    &1.4  &509&unresolved&&                25.26\\  
 
\hline                                              
\end{tabular}                                       
\\                                                  
Column 1: the name of the object. Column 2: the value of the injection spectral index. Column 3: spectral break frequencies. Column 4: reduced ${\chi}^2$ values for the goodness of the fit. Column 5: magnetic fields in nT. Column 6: the synchrotron age in Myr. Column 7: the low frequency turn over. Column 8: distance between two resolved components. Column 9: velocity separation between the two optical double peak emission line components. Column 10: log of radio power at 1.4 GHz. $\dagger$ For these sources we do not have 
good frequency coverage to fit the SYNAGE software. The estimated power law spectral indices are 0.77, 0.30, 0.25, 0.45 for 14070$+$0525, IRAS 16156$+$0146, 17044$+$6720 and 23060$+$0505 respectively.

\label{table4}
\end{table*}
\section{DISCUSSION}

\noindent
{\bf GPS, CSS, CSO sources and ULIRGs:}
The ultimate goal of this project is to study ULIRGs at low frequencies and identify any signatures of core-jet structures or extensions. This can help determine whether there is an underlying evolutionary connection between ULIRGs and  GPS/CSS/CSO sources. To accomplish this we have studied how the radio flux density of a sample of ULIRGs is distributed over a wide range of frequencies and tried to estimate the break and turn-over frequencies. We have also examined the resolved radio morphologies at several frequencies of a few sources using both new observations and archival data. The derived spectral break frequencies, spectral ages, low frequency turn overs and radio power at 1.4 GHz are given in Table \ref{table4}. 

It is well known that galaxy mergers trigger star formation along with “quasar-mode accretion” onto a central supermassive black hole (SMBH) \citep{2007MNRAS.376.1849H, 2016AN....337...36C}. Thus it is not surprising that ULIRGs are often starbursting merger remnants and host strong AGN activity at their centers. There is also evidence that radio galaxies are associated with galaxy mergers \citep{chiaberge.etal.2015}. Hence, ULIRGs may represent the early stages of the evolution of mergers into radio galaxies in the local Universe.  An example of such a ULIRG is F00183-7112. It shows marginal evidence of a turnover at low frequencies which is similar to GPS and CSS sources. The VLBI imaging of this source reveals a young radio galaxy at its center \citep{2012MNRAS.422.1453N}. Furthermore, it is widely accepted that GPS, CSS and CSO sources are the precursors of large radio galaxies. Along this evolutionary track, GPS, CSS and CSO are expected to represent the scaled-down versions of extended radio sources \citep{1995A&A...302..317F, 1999A&A...345..769M}. It is also thought that most radio galaxies start their journey as GPS sources and gradually evolve into CSS/CSO sources, and then finally they may evolve into a large radio galaxy or a radio loud quasar. 
Thus, if the detection of GPS, CSS or CSO sources in ULIRGs is statistically significant, then ULIRGs are very important to study the link between extreme starburst activity, the early stages of AGN evolution and the triggering of radio jets. Also, the association of GPS, CSS and CSO sources with ULIRGs indicates that the progenitors of the low redshift 
large radio sources are possibly ULIRGs.

\noindent
{\bf Timescale for radio jet formation:~}An important question about the evolution of 
ULIRGs into radio galaxies is what is the timescale over which it happens? 
One of the timescales is the SMBHs merger timescale, which can vary over several 
dynamical timescales ($\sim$10$^8$ to 10$^9$yr) and the binary orbit may also 
stall when the SMBH separation reaches a parsec \citep{burke-spolaor.2011}. 
The radio jets may form in one or both merging nuclei, 
a good example being the binary AGN system 0402+379 in which one AGN has a 
double lobed structure while the other one has a single radio core \citep{rodriguez.etal.2006}. 
Alternatively, the radio jets may form only after the SMBHs have merged. 
Radio jets have sizes that extend from a few pc to a Mpc \citep{blandford.etal.2019}. 
The time taken for a jet with a speed of 0.1c to travel 100 kpc is $\sim$10$^6$ to 10$^7$yr, 
although the radio jet may also stall at smaller radii due to inhomogeneities in the 
interstellar or intergalactic medium. Hence, the deciding timescales for the evolution of 
ULIRGs into radio loud galaxies with kpc scale radio jets is the galaxy merger timescales and 
the SMBH in-spiral times in the newly formed galaxy. Since these time scales are shorter compared
to the time available ($\sim$1.3 Gyr) for the ULRIGS in this sample ( z$\sim$0.1) to evolve, there is
no timescale problem for the ULIRGs to evolve into radio galaxies in the local universe.

\noindent
{\bf Spectral age :}
We carried out spectral ageing analysis for 9 ULIRGs. We note that there is a low frequency turn over ($\lsim$500 MHz) for the sources IRAS 00188$-$0856, IRAS 17028$+$5817,  IRAS 23389$+$0300, IRAS 12112$+$0305 and IRAS 17179$+$5444. So these ULIRGs follow a similar trend like CSO sources. We also estimated their spectral ages using the multifrequency radio data. We find that the median spectral age of these sources is 1.4 Myr (Table \ref{table4}). The source IRAS 1305$-$1739 shows relatively high spectral age value. Since the high resolution data is not available for this source, the projected linear size may be overestimated for this case. Overall derived spectral age of these ULIRGs are less than that of the large or relic radio sources with spectral ages 10$^{7}$-10$^{8}$ years \citep{2009BASI...37...63S, 2019MNRAS.486.5158N}. Hence, the spectral ages that we have derived are consistent with the hypothesis that ULIRGs are young radio sources.   


\noindent
{\bf Resolved systems:} 
There are three resolved sources in our study. The first resolved system, IRAS 01572$+$0009, has a two-sided lobe structure with a bright core. This symmetric source is hosted by a quasar and its radio structure is very similar to classical double lobed sources. 
The estimated radio power of this ULIRG is below the break between FRI and FRII sources.
The second ULIRG, IRAS 23389$+$0300, has a symmetric double lobe structure with a low frequency turn over. No core emission was found for this ULIRG. The radio power is above the FRI/FRII break. The radio power and morphology indicates that both of these sources represent scaled-down versions of powerful radio galaxies. For the third resolved system, IRAS 12112$+$0305, we note that both a double lobe structure and a low frequency turnover exist. 
Its radio power lies below the FRI/FRII break. The double structure is asymmetric in nature. The 
brighter northern component has a flatter spectral index than the weaker southern component. The optical data suggests that this source is a double nuclei system. Fig. \ref{IRAS12112_fig1} (lower panel) clearly shows that the radio emission is associated with two star like optical nuclei. It appears to be two interacting nuclei with associated radio emission. So the double radio structure may not represent the early stages of a classical radio galaxy, but instead maybe a dual AGN system which may evolve into a radio galaxy after the SMBHs merge. Such mergers are thought to trigger powerful radio jets \citep{begelman.etal.1980}.  
         
 \noindent
 {\bf Double peaked emission lines:} There is a significant amount of evidence linking double peaked AGN emission (DPAGN) lines with radio loud galaxies \citep{eracleous.etal.2003}. If we assume that ULIRGS are gas rich merger remnants evolving into radio loud galaxies, then it is not surprising that some of them should show DPAGN emission lines in their optical spectra. We studied the host galaxy optical spectra of our sample ULIRGs to investigate their association with DPAGN and double nuclei systems. The derived velocity separation between double peaks resulting from the different fitting procedure are given in Table \ref{table4}
 
 Four ULIRGs in our sample, IRAS 1439$+$5332, IRAS 15001$+$1433 IRAS 12112$+$0305 and IRAS 23060$+$0505 show double peaked optical emission lines. For IRAS 15001$+$1433 and IRAS 12112$+$0305 we note that there are narrow double peaks in the H$_\alpha$, [NII] and [SII] emission lines. Although the [OIII] $\lambda$ $\lambda$5007, 4959 doublet are prominent in both spectra,  they hardly show any splitting. 
 Also, the pPXF estimated stellar velocity dispersion of the host is 237$\pm$22 km s$^{-1}$  and 111$\pm$24 km s$^{-1}$ for the sources IRAS 15001$+$1433 and IRAS 12112$+$0305 respectively.  The redshift-corrected  host galaxy spectra (in blue), their best-fit model spectra for the underlying stellar population produced by pPXF (in green) and the residual spectra with the pure emission lines (in red) are shown in Figs. 
 \ref{IRAS1500_fig2} (upper panel) and \ref{IRAS12112_fig2} (upper panel).
  In  Fig. \ref{IRAS1500_fig2} (middle and lower panel) and \ref{IRAS12112_fig2} (middle and lower panel) the 
 total model flux of each spectral line is shown in red. The left green and right yellow gaussian fits represent the two kinematic components. The estimated mean velocity separations between the two double peaks are 245.5$\pm$17 km s$^{-1}$ and 277$\pm$4.3 km s$^{-1}$ for IRAS 15001$+$1433 and IRAS 12112$+$0305 respectively. The splitting in emission lines may arise from  dual AGN, jet-ISM interaction, outflows or NLR rotating discs. However, in the case of IRAS 12112$+$0305, the SDSS fiber radius of 3$''$ is less than the separation between the two radio cores. So, the line split may not be due to the two radio components. 
 Also the double-peaks are not symmetric, so, the NLR disc scenario is not applicable for this case. Hence, the most probable origin for the double-peaks is outflows for this source. On the other hand, IRAS 15001$+$1433 is unresolved in the 8 GHz map which has a resolution 0.325$ ''\times$0.257$''$, which is much smaller than the SDSS fibre diameter. The source has an extended radio structure at lower frequencies of 1.28 GHz, but its double-peak emission lines are not symmetric which suggests that the DPAGN lines are not due to a rotating disk. Hence, AGN/ star formation outflows or dual nuclei of separation smaller than the 8 GHz beam maybe the origin for the double peaks in IRAS 15001$+$1433.
 
 On the other hand sources IRAS 1439$+$5332 and IRAS 23060$+$0505 show split for [OIII] doublet with outflow signatures. Unlike other two sources these two do not show any splitting for H$_\alpha$, [NII] and [SII] lines. The multi component Gaussian fittings of [OIII] $\lambda\lambda$ 4960, 5007 lines of IRAS 1439$+$5332 and IRAS 23060$+$0505 are shown in Fig. \ref{OIII_fig1}. 
 For both sources the redshift-corrected host galaxy spectra is in blue. The model flux of spectral line is shown in red while the green and yellow color gaussian fits correspond to two kinematic components. For each of the lines the  gaussian with  magenta color represents the outflow associated with one of the corresponding systems. The estimated velocity separations between the two [OIII] double peaks are 688$\pm$8.8 km s$^{-1}$ and 284$\pm$5.7 km s$^{-1}$ for IRAS 1439$+$5332 and IRAS 23060$+$0505 respectively. According to \citet{2002ApJS..143..315V} the optical system of IRAS 1439$+$5332 has two components with a 1.29'' separation which is less than the SDSS fibre diameter. Therefore, the line splitting can be produced by the presence of outflows or dual nuclei. IRAS 23060$+$0505 hosts a compact red quasar which may go through a merging process. So, this line split in this case can also be due to the  outflows or dual nuclei.

\label{sec:discussion}
\section{CONCLUSIONS} 
\begin{enumerate}
\item
In this study we have examined the radio spectral energy distribution of ULIRGs and examined their optical spectra when available.
The source IRAS 15001$+$1433 shows minor extended radio emission. This emission may come from a core-jet radio structure or outflows. The radio powers of these sources at 1.4 GHz are  close to the radio power that separates the FRI and FRII sources. 
\item 
Using archival data as well as current observations it was possible to study the integrated radio spectra for 9 sources. We identified the low frequency turn over for the sources IRAS 00188$-$0856, IRAS 1728$+$5817, IRAS 23389$+$0300, IRAS 12112$+$0305 and IRAS 17179$+$5444. Our spectral ageing analysis shows that the sources are younger than the extended large radio sources or remnant radio sources. 
\item
Archival VLA high frequency data revealed classical double structure for 3 sources (IRAS 01572$+$0009, IRAS 23389$+$0300 and IRAS 12112$+$0305). The separation between two components vary from $\sim$ 0.74 - 4.4 kpc. So, our study indicates that the spectral age, spectral shape and radio morphology of these ULIRGs are similar to CSS, GPS and CSO sources. For these sources ``youth'' hypothesis is more consistent than the ``frustration'' scenario \citep{2015ApJ...809..168C}. 
\item 
We identified double peak emission lines with the sources IRAS 14394$+$5332, IRAS 23060$+$0505, IRAS 15001$+$1433 and IRAS 12112$+$0305. For 14394$+$5332, IRAS 23060$+$0505 and IRAS 15001$+$1433, outflows or dual nuclei may cause the line splitting. While, IRAS 12112$+$0305 hosts optically resolved system of two AGNs. Only the northern bright  component shows the line splitting. Hence, most possible reason for the double peak in IRAS 12112$+$0305 is outflow. 

\item
In future we will propose to map the low and high frequecny radio continuum emission from a large sample of ULIRGs that are bright in NVSS and TGSS surveys. This study will help us to further understand their evolution into radio loud galaxies.

\end{enumerate}

\label{sec:Conclusions}
\section*{Data Availability}

The data underlying this article can be shared on reasonable request to the corresponding author. The GMRT raw data can be obtained from the following link: https://naps.ncra.tifr.res.in/goa/data/search. 
\section*{Acknowledgments}

The authors would like to thank an anonymous referee for constructive comments and insightful suggestions which have helped to improve this manuscript. SN acknowledges support by the Science $\&$ Engineering Research Board, a statutory body of Department of Science $\&$  Technology (DST), Government of India (FILE NO. PDF/2018/002833). We acknowledge the support of the Department of Atomic Energy, Government of India, under the project 12-R\&D-TFR-5.02-0700. We thank the GMRT staff for technical support during the observations. GMRT is run by the National Centre for Radio Astrophysics of the Tata Institute of Fundamental Research.  This research has made use of the NASA/IPAC Extragalactic Database (NED) which is operated by the Jet Propulsion Laboratory, California Institute of Technology, under contract with the National Aeronautics and Space Administration.  Funding for the SDSS and SDSS-II has been provided by the Alfred P. Sloan Foundation, the Participating Institutions,
the National Science Foundation, the U.S. Department  Energy, the National Aeronautics and Space Administration, the Japanese Monbukagakusho, the Max Planck Society, and the Higher Education Funding Council for England. The National Radio Astronomy Observatory is a facility of the National Science Foundation operated under cooperative agreement by Associated Universities, Inc. AIPS is produced and maintained by the National Radio Astronomy Observatory, a facility of the National Science Foundation operated under cooperative agreement by Associated Universities, Inc.
\appendix
\section{Journal abbreviations}
\label{sec:abbreviations}
\begin{figure*}
\vbox{
   \hbox{
      \psfig{file=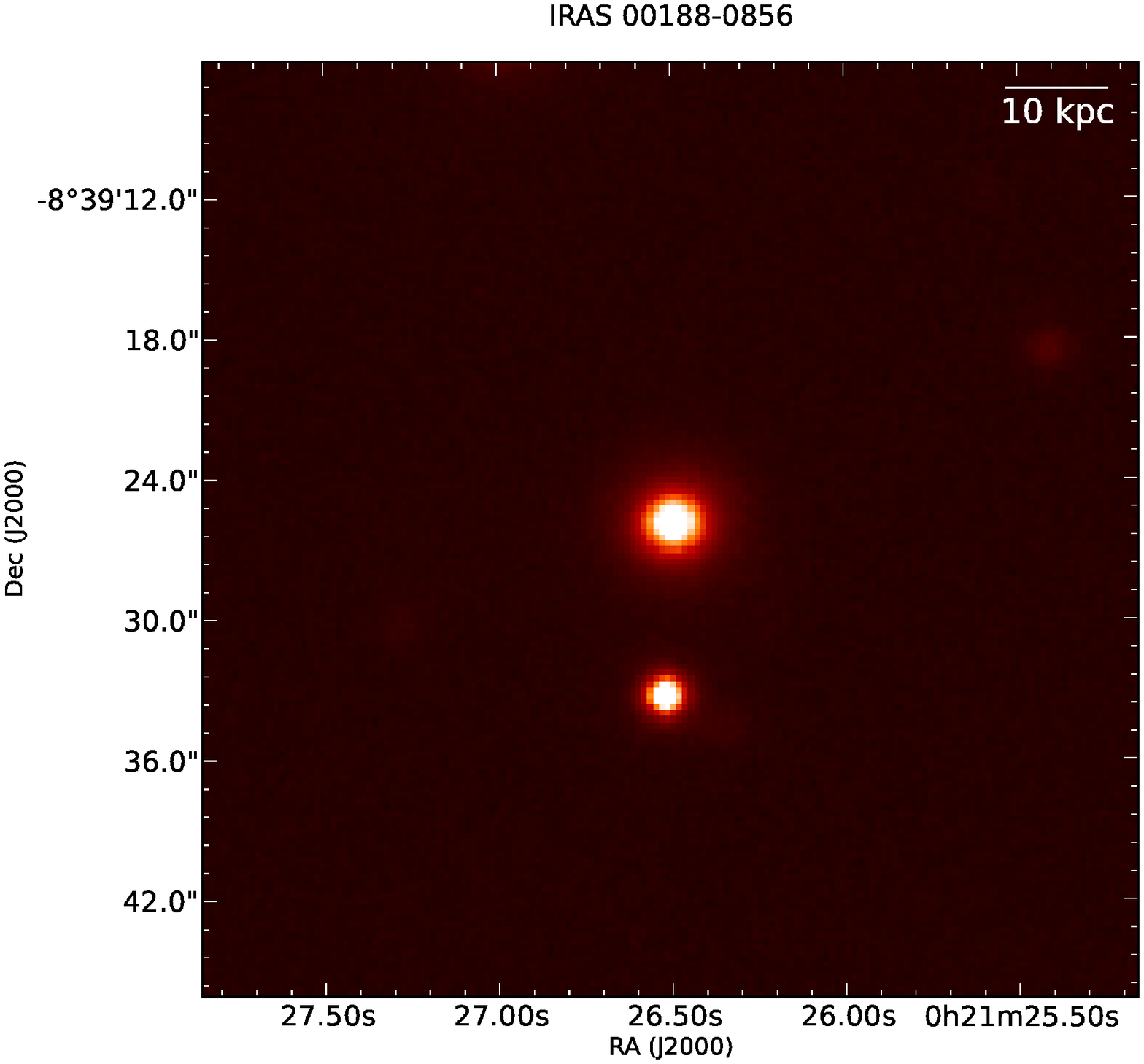,width=2.4in,angle=90}%
     
      \psfig{file=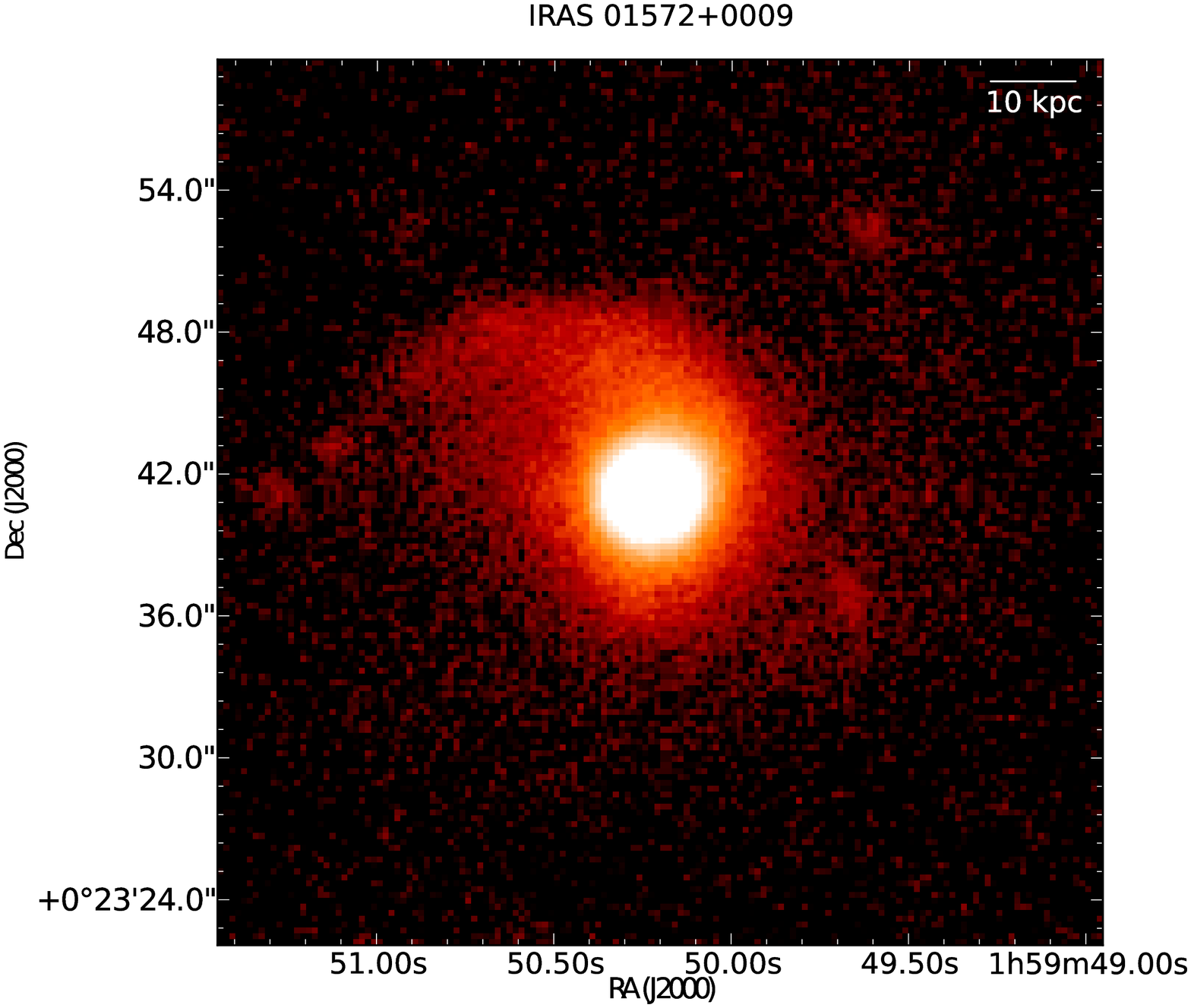,width=2.8in,angle=0}%
            }
  \vspace{0.1cm}
 \hbox{
      \psfig{file=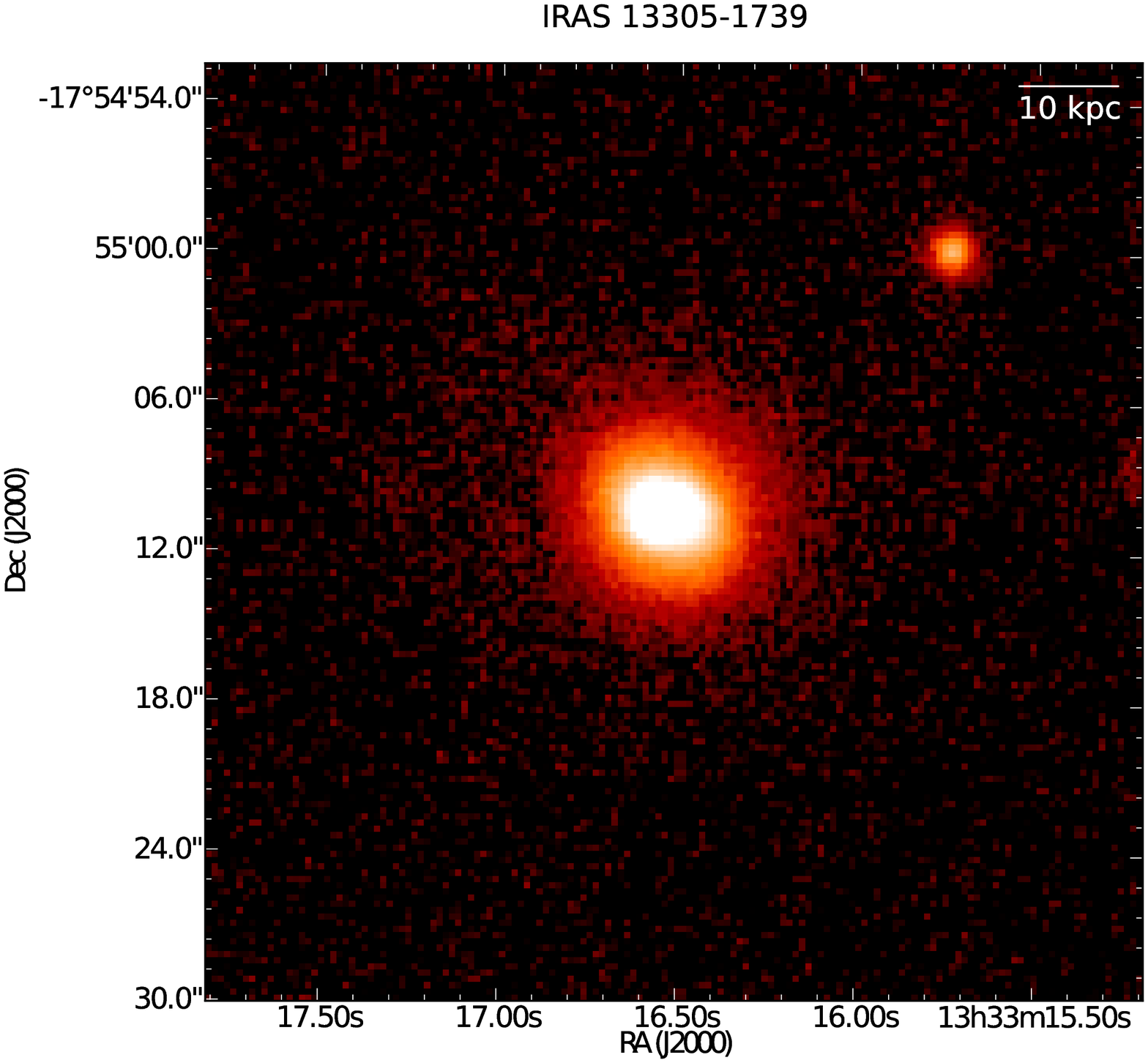,width=2.6in,angle=0}%
      \psfig{file=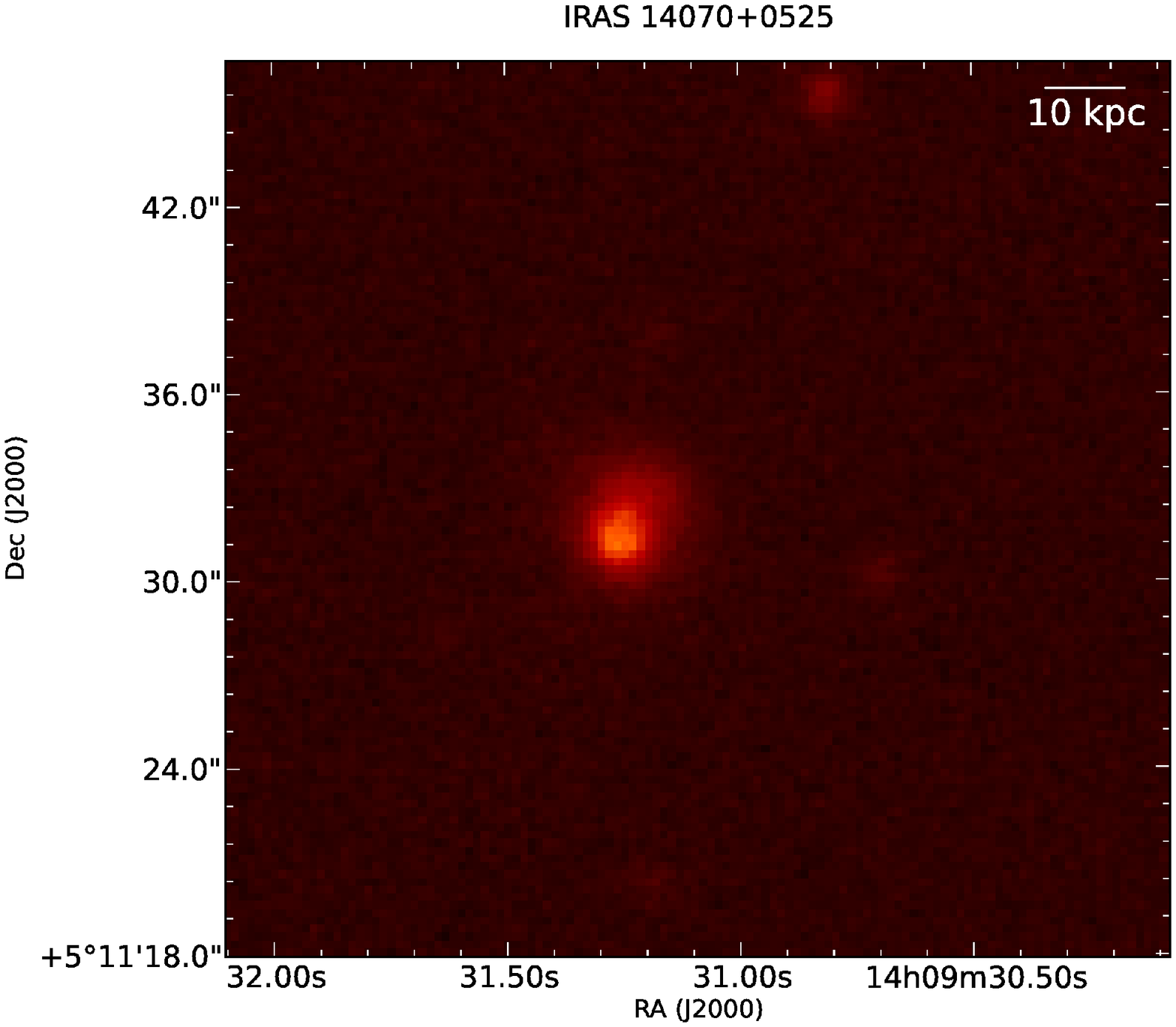,width=2.4in,angle=90}%
                    }
\vspace{0.1cm}
 \hbox{
      \psfig{file=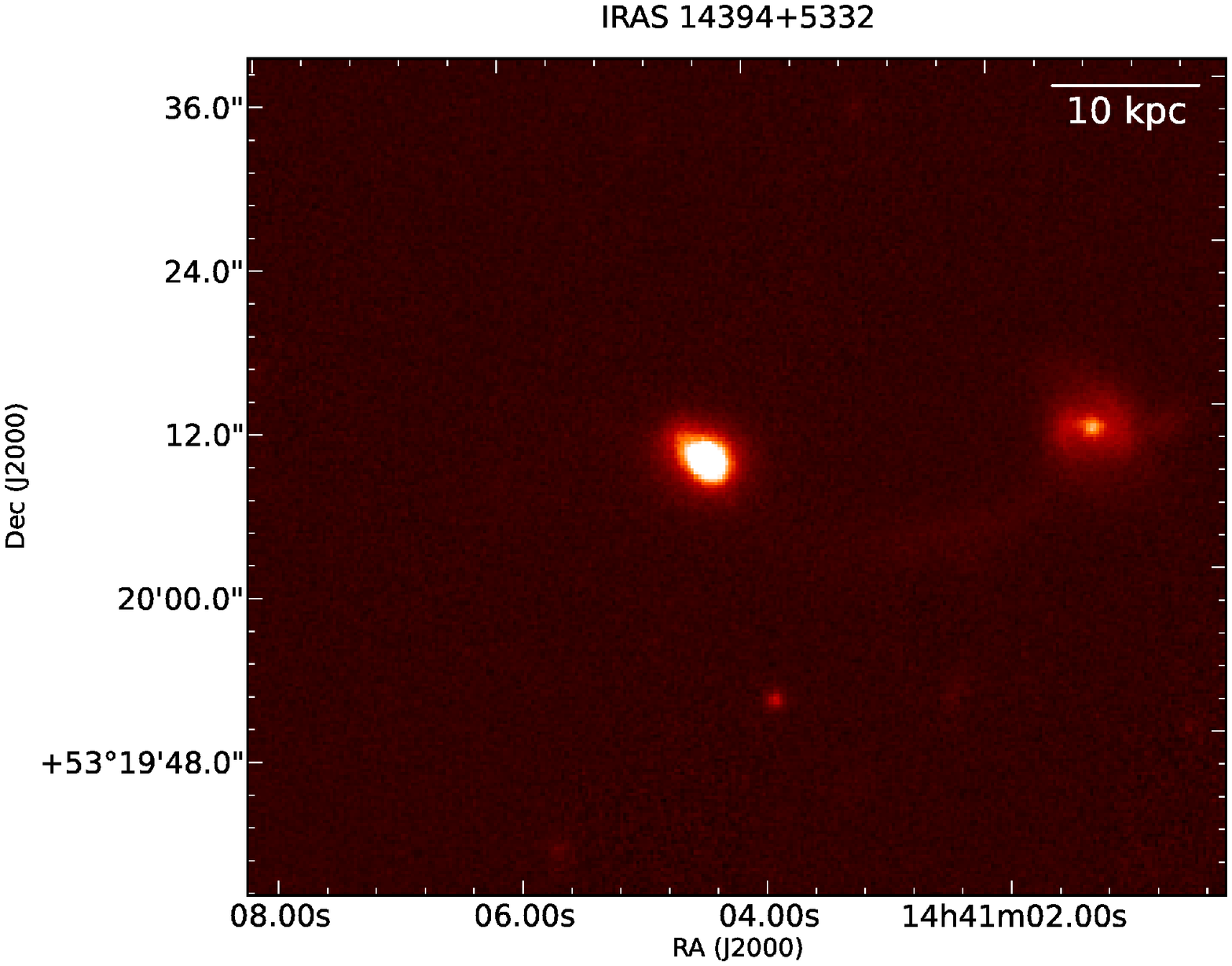,width=2.2in,angle=90}%
      \psfig{file=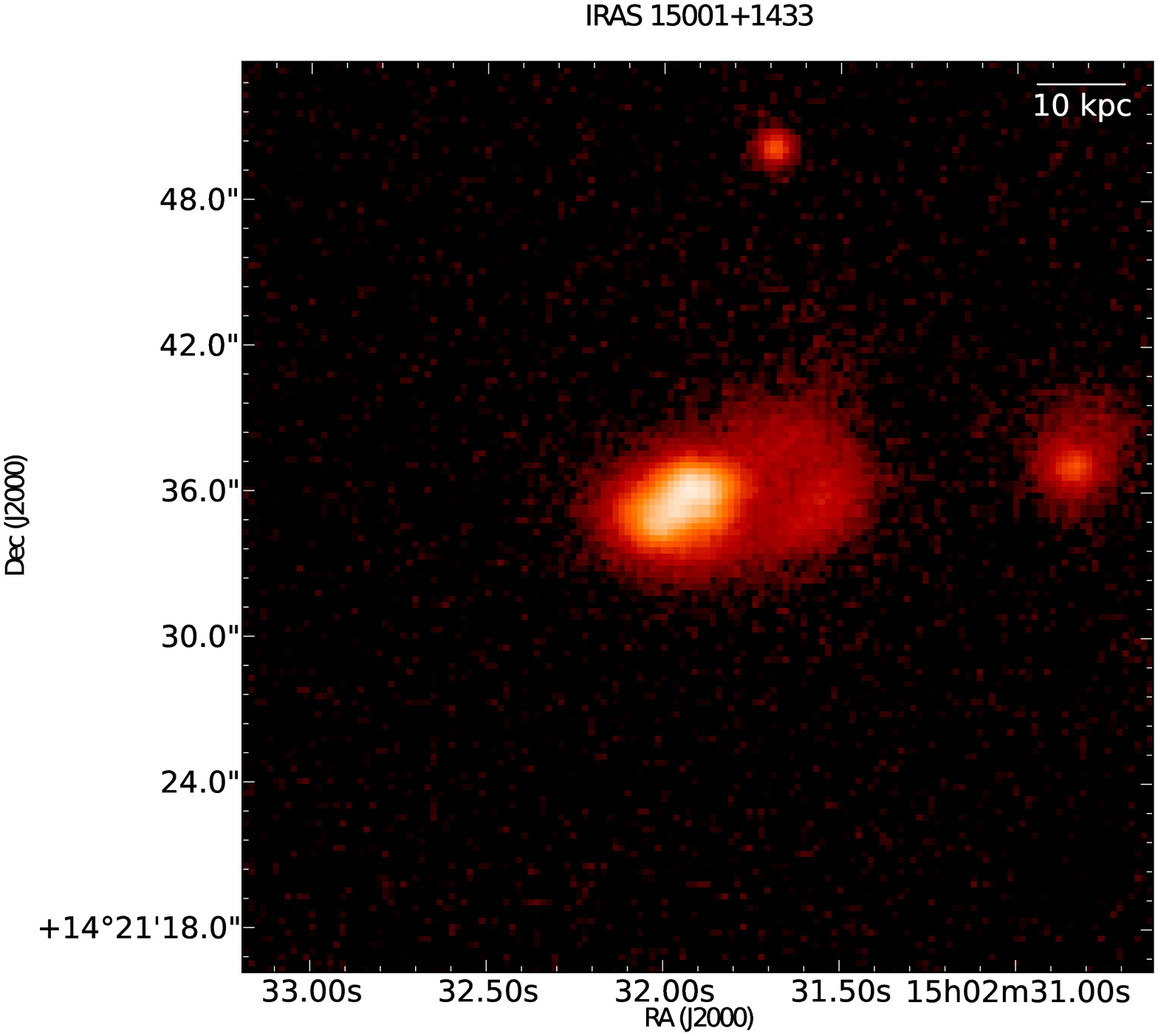,width=2.6in,angle=0}%
       }
}
\caption[] {Optical Pan-STARRS  r band host galaxy images of  ULIRGs.}
\label{host_galaxy1}
\end{figure*}

\begin{figure*}
\vbox{
   \hbox{
      \psfig{file=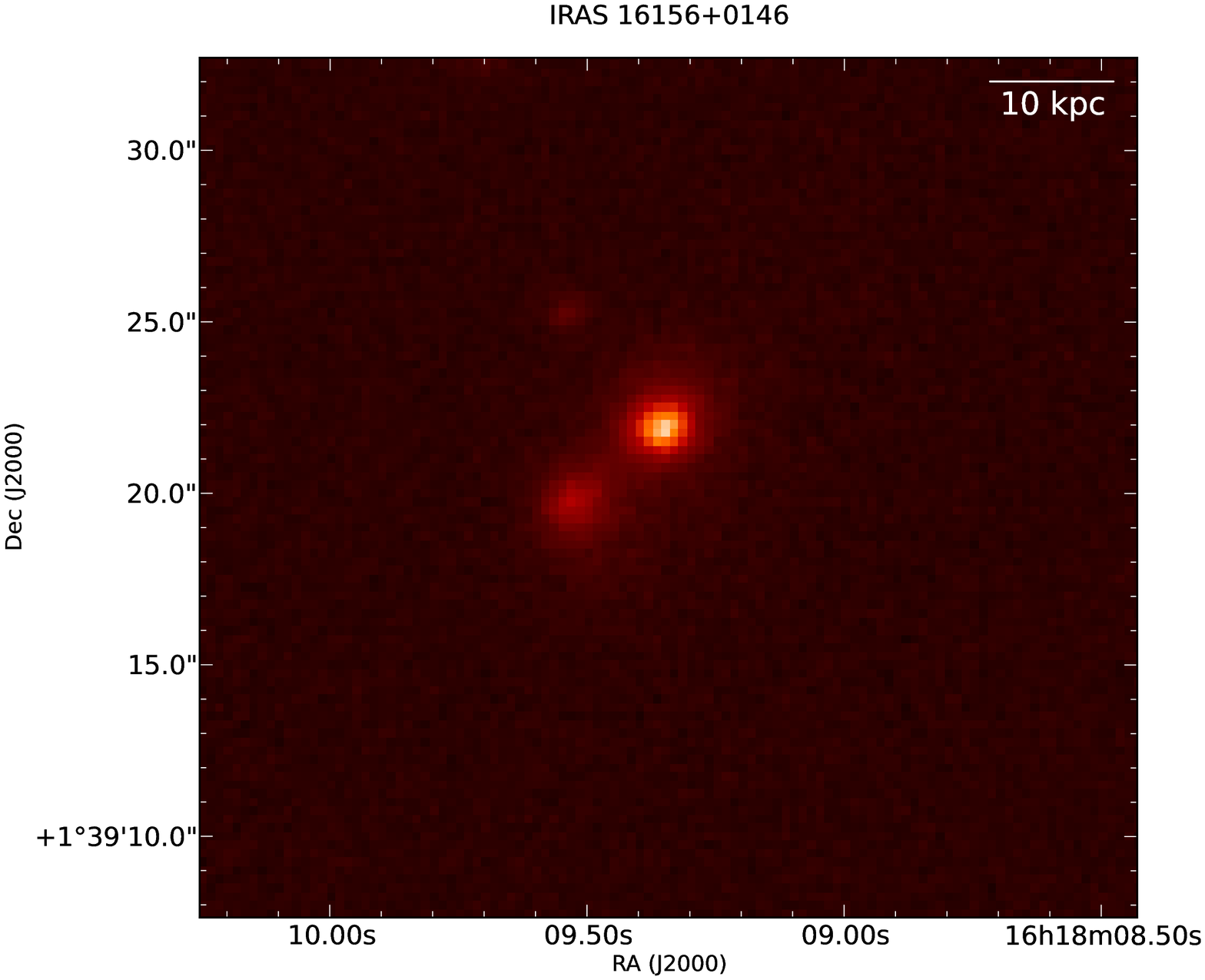,width=2.8in,angle=0}%
     \psfig{file=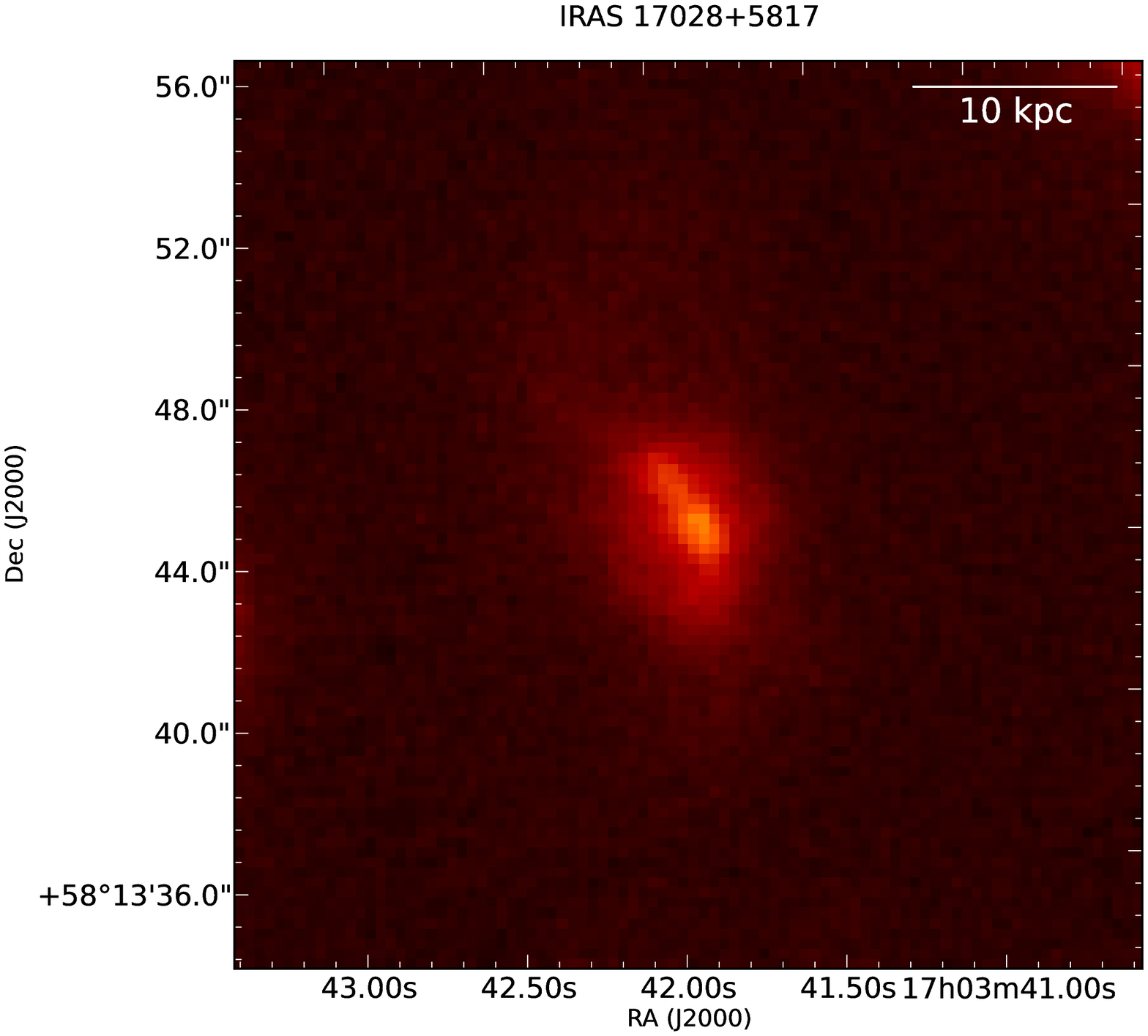,width=2.8in,angle=0}%
      }
  \vspace{0.1cm}
 \hbox{
      \psfig{file=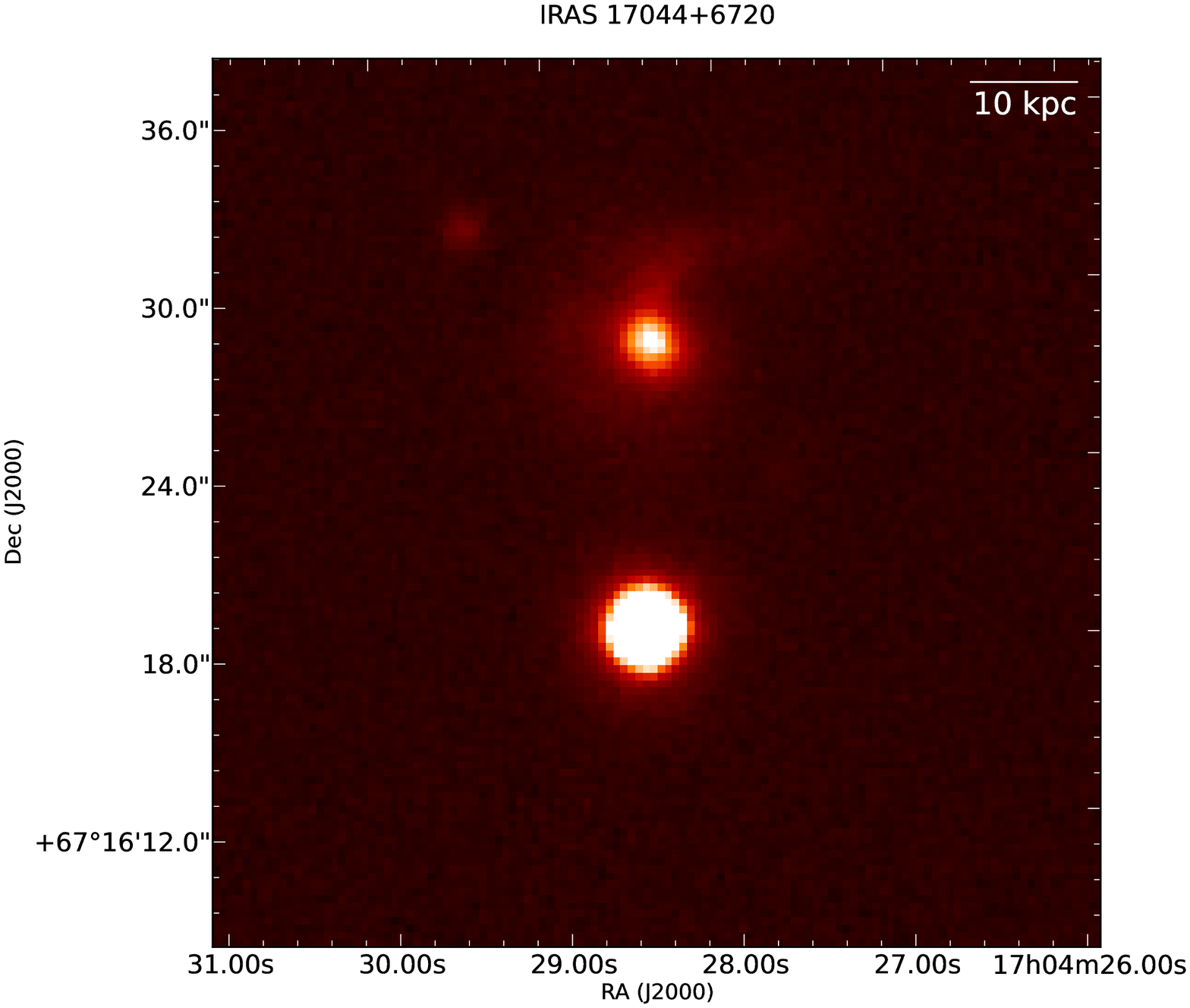,width=2.8in,angle=0}%
      \psfig{file=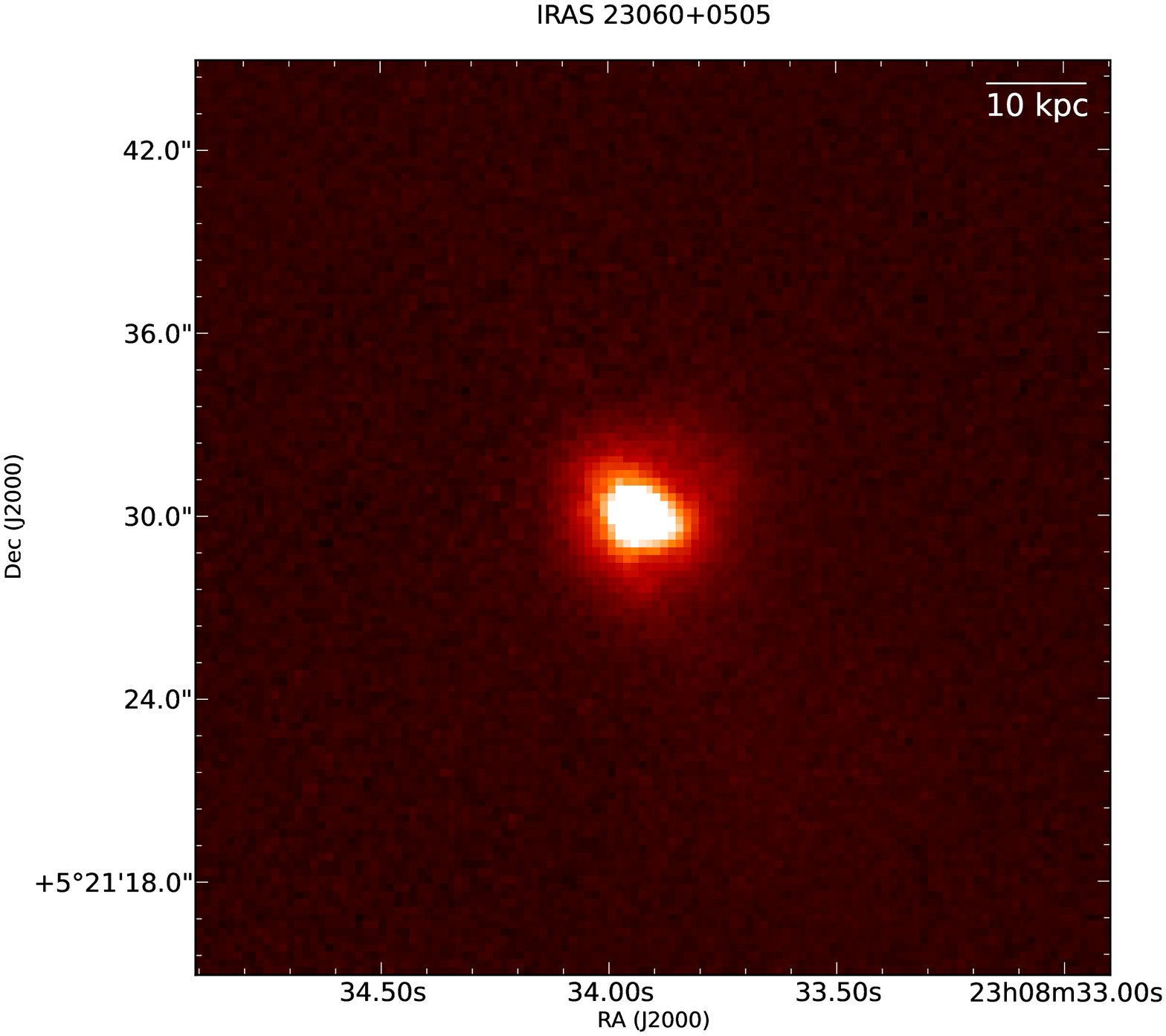,width=2.8in,angle=0}%
                    }
\vspace{0.1cm}
 \hbox{
      \psfig{file=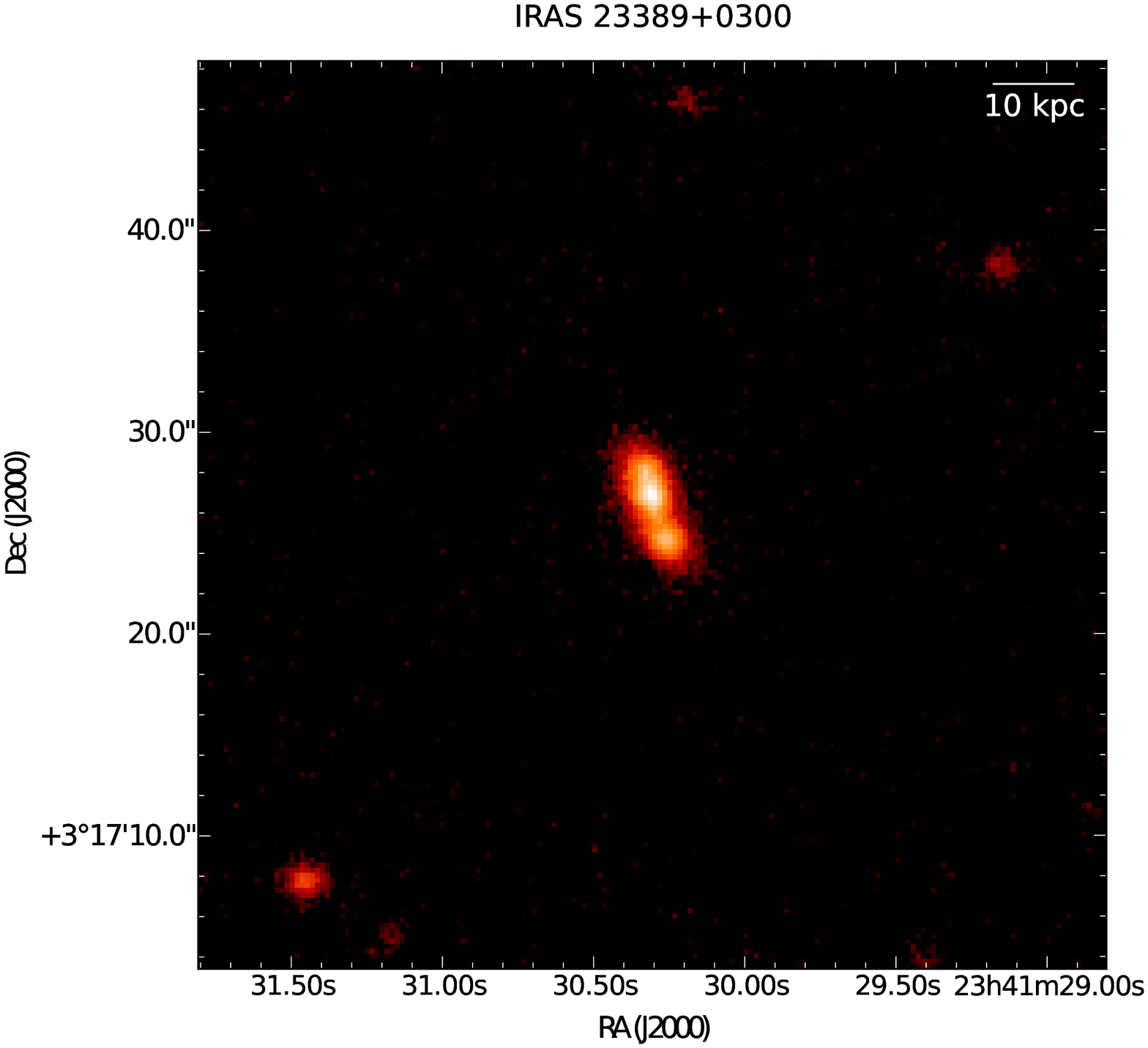,width=2.8in,angle=0}%
      \psfig{file=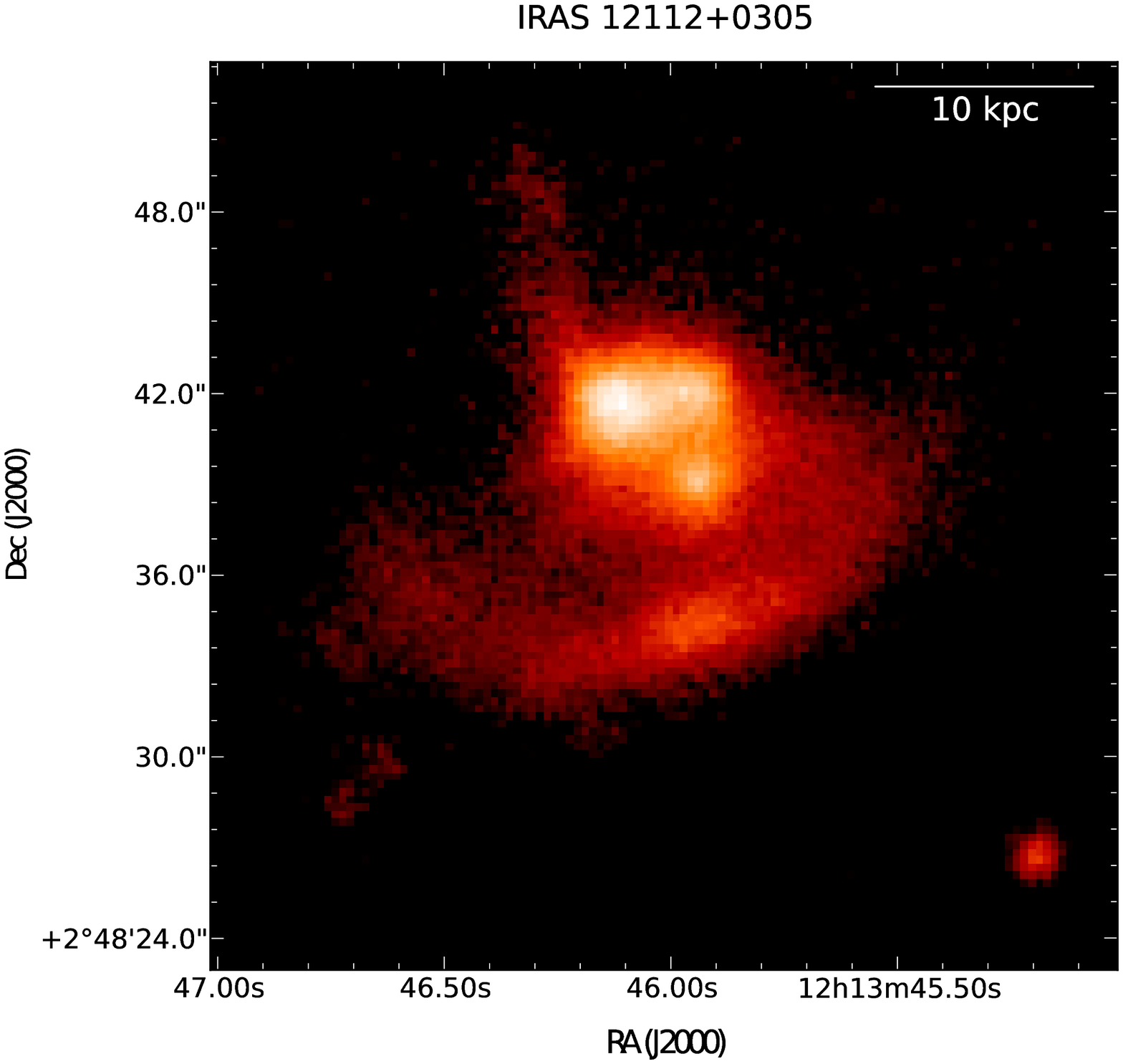,width=2.8in,angle=0}%

     }
}
\caption[]{Optical Pan-STARRS  r band host galaxy images of ULIRGs.}
\label{host_galaxy2}
\end{figure*}

\begin{figure}
\vbox{
  \hbox{
      
     \psfig{file=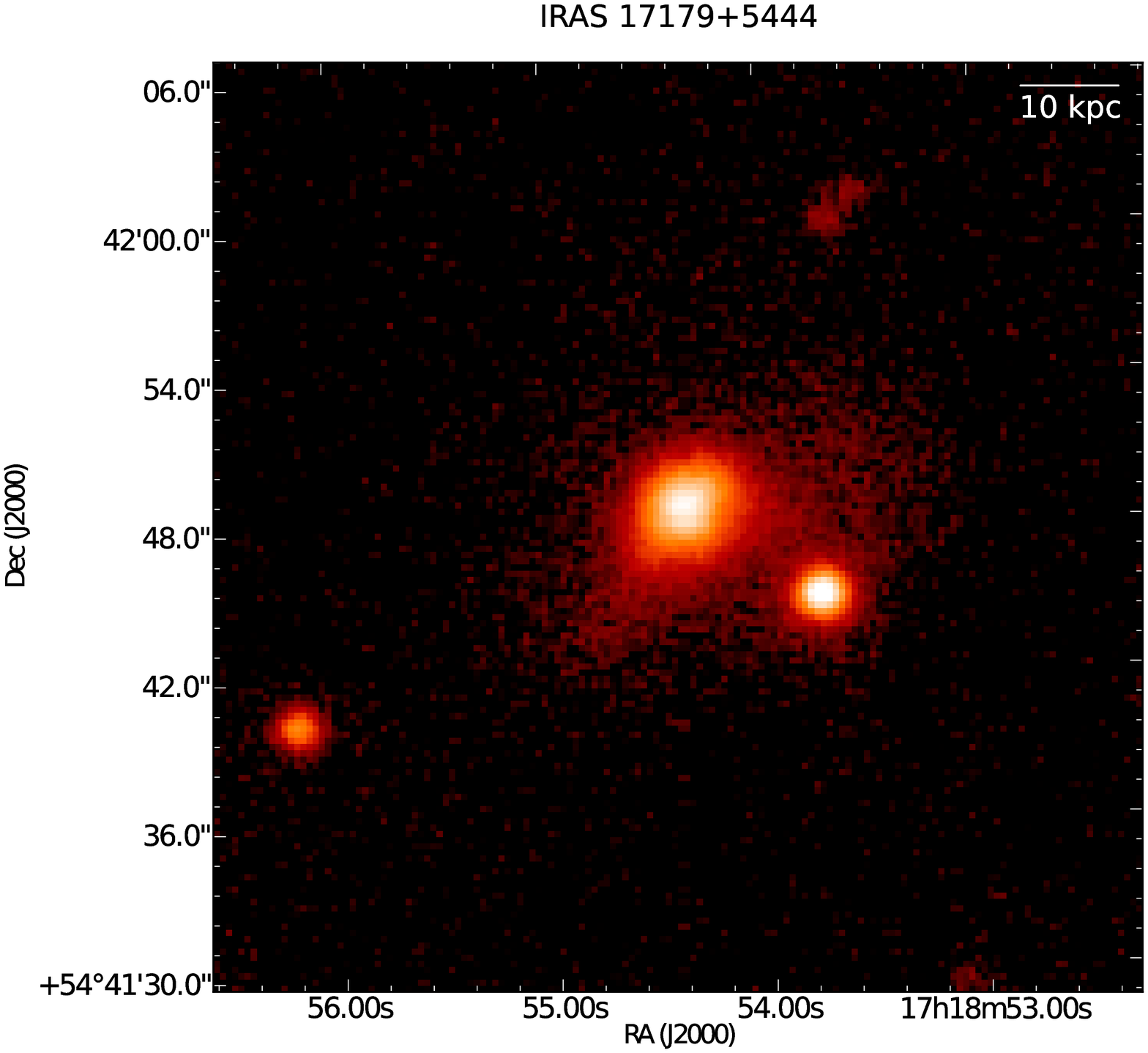,width=2.8in,angle=0}%

     }
}
\caption[]{Optical Pan-STARRS  r band host galaxy images of ULIRG.}
\label{host_galaxy3}
\end{figure}

Pan-STARRS  r band host galaxy images of all ULIRGs are presented in Fig. \ref{host_galaxy1}, Fig. \ref{host_galaxy2} and Fig. \ref{host_galaxy3}.






\end{document}